\documentclass[a4paper]{article}

\usepackage[T1]{fontenc}
\usepackage{lmodern}
\usepackage{newtxtext}
\usepackage{url}
\usepackage{booktabs}

\usepackage{graphicx}
\usepackage{xcolor}
\usepackage{float}
\usepackage{amsmath,amssymb,bm,amsthm}
\usepackage{mathtools}
\usepackage{array}
\usepackage{tikz}
\usetikzlibrary{arrows.meta,positioning}
\usepackage{braket}
\usepackage{cite}
\usepackage{comment}
\usepackage{caption}
\usepackage{subcaption}

\usepackage[setpagesize=false, bookmarks=true,
bookmarksdepth=2,
bookmarksnumbered=true,
colorlinks=false,
hidelinks]{hyperref}
\usepackage{doi} 
\usepackage{cleveref} 

\newcommand{\pd}[2]{\dfrac{\partial {#1}}{\partial {#2}}}
\newcommand{\pdd}[2]{\dfrac{\partial^2 {#1}}{\partial {#2}^2}}

\newcommand{\dsum}{\displaystyle \sum}

\newcommand{\pr}[1]{\left ( {#1} \right )}
\newcommand{\br}[1]{\left [ {#1} \right ]}
\newcommand{\brc}[1]{\left \{ {#1} \right \}}

\newcommand{\expec}[2]{\mathbb{E} _{#1} \left [ {#2} \right ]}

\newcommand{\KL}[2]{\mathrm{KL} ( {#1} || {#2} ) }
\newcommand{\tra}[1]{ {#1} ^{\top} }
\newcommand{\trace}[2]{ \mathrm{Tr}_{#1} \left[ #2 \right] }

\theoremstyle{definition}
\newtheorem{dfn}{Definition}[section]
\newtheorem{result}{Result}[section]

\newtheorem{asmp}{Assumption}[section]

\makeatletter\@addtoreset{equation}{section}

\usepackage[margin=20truemm]{geometry}
\title{Entropy Production for Discrete-Time Markov Processes}

\date{\today}
 
\author{Masanao Igarashi
    \thanks{igarashi.masanao@gmail.com \\ This manuscript is the English version of the author's master thesis submitted to Hokkaido University.}
}

\begin{document}
\maketitle
\thispagestyle{empty}
\begin{abstract}
We study the multiple definitions of the entropy production for discrete-time Markov processes in single systems and composite systems. These definitions have been studied in single systems, but less so in composite systems. With a clear distinction, we review the equivalence condition and the meaning of the multiple definitions and show that all definitions satisfy the important property for the entropy production, such as non-negativity. We also show that the inequalities between total entropy production and marginal entropy production holds for a definition but doesn't for another definition. Furthermore, we verify that fact by calculating entropy productions for Gaussian process and numerically show the result. Finally, we find appropriate use of each definition taking all results into account.
\end{abstract}

\tableofcontents

\section{Introduction}
\label{intro}
Thermodynamics deals with the equilibrium states of macroscopic systems consisting of many components (e.g., gases and liquids) and the transitions between these states. A key principle of this field is the second law of thermodynamics, which states that only processes with non-decreasing entropy can take place in isolated systems. This principle also plays an important role in describing non-equilibrium processes in open systems that exchange heat, work, and matter with the outside world, and in closed systems where that exchange heat and work with the outside world. These systems are in contact with the outside (the heat bath), and the entropy production is introduced as a measure of irreversible processes. The entropy production is expressed as the sum of the entropy change of the system and the entropy change of the heat bath in which the system is placed, and the second law of thermodynamics is formulated in such a way that the entropy production takes a non-negative value. For example, in both open and closed systems, the entropy of the system can be kept constant by discharging the entropy produced by irreversible processes to the heat bath and that leads to realizing non-equilibrium steady states. This sort of system forms a so-called dissipative structure, which has been proposed as a basic principle of chemical reactions and biological processes \cite{kondepudiModernThermodynamicsHeat2014}.

In recent years, the field of stochastic thermodynamics has evolved as a way of formulating the thermodynamic quantities of non-equilibrium processes within a framework of stochastic differential equations and information theory \cite{schnakenbergNetworkTheoryMicroscopic1976,sekimotoStochasticEnergetics2010,seifertStochasticThermodynamicsFluctuation2012,vandenbroeckEnsembleTrajectoryThermodynamics2015}, and nowadays, the correspondence between the second law of thermodynamics and the irreversibility of time series is clearly recognized. The entropy and heat are defined in a stochastic system based on differential equations such as the Fokker-Planck equation, which describes the probability distribution of the quantity such as position of Brownian particles in a heat bath, or the master equation (which is more generalized). The entropy production is introduced as the overall change of entropy in the stochastic dynamics (entropy change of the system + entropy change of the heat bath in which the system is placed). These studies also show that entropy production is linked to the irreversibility of dynamics based on the so-called fluctuation theorem, and that the second law of thermodynamics holds for expectation values as the non-negativity of entropy production.

At first, these results were limited to systems described by a single random variable (called single systems), but they have also been extended to subsystems of multiple interacting systems described by multiple random variables (called composite systems). The stochastic thermodynamics of subsystems is specifically referred to as information thermodynamics \cite{parrondoThermodynamicsInformation2015}, and is best exemplified by Maxwell's demon and the derivation of second-law-like limits for systems that are adaptive to time-varying environments, including living organisms \cite{sagawaGeneralizedJarzynskiEquality2010,stillThermodynamicsPrediction2012,baratoEfficiencyCellularInformation2014,brittainWhatWeLearn2017,dianaMutualEntropyProduction2014,goldtStochasticThermodynamicsLearning2017,goldtThermodynamicEfficiencyLearning2017,hartichStochasticThermodynamicsBipartite2014,hartichSensoryCapacityInformation2016,horowitzThermodynamicsContinuousInformation2014,horowitzSecondlawlikeInequalitiesInformation2014,matsumotoRoleSufficientStatistics2018,sartoriThermodynamicCostsInformation2014,horowitzMultipartiteInformationFlow2015}. These results have also been verified experimentally due to advances in single-molecule manipulation techniques \cite{cilibertoExperimentsStochasticThermodynamics2017}.

These advances in stochastic thermodynamics have led to the idea that entropy production is a useful indicator of irreversibility in stochastic dynamics. In recent years, it has also been applied to the stochastic dynamics of information-theoretic systems that are not in contact with a heat bath and do not exchange energy, such as neural networks \cite{cofreInformationEntropyProduction2018,lynnBrokenDetailedBalance2021} and the analysis of time series data \cite{roldanEntropyProductionKullbackLeibler2012,auconiInformationThermodynamicsTime2019}.

Many studies of stochastic thermodynamics have dealt with continuous-time Markov processes based on the master equation and instantaneous value of entropy production, called entropy production rate. In the following, we focus our discussion on the entropy prodcuction rate and denote it as entropy production for simplicity. As mentioned above, stochastic thermodynamics also has important applications in discrete-time systems such as neural networks and time-series data analysis and some researchers dealing with discrete-time processes \cite{gaspardTimeReversedDynamicalEntropy2004,leeDerivationMarkovProcesses2018,crooksMarginalConditionalSecond2019,liuThermodynamicUncertaintyRelation2020}. Here, there are multiple definitions of entropy production: one is entropy production and the other is dissipation function\cite{evans_fluctuation_2002}. Note that unless there is a possibility of confusion, we refer to entropy production as a general term for both entropy production and dissipation function. Although these definitions were known, not many studies have clearly distinguished them and discussed the differences\cite{seifertStochasticThermodynamicsFluctuation2012,yang_unified_2020}.
Among them, Yang and Qian discuss the difference in detail\cite{yang_unified_2020} but their study considered only a single system. This leaves several unanswered questions, including:
\begin{itemize}
    \item Do known inequalities---such as those derived from the decomposition of entropy production, including the second law of thermodynamics and the second law of information thermodynamics---hold for any definition?
    \item In a composite system, do the known inequalities of entropy production for the whole system and subsystems hold?
\end{itemize}
In this paper, we aim to clarify these points regarding multiple forms of entropy production in the discrete-time Markov processes of single systems and composite systems. We confirm these results by applying them to a concrete probabilistic model. Furthermore, by taking all these results into account, we clarify the appropriate use of each definition.

This paper is organized as follows. In chapter \ref{EntPro}, we introduce the entropy production. First, we describe the definitions of entropy production for a single system. Then, for composite systems, we present the definitions of total entropy production for an entire composite system, partial entropy prodcuction, marginal entropy production, and conditional entropy production. In chapter \ref{DecEntPro}, we discuss various inequalities obtained by decomposing the multiple forms of entropy production defined in chapter \ref{EntPro} and examining their implications. In chapter \ref{IneEntPro}, we discuss the inequalities for total entropy production and marginal entropy production, and for partial entropy production and marginal entropy production. In chapter \ref{Ex}, as a concrete example, we calculate the total entropy production, marginal entropy production, and partial entropy production for Gaussian process, and we confirm the inequalities described in chapter \ref{IneEntPro}. Chapter \ref{Sum} concludes with a summary of this paper and future works.

\section{Entropy Production}
\label{EntPro}
This chapter presents an introduction to the entropy production, which is the subject of this research. Section \ref{sec:single_system} shows the multiple definitions, meanings, and properties of entropy production in discrete-time Markov processes for single systems that can be described by a single random variable. Section \ref{sec:composite_system} shows the multiple definitions, meanings, and properties of entropy production in discrete-time Markov processes for composite systems that can be described by multiple random variables.

Although Chapters \ref{intro} to \ref{IneEntPro} deal with discrete variables to simplify the description, the same reasoning can also be applied to continuous variables by replacing the summations with integrals.

\subsection{Single Systems}
\label{sec:single_system}
In this section, we examine the situation where a single system evolves over time with a discrete-time Markov process, and we discuss the definition of multiple entropy production, its properties, and its treatment in previous studies.

we will use the terminology $\mathbf{S}_{t}$ to denote a random variable representing the state of a system that varies with time $t$, $\mathbf{s}_{t}$ to denote the value of this random variable, and $P_{\mathbf{S}_{t}} (\mathbf{s}_{t})$ to denote the probability that the value of $\mathbf{S}_{t}$ is equal to $\mathbf{s}_{t}$. Similarly, in the following, we will use uppercase letters to represent random variables and lowercase letters to represent values taken by these random variables.

\subsubsection{Setup}
The joint probability that the state is $\mathbf{s}_{t}$ at time $t$ and $\mathbf{s}_{t+1}$ at time $t+1$ is expressed as shown below.
\begin{align}
    P_{\mathbf{S}_{t+1}, \mathbf{S}_{t}} (\mathbf{s}_{t+1}, \mathbf{s}_{t})
\end{align}
The probability that the state is $\mathbf{s}_{t}$ at time $t$ and the probability that the state is $\mathbf{s}_{t+1}$ at time $t+1$ are respectively obtained as follows.
\begin{align}
    P_{\mathbf{S}_{t}} (\mathbf{s}_{t}) =& \dsum_{\mathbf{s}_{t+1}} P_{\mathbf{S}_{t+1}, \mathbf{S}_{t}} (\mathbf{s}_{t+1}, \mathbf{s}_{t}) \\
    P_{\mathbf{S}_{t+1}} (\mathbf{s}_{t+1}) =& \dsum_{\mathbf{s}_{t}} P_{\mathbf{S}_{t+1}, \mathbf{S}_{t}} (\mathbf{s}_{t+1}, \mathbf{s}_{t})
\end{align}
These calculations are called marginalizations of $\mathbf{S}_{t+1}$ and $\mathbf{S}_{t}$ respectively, and the probability distributions obtained by the marginalization of $P_{\mathbf{S}_{t}}$ and $P_{\mathbf{S}_{t+1}}$ are called the marginal distributions of $\mathbf{S}_{t}$ and $\mathbf{S}_{t+1}$ respectively.

\begin{figure}[bt]
    \centering
    \includegraphics[width=70mm]{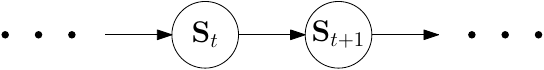}
    \caption{Schematic diagram of a discrete-time Markov process in a single system.}
    \label{fig:sin_mar}
\end{figure}

The conditional probability that the state is $\mathbf{s}_{t+1}$ at time $t+1$ under the condition that the state is $\mathbf{s}_{t}$ at time $t$ is expressed as follows.
\begin{align}
    P_{\mathbf{S}_{t+1}| \mathbf{S}_{t}} (\mathbf{s}_{t+1}|\mathbf{s}_{t}) = \dfrac{P_{\mathbf{S}_{t+1}, \mathbf{S}_{t}} (\mathbf{s}_{t+1}, \mathbf{s}_{t})}{P_{\mathbf{S}_{t}} (\mathbf{s}_{t})}
\end{align}
The simultaneous probability is represented by the product of the conditional probability and the marginal distribution at time $t$.
\begin{align}
    P_{\mathbf{S}_{t+1}, \mathbf{S}_{t}} (\mathbf{s}_{t+1}, \mathbf{s}_{t})
    = P_{\mathbf{S}_{t+1}| \mathbf{S}_{t}} (\mathbf{s}_{t+1}|\mathbf{s}_{t}) P_{\mathbf{S}_{t}} (\mathbf{s}_{t})
\end{align}
A process generated according to time evolution like the right side of this equation is called a forward process. Also, the conditional probability $P_{\mathbf{S}_{t+1}| \mathbf{S}_{t}} (\mathbf{s}_{t+1}|\mathbf{s}_{t})$ is called the transition probability of the forward process. A process in which the transition probability representing time evolution is represented by a conditional probability that depends only on the variables of the previous time slot is called a Markov process (Fig.~\ref{fig:sin_mar}).
The joint probability distribution and marginal distribution respectively satisfy the following normalization conditions.
\begin{align}
    \sum_{\mathbf{s}_{t+1},\mathbf{s}_{t}} P_{\mathbf{S}_{t+1}, \mathbf{S}_{t}} (\mathbf{s}_{t+1}, \mathbf{s}_{t}) =& 1, \\
    \sum_{\mathbf{s}_{t}} P_{\mathbf{S}_{t}} (\mathbf{s}_{t}) =& 1, \\
    \sum_{\mathbf{s}_{t+1}} P_{\mathbf{S}_{t+1}} (\mathbf{s}_{t+1}) =& 1.
\end{align}
In a stochastic process, a special situation can be considered in which the probability distribution does not vary over time. Under such circumstances, the system is said to be in a steady state and is defined as follows.
\begin{dfn}[Steady State]
    A system is said to be in a steady state if the following relationship holds for each state $\mathbf{s}$ at arbitrary times $t, t'$.
    \begin{align}
        P_{\mathbf{S}_{t}} (\mathbf{s}) =& P_{\mathbf{S}_{t'}} (\mathbf{s}) \quad (t \neq t') \\
        =& P^{\mathrm{ss}} (\mathbf{s})
        \label{def_steady}
    \end{align}
    In this case, the probability distribution is called a stationary distribution, which is denoted by $P^{\mathrm{ss}}$. A process that transitions between stationary states is called a stationary process.
\end{dfn}
When a system is in a steady state, this situation can also be expressed by the balance condition defined below.
\begin{dfn}[Balance Condition]
    The balance condition is given by the following relationship for any $\mathbf{s}_{t+1}$.
    \begin{align}
        \dsum_{\mathbf{s}_{t}} \br{P_{\mathbf{S}_{t+1}| \mathbf{S}_{t}} (\mathbf{s}_{t+1}| \mathbf{s}_{t}) P_{\mathbf{S}_{t}} (\mathbf{s}_{t}) - P_{\mathbf{S}_{t+1}| \mathbf{S}_{t}} (\mathbf{s}_{t}| \mathbf{s}_{t+1}) P_{\mathbf{S}_{t}} (\mathbf{s}_{t+1})} = 0
        \label{def_balance}
    \end{align}
\end{dfn}
If the balance condition holds, this is equivalent to the system being in a steady state, as demonstrated below. First, consider the time evolution of probability.
\begin{align}
    P_{\mathbf{S}_{t+1}} (\mathbf{s}_{t+1}) - P_{\mathbf{S}_{t}} (\mathbf{s}_{t+1})
    =& \dsum_{\mathbf{s}_{t}} P_{\mathbf{S}_{t+1}| \mathbf{S}_{t}} (\mathbf{s}_{t+1}| \mathbf{s}_{t}) P_{\mathbf{S}_{t}} (\mathbf{s}_{t}) - P_{\mathbf{S}_{t}} (\mathbf{s}_{t+1}) \\
    =& \dsum_{\mathbf{s}_{t}} P_{\mathbf{S}_{t+1}| \mathbf{S}_{t}} (\mathbf{s}_{t+1}| \mathbf{s}_{t}) P_{\mathbf{S}_{t}} (\mathbf{s}_{t}) - \dsum_{\mathbf{s}_{t}} P_{\mathbf{S}_{t+1}| \mathbf{S}_{t}} (\mathbf{s}_{t}| \mathbf{s}_{t+1}) P_{\mathbf{S}_{t}} (\mathbf{s}_{t+1}) \\
    =& \dsum_{\mathbf{s}_{t}} \br{P_{\mathbf{S}_{t+1}| \mathbf{S}_{t}} (\mathbf{s}_{t+1}| \mathbf{s}_{t}) P_{\mathbf{S}_{t}} (\mathbf{s}_{t}) - P_{\mathbf{S}_{t+1}| \mathbf{S}_{t}} (\mathbf{s}_{t}| \mathbf{s}_{t+1}) P_{\mathbf{S}_{t}} (\mathbf{s}_{t+1})} \\
    =& \dsum_{\mathbf{s}_{t}} \br{
    P_{\mathbf{S}_{t+1}, \mathbf{S}_{t}} (\mathbf{s}_{t+1}, \mathbf{s}_{t}) - P_{\mathbf{S}_{t+1}, \mathbf{S}_{t}} (\mathbf{s}_{t}, \mathbf{s}_{t+1})
    }
\end{align}
In the steady state, the left side of the equation is zero. Therefore, if the balance condition holds, this is equivalent to the system being in a steady state.

As a special case where the balance condition holds, it is possible to conceive of the following detailed balance condition where the the balance condition holds for all $\mathbf{s}_{t},\mathbf{s}_{t+1}$.
\begin{dfn}[Detailed Balance Condition]
    The detailed balance condition is expressed by the following relationship for any $\mathbf{s}_{t},\mathbf{s}_{t+1}$.
    \begin{align}
        P_{\mathbf{S}_{t+1}| \mathbf{S}_{t}} (\mathbf{s}_{t+1}| \mathbf{s}_{t}) P^{\mathrm{ss}} (\mathbf{s}_{t}) =& P_{\mathbf{S}_{t+1}| \mathbf{S}_{t}} (\mathbf{s}_{t}| \mathbf{s}_{t+1}) P^{\mathrm{ss}} (\mathbf{s}_{t+1}) \\
        \label{det_bal}
        P_{\mathbf{S}_{t+1}, \mathbf{S}_{t}} (\mathbf{s}_{t+1}, \mathbf{s}_{t}) =& P_{\mathbf{S}_{t+1}, \mathbf{S}_{t}} (\mathbf{s}_{t}, \mathbf{s}_{t+1})
    \end{align}
\end{dfn}
When the detailed balance condition is satisfied, the system is in a special kind of steady state called the equilibrium state.
\begin{dfn}[Equilibrium State]
    A system is said to be in an equilibrium state if at any time $t$ and for any $\mathbf{s}_{t+1},\mathbf{s}_{t}$, the following detailed balance condition holds.
    \begin{align}
        P_{\mathbf{S}_{t+1}| \mathbf{S}_{t}} (\mathbf{s}_{t+1}| \mathbf{s}_{t}) P^{\mathrm{eq}} (\mathbf{s}_{t}) = P_{\mathbf{S}_{t+1}| \mathbf{S}_{t}} (\mathbf{s}_{t}| \mathbf{s}_{t+1}) P^{\mathrm{eq}} (\mathbf{s}_{t+1})
        \label{def_eq}
    \end{align}
    The probability distribution at this time is called the equilibrium distribution, and is indicated by the special notation $P^{\mathrm{eq}}$. A process that transitions between equilibrium states is called a reversible process or an equilibrium process.
\end{dfn}
\subsubsection{Entropy Production}
As a measure of the irreversibility of the dynamics in a stochastically varying system, it seems appropriate to use a quantity expressing how much the probability distribution representing a process differs from the probability distribution representing the same process going backwards in time. In information theory, this is defined as the Kullback--Leibler divergence (KLD), which is used in information theory as a quantity expressing the difference between two probability distributions \cite{coverElementsInformationTheory2012}.

Two definitions have been used in previous studies, which are defined here as entropy production and dissipation function.
\begin{dfn}[Entropy Production]
Entropy production is defined as follows.
    \begin{align}
        \label{bep_sin}
        \Sigma \coloneq& \KL{ P_{\mathbf{S}_{t:t+1}} }{ Q_{\mathbf{S}_{t:t+1}} } \\
        =& \sum_{\mathbf{s}_{t:t+1}} P_{\mathbf{S}_{t:t+1}} (\mathbf{s}_{t:t+1})
        \log \frac{
        P_{\mathbf{S}_{t+1}|\mathbf{S}_{t}} (\mathbf{s}_{t+1}|\mathbf{s}_{t}) P_{\mathbf{S}_{t}} (\mathbf{s}_{t})
        }{
        P_{\mathbf{S}_{t+1}|\mathbf{S}_{t}} (\mathbf{s}_{t}|\mathbf{s}_{t+1}) P_{\mathbf{S}_{t+1}} (\mathbf{s}_{t+1})
        }
    \end{align}
where the probability distribution $Q_{\mathbf{S}_{t:t+1}}$ is defined as follows.
    \begin{align}
        Q_{\mathbf{S}_{t:t+1}} (\mathbf{s}_{t:t+1}) \coloneq P_{\mathbf{S}_{t+1}|\mathbf{S}_{t}} (\mathbf{s}_{t}|\mathbf{s}_{t+1}) P_{\mathbf{S}_{t+1}} (\mathbf{s}_{t+1})
    \end{align}
$Q_{\mathbf{S}_{t:t+1}}$ is a probability distribution that represents time evolution with transition probabilities that the values of the random variables in the transition probability of the forward process are swaped to the opposite temporal direction, and with the initial distribution set to the one at the later time $t+1$. In the following, we call this process the backward process, and we call the distribution representing the backward process the backward distribution.
\end{dfn}
Entropy production defined above is always non-negative.
\begin{result}[Non-Negativity of Entropy Production]
entropy production $\Sigma$ obeys the following rule:
    \begin{align}
        \Sigma \geq 0. \label{nonneg_bep}
    \end{align}
\end{result}
\begin{proof}
    This follows directly from the non-negativity of KLD \cite{coverElementsInformationTheory2012}. Based on the conditions under which KLD is zero, this relationship becomes an equality when the following equation holds for all $\mathbf{s}_{t},\mathbf{s}_{t+1}$.
    \begin{align}
        P_{\mathbf{S}_{t:t+1}} (\mathbf{s}_{t:t+1}) = Q_{\mathbf{S}_{t:t+1}} (\mathbf{s}_{t:t+1})
    \end{align}
    From the definition of the backward process, this can be expressed as shown below.
    \begin{align}
        P_{\mathbf{S}_{t+1}|\mathbf{S}_{t}} (\mathbf{s}_{t+1}|\mathbf{s}_{t}) P_{\mathbf{S}_{t}} (\mathbf{s}_{t}) =& P_{\mathbf{S}_{t+1}|\mathbf{S}_{t}} (\mathbf{s}_{t}|\mathbf{s}_{t+1}) P_{\mathbf{S}_{t+1}} (\mathbf{s}_{t+1}) \\
        \dfrac{P_{\mathbf{S}_{t+1},\mathbf{S}_{t}} (\mathbf{s}_{t+1},\mathbf{s}_{t})}{P_{\mathbf{S}_{t+1}} (\mathbf{s}_{t+1})} =& P_{\mathbf{S}_{t+1}|\mathbf{S}_{t}} (\mathbf{s}_{t}|\mathbf{s}_{t+1}) \\
        P_{\mathbf{S}_{t}|\mathbf{S}_{t+1}} (\mathbf{s}_{t}|\mathbf{s}_{t+1}) =& P_{\mathbf{S}_{t+1}|\mathbf{S}_{t}} (\mathbf{s}_{t}|\mathbf{s}_{t+1})
        \label{eqcon_bep_tot}
    \end{align}
    We used Bayes' theorem to transform the second line to the third line.

    From \eqref{eqcon_bep_tot}, the entropy production becomes zero when the transition probability of the backward process is equal to the posterior distribution of $\mathbf{S}_{t}$ obtained from Bayes' theorem.

    Here, when the system is in a steady state, the relationship $P_{\mathbf{S}_{t+1}}=P_{\mathbf{S}_{t}}=P^{\mathrm{ss}}$ can be used to rewrite this formula as
    \begin{align}
        P_{\mathbf{S}_{t+1}|\mathbf{S}_{t}} (\mathbf{s}_{t+1}|\mathbf{s}_{t}) P^{\mathrm{ss}} (\mathbf{s}_{t}) =& P_{\mathbf{S}_{t+1}|\mathbf{S}_{t}} (\mathbf{s}_{t}|\mathbf{s}_{t+1}) P^{\mathrm{ss}} (\mathbf{s}_{t+1})
    \end{align}
    and since \eqref{eqcon_bep_tot} become equivalent to the detailed balance condition \eqref{det_bal}.
\end{proof}
\subsubsection{Dissipation Function}
Another definition of entropy production is introduced here.
\begin{dfn}[Dissipation Function]
    Dissipation function is defined as follows.
    \begin{align}
        \label{trep_sin}
        \tilde{\Sigma} &\coloneq \KL{P_{\mathbf{S}_{t:t+1}}}{\tilde{Q}_{\mathbf{S}_{t:t+1}}} \\
        =& \sum_{\mathbf{s}_{t:t+1}} P_{\mathbf{S}_{t:t+1}} (\mathbf{s}_{t:t+1})
        \log \frac{
        P_{\mathbf{S}_{t+1}|\mathbf{S}_{t}} (\mathbf{s}_{t+1}|\mathbf{s}_{t}) P_{\mathbf{S}_{t}} (\mathbf{s}_{t})
        }{
        P_{\mathbf{S}_{t+1}|\mathbf{S}_{t}} (\mathbf{s}_{t}|\mathbf{s}_{t+1}) P_{\mathbf{S}_{t}} (\mathbf{s}_{t+1})
        } \\
        =& \sum_{\mathbf{s}_{t:t+1}} \br{
        P_{\mathbf{S}_{t+1}| \mathbf{S}_{t}} (\mathbf{s}_{t+1}| \mathbf{s}_{t}) P_{\mathbf{S}_{t}} (\mathbf{s}_{t}) - P_{\mathbf{S}_{t+1}| \mathbf{S}_{t}} (\mathbf{s}_{t}| \mathbf{s}_{t+1}) P_{\mathbf{S}_{t}} (\mathbf{s}_{t+1})
        }
        \log \frac{
        P_{\mathbf{S}_{t+1}| \mathbf{S}_{t}} (\mathbf{s}_{t+1}| \mathbf{s}_{t}) P_{\mathbf{S}_{t}} (\mathbf{s}_{t})
        }{
        P_{\mathbf{S}_{t+1}| \mathbf{S}_{t}} (\mathbf{s}_{t}| \mathbf{s}_{t+1}) P_{\mathbf{S}_{t}} (\mathbf{s}_{t+1})
        }
    \end{align}
    where the probability distribution $\tilde{Q}_{\mathbf{S}_{t:t+1}}$ is defined as follows.
    \begin{align}
        \tilde{Q}_{\mathbf{S}_{t:t+1}} (\mathbf{s}_{t:t+1}) \coloneq& P_{\mathbf{S}_{t+1},\mathbf{S}_{t}} (\mathbf{s}_{t},\mathbf{s}_{t+1})
    \end{align}
    $\tilde{Q}_{\mathbf{S}_{t:t+1}}$ is a probability distribution obtained by swapping the values of random variables in the probability distribution of the forward process to the opposite temporal direction. In the following, we call this process the time-reversed process, and we call the distribution representing the time-reversed process the time-reversed distribution.
\end{dfn}
Dissipation function defined above is always non-negative.
\begin{result}[Non-Negativity of Dissipation Function]
For dissipation function $\Sigma$ the following relationship holds.
    \begin{align}
        \tilde{\Sigma} \geq 0. \label{nonneg_trep}
    \end{align}
\end{result}
\begin{proof}
    This follows directly from the non-negativity of KLD. Based on the conditions under which KLD is zero, this relationship becomes an equality when the following equation holds for all $\mathbf{s}_{t},\mathbf{s}_{t+1}$.
    \begin{align}
        P_{\mathbf{S}_{t:t+1}} (\mathbf{s}_{t:t+1}) = \tilde{Q}_{\mathbf{S}_{t:t+1}} (\mathbf{s}_{t:t+1})
    \end{align}
    From the definition of a time-reversed probability distribution, this can be represented as follows.
    \begin{align}
        P_{\mathbf{S}_{t+1},\mathbf{S}_{t}} (\mathbf{s}_{t+1},\mathbf{s}_{t}) =& P_{\mathbf{S}_{t+1},\mathbf{S}_{t}} (\mathbf{s}_{t},\mathbf{s}_{t+1}) \\
        P_{\mathbf{S}_{t+1}|\mathbf{S}_{t}} (\mathbf{s}_{t+1}|\mathbf{s}_{t}) P_{\mathbf{S}_{t}} (\mathbf{s}_{t}) =& P_{\mathbf{S}_{t+1}|\mathbf{S}_{t}} (\mathbf{s}_{t}|\mathbf{s}_{t+1}) P_{\mathbf{S}_{t}} (\mathbf{s}_{t+1})
        \label{eqcon_trep_tot}
    \end{align}
    From \eqref{eqcon_trep_tot}, the dissipation function is zero when the detailed balance condition of \eqref{det_bal} holds.
\end{proof}
\subsubsection{Difference Between Entropy Production and Dissipation Function}
We discuss the difference between entropy production and dissipation function below for the case where the forward process has a two-dimensional Gaussian distribution $P_{\mathbf{S}_{t+1} | \mathbf{S}_{t}} (\mathbf{s}_{t+1} | \mathbf{s}_{t})P_{\mathbf{S}_{t}} (\mathbf{s}_{t})=P_{\mathbf{S}_{t+1}, \mathbf{S}_{t}} (\mathbf{s}_{t+1}, \mathbf{s}_{t})$.

\begin{figure}[H]
	\begin{minipage}[b]{0.45\linewidth}
		\centering
		\includegraphics[width=70mm]{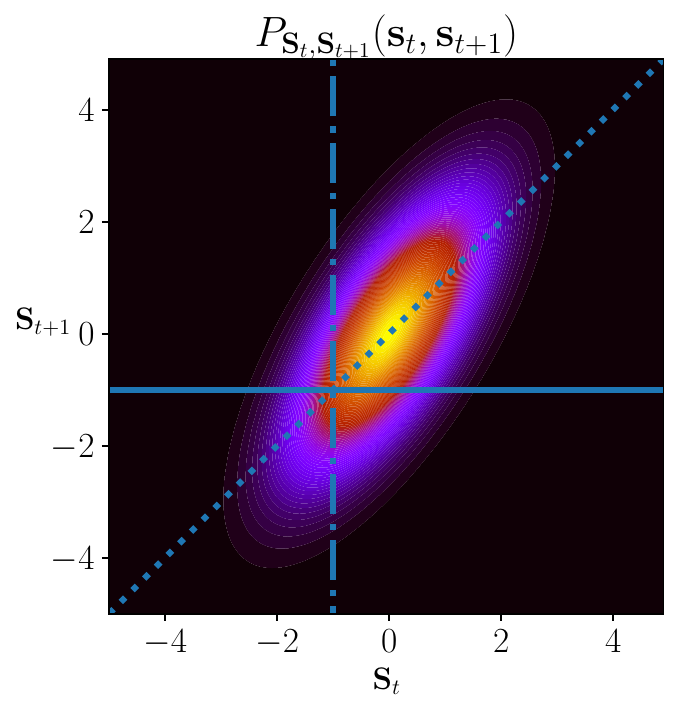}
		\subcaption{}\label{fig:bin-bep}
	\end{minipage}
  \begin{minipage}[b]{0.45\linewidth}
    \centering
    \includegraphics[width=70mm]{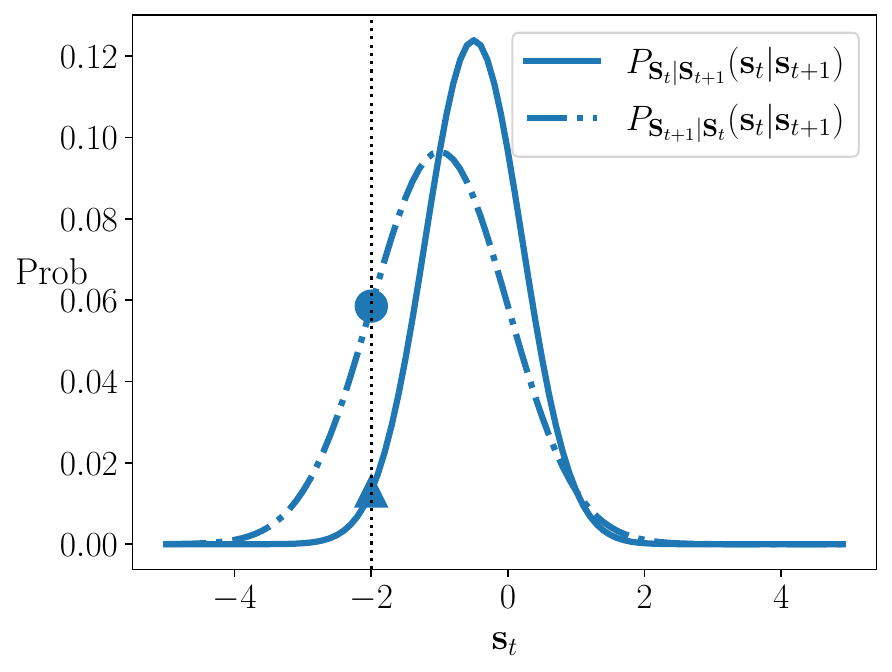}
    \subcaption{}\label{fig:bin-bep-inset}
  \end{minipage}
  \caption{Meaning of the entropy production in a two-dimensional Gaussian distribution. (a) Transition probability of the backward process $P_{\mathbf{S}_{t+1} | \mathbf{S}_{t}} (\mathbf{s}_{t} | \mathbf{s}_{t+1})$ (chain line) in a two-dimensional Gaussian distribution and posterior probability $P_{\mathbf{S}_{t} | \mathbf{S}_{t+1}} (\mathbf{s}_{t} | \mathbf{s}_{t+1})$ of $\mathbf{S}_{t}$ (solid line); (b) These probabilities expressed as functions of the same variable $\mathbf{s}_{t}$.}
\end{figure}

First, the entropy production is as shown in Figs.~\ref{fig:bin-bep} and \ref{fig:bin-bep-inset}. In Fig.~\ref{fig:bin-bep}, the horizontal axis represents $\mathbf{S}_{t}$, the vertical axis represents $\mathbf{S}_{t+1}$, and the color map represents the value of the probability $P_{\mathbf{S}_{t+1}, \mathbf{S}_{t}} (\mathbf{s}_{t+1}, \mathbf{s}_{t})$. The chain and solid lines respectively represent the transition probability $P_{\mathbf{S}_{t+1} | \mathbf{S}_{t}} (\mathbf{s}_{t} | \mathbf{s}_{t+1})$ of the backward process and the posterior probability $P_{\mathbf{S}_{t} | \mathbf{S}_{t+1}} (\mathbf{s}_{t} | \mathbf{s}_{t+1})$ of $\mathbf{S}_{t}$ at a specific $\mathbf{s}_{t+1}$, and the blue dotted line represents a line with slope 1. Figure \ref{fig:bin-bep-inset} shows the transition probability distribution of the backward process and the posterior distribution of $\mathbf{S}_{t}$ arranged with $\mathbf{s}_{t}$ as a variable, and the green and red dots show the pair of values corresponding to a specific $\mathbf{s}_{t}$. Since entropy production is a quantity that compares the circle and triangle markers in Fig.~\ref{fig:bin-bep-inset}, the difference between the posterior distribution of ${S}_{t}$ and the transition probability of the backward process is expressed by Bayesian inference. In this case, if the distribution of $\mathbf{S}_{t}$ obtained according to the transition probability of the forward process starting from $\mathbf{S}_{t+1}$ is equivalent to performing Bayesian inference, then the entropy production will be zero. In other words, when the forward evolution of a time series is equivalent to making Bayesian optimal inferences about past states, this discrete process is defined as a process with no time directionality, so entropy production is a quantified measure of deviation from this sort of process.

\begin{figure}[H]
    \centering
    \includegraphics[width=70mm]{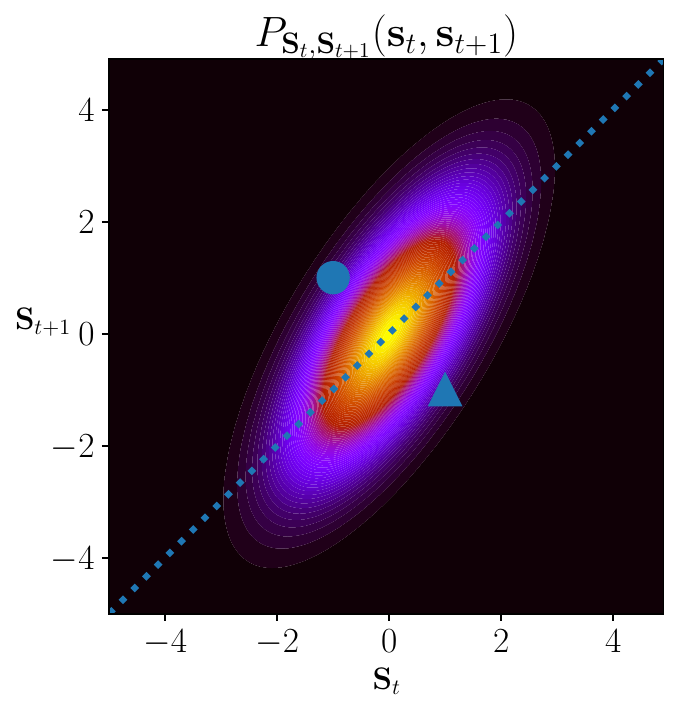}
    \caption{Meaning of the dissipation function in a two-dimensional Gaussian distribution.}
    \label{fig:bin-trep}
\end{figure}

Next, dissipation function is shown in Fig.~\ref{fig:bin-trep}, where the horizontal axis represents $\mathbf{S}_{t}$, the vertical axis represents $\mathbf{S}_{t+1}$, and the color map represents the values of the probabilities $P_{\mathbf{S}_{t+1}, \mathbf{S}_{t}} (\mathbf{s}_{t+1}, \mathbf{s}_{t})$. With regard to the blue dotted line with a slope 1, the circle and triangle markers arranged symmetrically represent $P_{\mathbf{S}_{t+1}, \mathbf{S}_{t}} (\mathbf{s}_{t+1}, \mathbf{s}_{t})$ and $P_{\mathbf{S}_{t+1}, \mathbf{S}_{t}} (\mathbf{s}_{t}, \mathbf{s}_{t+1})$ for a particular $\mathbf{s}_{t+1},\mathbf{s}_{t}$, respectively. Dissipation function is a quantity that compares the circle and triangle markers in Fig.~\ref{fig:bin-trep}, and $P_{\mathbf{S}_{t+1}, \mathbf{S}_{t}} (\mathbf{s}_{t+1}, \mathbf{s}_{t})=P_{\mathbf{S}_{t+1}, \mathbf{S}_{t}} (\mathbf{s}_{t}, \mathbf{s}_{t+1})$ is equivalent to the detailed balance condition \eqref{det_bal}, which represents the degree of violation of the detailed balance. In other words, it can be regarded as a measure of the irreversibility of non-equilibrium stochastic dynamics.

Note that Yang and Qian discuss the difference that differ from ours: the entropy production compares the probability of the forward process with the time-reversed probability of the ``protocol-reversed'' process, which is at the macroscopic/thermodynamics level. On the other hand, the dissipation function compares the probability of the forward process with the probability of time-reversed process, which is at the microscopic/trajectory level\cite{yang_unified_2020}.
\subsubsection{Relationship Between Entropy Production and Dissipation Function}
\subsubsection*{Conditions for equivalence}
Under certain conditions, the entropy production and dissipation function defined above are equivalent. To start with, the difference between the two can be expressed as follows.
\begin{align}
    \tilde{\Sigma} - \Sigma
    =& \KL{ P_{\mathbf{S}_{t:t+1}} }{ \tilde{Q}_{\mathbf{S}_{t:t+1}} } - \KL{P_{\mathbf{S}_{t:t+1}}}{Q_{\mathbf{S}_{t:t+1}}} \\
    =& \expec{\mathbf{S}_{t:t+1}}{
    \log \dfrac{Q_{\mathbf{S}_{t:t+1}}}{\tilde{Q}_{\mathbf{S}_{t:t+1}}}
    } \\
    =& \sum_{\mathbf{s}} P_{\mathbf{S}_{t+1}} (\mathbf{s}) \log \dfrac{P_{\mathbf{S}_{t+1}} (\mathbf{s})}{P_{\mathbf{S}_{t}} (\mathbf{s})} \\
    =& \KL{P_{\mathbf{S}_{t+1}}}{P_{\mathbf{S}_{t}}}
\end{align}
Therefore, the following relationship holds:
\begin{result}[Inequality for Entropy Production and Dissipation Function]
    The following inequality holds between entropy production and dissipation function:
    \begin{align}
        \tilde{\Sigma} \geq \Sigma.
    \end{align}
\end{result}
\begin{proof}
    This holds because $\tilde{\Sigma} - \Sigma = \KL{P_{\mathbf{S}_{t+1}}}{P_{\mathbf{S}_{t}}}$ and due to the non-negativity of KLD. When equality holds, i.e., under conditions where the entropy production and dissipation function are equivalent, the equality $P_{\mathbf{S}_{t+1}}(\mathbf{s})=P_{\mathbf{S}_{t}} (\mathbf{s})$ holds for any $\mathbf{s}$. In other words, in the steady state, entropy production and dissipation function are equivalent.
\end{proof}
\subsubsection*{Continuous-time limit}
These definitions are also equivalent when considering the continuous-time limit. If time is a continuous variable in $\tilde{\Sigma} - \Sigma=\KL{P_{\mathbf{S}_{t+1}}}{P_{\mathbf{S}_{t}}}$ and is discretized by a time interval $\tau$, it can be expanded as follows.
\begin{align}
    &\KL{P_{\mathbf{S}_{t+\tau}}}{P_{\mathbf{S}_{t}}} \\
    =& \sum_{\mathbf{s}} P_{\mathbf{S}_{t+\tau}} (\mathbf{s}) \log \dfrac{P_{\mathbf{S}_{t+\tau}} (\mathbf{s})}{P_{\mathbf{S}_{t}} (\mathbf{s})} \\
    =& \sum_{\mathbf{s}} P_{\mathbf{S}_{t+\tau}} (\mathbf{s}) \log \dfrac{
    P_{\mathbf{S}_{t}} (\mathbf{s}) + \tau \pd{P_{\mathbf{S}_{t}} (\mathbf{s})}{t} + \dfrac{\tau^2}{2} \pdd{P_{\mathbf{S}_{t}} (\mathbf{s})}{t} + O(\tau^3)
    }{P_{\mathbf{S}_{t}} (\mathbf{s})} \\
    =& \sum_{\mathbf{s}} P_{\mathbf{S}_{t+\tau}} (\mathbf{s}) \log \pr{1 + \tau \dfrac{1}{P_{\mathbf{S}_{t}} (\mathbf{s})} \pd{P_{\mathbf{S}_{t}} (\mathbf{s})}{t} + \dfrac{\tau^2}{2} \dfrac{1}{P_{\mathbf{S}_{t}} (\mathbf{s})} \pdd{P_{\mathbf{S}_{t}} (\mathbf{s})}{t} + O(\tau^3)} \\
    =& \sum_{\mathbf{s}} \pr{P_{\mathbf{S}_{t}} (\mathbf{s}) + \tau \pd{P_{\mathbf{S}_{t}} (\mathbf{s})}{t} + O(\tau^2)}
    \pr{\tau \dfrac{1}{P_{\mathbf{S}_{t}} (\mathbf{s})} \pd{P_{\mathbf{S}_{t}} (\mathbf{s})}{t} + \dfrac{\tau^2}{2} \dfrac{1}{P_{\mathbf{S}_{t}} (\mathbf{s})} \pdd{P_{\mathbf{S}_{t}} (\mathbf{s})}{t} + O(\tau^3)} \\
    =& \sum_{\mathbf{s}} \pr{
    \tau \pd{P_{\mathbf{S}_{t}} (\mathbf{s})}{t}
    + \dfrac{\tau^2}{2} \pdd{P_{\mathbf{S}_{t}} (\mathbf{s})}{t}
    + \tau^2 \dfrac{1}{P_{\mathbf{S}_{t}} (\mathbf{s})} \pdd{P_{\mathbf{S}_{t}} (\mathbf{s})}{t} + O(\tau^3)
    } \\
    =& \sum_{\mathbf{s}} \tau^2 \dfrac{1}{P_{\mathbf{S}_{t}} (\mathbf{s})} \pdd{P_{\mathbf{S}_{t}} (\mathbf{s})}{t} + O(\tau^3) \\
    =& O(\tau^2)
\end{align}
where the seventh line is obtained from the sixth by applying the normalization condition $\sum_{\mathbf{s}} P_{\mathbf{S}_{t}} (\mathbf{s}) = 1$. Therefore, this becomes zero at the continuous-time limit $ \tau \rightarrow 0 $.

\subsubsection{Previous Studies}
Previous studies focused on either entropy production or dissipation function in discrete-time systems. We showed some cases in Table \ref{table:EP_table}. When a continuous-time stochastic process is discretized with a finite time interval, its definition is taken to be the one in discrete-time.
\begin{table}[h]
 \caption{Definition types of entropy production and previous studies.}
 \label{table:EP_table}
 \centering
  \begin{tabular}{cc}
   \toprule
   Types  &Previous studies \\
   \midrule
   $\Sigma$
   & Lee (2018), Ito \textit{et al.} (2020), Liu \textit{et al.} (2020) \\
   $\tilde{\Sigma}$ &Gaspard (2004), Touchette (2009), Proesmans \textit{et al.} (2017) \\
   \midrule
  \end{tabular}
\end{table}
\subsection{Composite Systems}
\label{sec:composite_system}
In this section, we present multiple definitions of entropy for a composite system XY consisting of two interacting systems X and Y that evolve in time as a whole based on a discrete-time Markov process, and we discuss the properties of such systems and their treatment in previous studies.

\subsubsection{Setup}
\begin{figure}[bt]
    \centering
    \includegraphics[width=70mm]{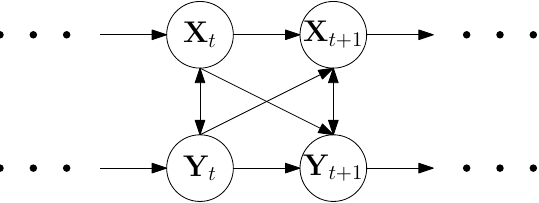}
    \caption{Schematic diagram of a discrete-time Markov process in a composite system.}
    \label{fig:com_mar}
\end{figure}

The random variables representing the state of systems X and Y at time $t$ are $\mathbf{X}_{t}$ and $\mathbf{Y}_{t}$, respectively. By setting $\mathbf{S}_{t}=\brc{\mathbf{X}_{t},\mathbf{Y}_{t}}$ in section \ref{sec:single_system}, the probability distribution representing the process from time $t$ to $t+1$ is defined in the same way. The steady state \eqref{def_steady}, the balance condition \eqref{def_balance}, the detailed balance condition \eqref{det_bal}, and the equilibrium state \eqref{def_eq} are also defined in the same way. In particular, the detailed balance of a composite system can be expressed as follows.
\begin{dfn}[Detailed Balance Condition of a Composite System]
    The detailed balance condition in a composite system is expressed by the following relation for any $\mathbf{x}_{t},\mathbf{y}_{t},\mathbf{x}_{t+1},\mathbf{y}_{t+1}$.
    \begin{align}
        P_{\mathbf{X}_{t+1},\mathbf{Y}_{t+1}| \mathbf{X}_t,\mathbf{Y}_t} (\mathbf{x}_{t+1},\mathbf{y}_{t+1}| \mathbf{x}_t,\mathbf{y}_t) P_{\mathbf{X}_t,\mathbf{Y}_t} (\mathbf{x}_t,\mathbf{y}_t) =& P_{\mathbf{X}_{t+1},\mathbf{Y}_{t+1}| \mathbf{X}_t,\mathbf{Y}_t} (\mathbf{x}_{t},\mathbf{y}_{t}| \mathbf{x}_{t+1},\mathbf{y}_{t+1}) P_{\mathbf{X}_{t},\mathbf{Y}_{t}} (\mathbf{x}_{t+1},\mathbf{y}_{t+1})
        \label{det_bal_com}
    \end{align}
\end{dfn}
In addition, stochastic thermodynamics often requires that the time evolution of composite system XY is bipartite \cite{hartichStochasticThermodynamicsBipartite2014,hartichSensoryCapacityInformation2016,matsumotoRoleSufficientStatistics2018,horowitzSecondlawlikeInequalitiesInformation2014,horowitzThermodynamicsContinuousInformation2014,baratoEfficiencyCellularInformation2014,sartoriThermodynamicCostsInformation2014,brittainWhatWeLearn2017,goldtThermodynamicEfficiencyLearning2017,crooksMarginalConditionalSecond2019,stillThermodynamicsPrediction2012,dianaMutualEntropyProduction2014}. Bipartiteness is the property whereby, given the current values of X and Y, the next values of X and Y are conditionally independent, indicating that the time evolution of the composite system XY can be divided into separate time evolutions for X and Y.
\begin{dfn}[Bipartiteness]
    For any $\mathbf{x}_{t},\mathbf{y}_{t},\mathbf{x}_{t+1},\mathbf{y}_{t+1}$, the property of bipartiteness is expressed as follows.\footnote{Unlike the conditional independence definition of this thesis, bipartiteness can also be defined as satisfying the condition that the transition probability is zero when both X and Y change, and when only X or Y changes, the transition probability depends on the current values of X and Y and the next value of whichever of these variables changes \cite{hartichStochasticThermodynamicsBipartite2014,hartichSensoryCapacityInformation2016,matsumotoRoleSufficientStatistics2018,horowitzSecondlawlikeInequalitiesInformation2014,horowitzThermodynamicsContinuousInformation2014,baratoEfficiencyCellularInformation2014,sartoriThermodynamicCostsInformation2014,brittainWhatWeLearn2017,dianaMutualEntropyProduction2014}. This is the definition that is included in the definition based on conditional independence \cite{matsumotoRoleSufficientStatistics2018}.}
    \begin{align}
        P_{\mathbf{X}_{t+1},\mathbf{Y}_{t+1}| \mathbf{X}_t,\mathbf{Y}_t} (\mathbf{x}_{t+1},\mathbf{y}_{t+1}| \mathbf{x}_t,\mathbf{y}_t)
        = P_{\mathbf{X}_{t+1}| \mathbf{X}_t,\mathbf{Y}_t} (\mathbf{x}_{t+1}| \mathbf{x}_t,\mathbf{y}_t)
        P_{\mathbf{Y}_{t+1}| \mathbf{X}_t,\mathbf{Y}_t} (\mathbf{y}_{t+1}| \mathbf{x}_t,\mathbf{y}_t)
        \label{def_bip}
    \end{align}
\end{dfn}
This section presents a more general examination of the time evolution of non-bipartite composite systems that do not satisfy bipartiteness \cite{itoUnifiedFrameworkEntropy2020,auconiInformationThermodynamicsTime2019,chetriteInformationThermodynamicsInteracting2019} (Fig.~\ref{fig:com_mar}). In other words, the time evolution of a subsystem X(Y) is described by the conditional probability distribution $P_{ \mathbf{X}_{t+1} | \mathbf{Y}_{t+1}, \mathbf{X}_{t}, \mathbf{Y}_{t} }(P_{ \mathbf{Y}_{t+1} | \mathbf{Y}_{t+1}, \mathbf{X}_{t}, \mathbf{Y}_{t} })$.
\subsubsection{Total Entropy Production}
\label{def_comEP}
In entropy production \eqref{bep_sin} and dissipation function \eqref{trep_sin} as described in section \ref{sec:single_system}, the entropy production of the entire composite system can be defined by setting $\mathbf{S}_{t}=\brc{\mathbf{X}_{t},\mathbf{Y}_{t}}$. This is called the total entropy production, which has two definitions based on total entropy production and total dissipation function. In the same way as for a single system, it is possible to show that the entropy production is non-negative in both cases.
\begin{dfn}[Total Entropy Production]
    \begin{align}
        \label{bep_tot}
        \Sigma_{\mathrm{XY}} ^{\mathrm{tot}} \coloneq \KL{ P_{ \mathbf{X}_{t:t+1}, \mathbf{Y}_{t:t+1} } }{ Q_{ \mathbf{X}_{t:t+1}, \mathbf{Y}_{t:t+1} } }
    \end{align}
    where is $Q_{ \mathbf{X}_{t:t+1}, \mathbf{Y}_{t:t+1} }$ is defined as follows.
    \begin{align}
        Q_{ \mathbf{X}_{t:t+1}, \mathbf{Y}_{t:t+1} } (\mathbf{x}_{t:t+1}, \mathbf{y}_{t:t+1})
        \coloneq P_{ \mathbf{X}_{t+1}, \mathbf{Y}_{t+1} | \mathbf{X}_{t}, \mathbf{Y}_{t} } ( \mathbf{x}_{t}, \mathbf{y}_{t} | \mathbf{x}_{t+1}, \mathbf{y}_{t+1} )
        P_{ \mathbf{X}_{t+1}, \mathbf{Y}_{t+1} } (\mathbf{x}_{t+1}, \mathbf{y}_{t+1})
    \end{align}
    $Q_{ \mathbf{X}_{t:t+1}, \mathbf{Y}_{t:t+1} }$ is a probability distribution that represents the backward process of the composite system, which is defined in the same way as the single system. We also call the backward distribution of the composite system.
\end{dfn}
\begin{dfn}[Total Dissipation Function]
    \begin{align}
        \label{trep_tot}
        \tilde{\Sigma}_{\mathrm{XY}} ^{\mathrm{tot}}\coloneq \KL{ P_{ \mathbf{X}_{t:t+1}, \mathbf{Y}_{t:t+1} } }{ \tilde{Q}_{ \mathbf{X}_{t:t+1}, \mathbf{Y}_{t:t+1} } }
    \end{align}
    where $\tilde{Q}_{ \mathbf{X}_{t:t+1}, \mathbf{Y}_{t:t+1} }$ is defined as follows:
    \begin{align}
        \tilde{Q}_{ \mathbf{X}_{t:t+1}, \mathbf{Y}_{t:t+1} } (\mathbf{x}_{t:t+1}, \mathbf{y}_{t:t+1})
        \coloneq& P_{ \mathbf{X}_{t+1}, \mathbf{Y}_{t+1} | \mathbf{X}_{t}, \mathbf{Y}_{t} } ( \mathbf{x}_{t}, \mathbf{y}_{t} | \mathbf{x}_{t+1}, \mathbf{y}_{t+1} )
        P_{ \mathbf{X}_{t}, \mathbf{Y}_{t} } (\mathbf{x}_{t+1}, \mathbf{y}_{t+1}) \\
        =& P_{ \mathbf{X}_{t+1}, \mathbf{Y}_{t+1}, \mathbf{X}_{t}, \mathbf{Y}_{t} } ( \mathbf{x}_{t}, \mathbf{y}_{t}, \mathbf{x}_{t+1}, \mathbf{y}_{t+1} )
    \end{align}
    $\tilde{Q}_{ \mathbf{X}_{t:t+1}, \mathbf{Y}_{t:t+1} }$ is a probability distribution that represents the time-reversed process of the composite system, which is defined in the same way as the single system. We also call the time-reversed distribution of the composite system.
\end{dfn}
For both of total entropy production and dissipation function, they are non-negative as in the case of a single system.
\begin{result}[Non-Negativity of Total Entropy Production]
For total entropy production, $\Sigma_{\mathrm{XY}} ^{\mathrm{tot}}$ and total dissipation function $\tilde{\Sigma}_{\mathrm{XY}} ^{\mathrm{tot}}$,
    \begin{align}
        \Sigma_{\mathrm{XY}} ^{\mathrm{tot}} &\geq 0, \\
        \tilde{\Sigma}_{\mathrm{XY}} ^{\mathrm{tot}} &\geq 0.
    \end{align}
\end{result}
\begin{proof}
    This follows directly from the non-negativity of KLD. Based on the conditions under which KLD is zero, this relationship becomes an equality in the case of entropy production when the following condition holds for all $\mathbf{x}_{t},\mathbf{y}_{t},\mathbf{x}_{t+1},\mathbf{y}_{t+1}$:
    \begin{align}
        P_{ \mathbf{X}_{t:t+1}, \mathbf{Y}_{t:t+1} } ( \mathbf{x}_{t:t+1}, \mathbf{y}_{t:t+1} )= Q_{ \mathbf{X}_{t:t+1}, \mathbf{Y}_{t:t+1} }( \mathbf{x}_{t:t+1}, \mathbf{y}_{t:t+1})
    \end{align}
    As in \eqref{eqcon_bep_tot}, this results in the following relationship:
    \begin{align}
        P_{\mathbf{X}_{t}, \mathbf{Y}_{t}|\mathbf{X}_{t+1}, \mathbf{Y}_{t+1}} (\mathbf{x}_{t}, \mathbf{y}_{t}|\mathbf{x}_{t+1}, \mathbf{y}_{t+1})
        =& P_{\mathbf{X}_{t+1}, \mathbf{Y}_{t+1}|\mathbf{X}_{t}, \mathbf{Y}_{t}} (\mathbf{x}_{t}, \mathbf{y}_{t}|\mathbf{x}_{t+1}, \mathbf{y}_{t+1})
    \end{align}
    For the dissipation function, the following relationship holds for all $\mathbf{x}_{t},\mathbf{y}_{t},\mathbf{x}_{t+1},\mathbf{y}_{t+1}$:
    \begin{align}
        P_{ \mathbf{X}_{t:t+1}, \mathbf{Y}_{t:t+1} } ( \mathbf{x}_{t:t+1}, \mathbf{y}_{t:t+1})= \tilde{Q}_{ \mathbf{X}_{t:t+1}, \mathbf{Y}_{t:t+1} }( \mathbf{x}_{t:t+1}, \mathbf{y}_{t:t+1})
    \end{align}
    As in \eqref{eqcon_trep_tot}, this results in the following relationship:
    \begin{align}
        P_{\mathbf{X}_{t+1}, \mathbf{Y}_{t+1}|\mathbf{X}_{t}, \mathbf{Y}_{t}} (\mathbf{x}_{t+1}, \mathbf{y}_{t+1}|\mathbf{x}_{t}, \mathbf{y}_{t}) P_{\mathbf{X}_{t}, \mathbf{Y}_{t}} (\mathbf{x}_{t}, \mathbf{y}_{t}) =& P_{\mathbf{X}_{t+1}, \mathbf{Y}_{t+1}|\mathbf{X}_{t}, \mathbf{Y}_{t}} (\mathbf{x}_{t}, \mathbf{y}_{t}|\mathbf{x}_{t+1}, \mathbf{y}_{t+1}) P_{\mathbf{X}_{t}, \mathbf{Y}_{t}} (\mathbf{x}_{t+1}, \mathbf{y}_{t+1})
    \end{align}.
\end{proof}
\subsubsection{Partial Entropy Production}
Partial entropy production can be defined for a subsystem X that forms part of a composite system XY (and can also be defined for Y by swapping the random variables of X and Y). As with total entropy production, two types can be defined: partial entropy production and partial dissipation function.
\begin{dfn}[Partial Entropy Production]
    \begin{align}
        \Sigma_{\mathrm{X}} ^{\mathrm{par}} \coloneq \KL{ P_{ \mathbf{X}_{t:t+1}, \mathbf{Y}_{t:t+1} } }{ Q_{ \mathbf{X}_{t:t+1}, \mathbf{Y}_{t:t+1} } ^\mathrm{par} }
        \label{bep_par}
    \end{align}
    where $ Q_{ \mathbf{X}_{t:t+1}, \mathbf{Y}_{t:t+1} } ^{\mathrm{par}}$ is defined as:
    \begin{align}
        Q_{ \mathbf{X}_{t:t+1}, \mathbf{Y}_{t:t+1} } ^\mathrm{par} (\mathbf{x}_{t:t+1}, \mathbf{y}_{t:t+1})
        \coloneq P_{ \mathbf{X}_{t+1} | \mathbf{Y}_{t+1}, \mathbf{X}_{t}, \mathbf{Y}_{t} } ( \mathbf{x}_{t} | \mathbf{y}_{t}, \mathbf{x}_{t+1}, \mathbf{y}_{t+1} )
        P_{ \mathbf{X}_{t+1}, \mathbf{Y}_{t+1}, \mathbf{Y}_{t} } (\mathbf{x}_{t+1}, \mathbf{y}_{t+1}, \mathbf{y}_{t})
    \end{align}
    $Q_{ \mathbf{X}_{t:t+1}, \mathbf{Y}_{t:t+1} }^\mathrm{par}$ is a probability distribution that represents time evolution with transition probabilities where only the values of subsystem X in the transition probability of the forward process are swapped to the opposite temporal direction, and where the initial distribution is set to the one at the later time $t+1$ for X. In the following, we call this a partial backward process of X. We also call the distribution of a partial backward process a partial backward distribution.
\end{dfn}
\begin{dfn}[Partial Dissipation Function]
    \begin{align}
        \tilde{\Sigma}_{\mathrm{X}} ^{\mathrm{par}} \coloneq \KL{ P_{ \mathbf{X}_{t:t+1}, \mathbf{Y}_{t:t+1} } }{ \tilde{Q}_{ \mathbf{X}_{t:t+1}, \mathbf{Y}_{t:t+1} } ^{\mathrm{par}} }
        \label{trep_par}
    \end{align}
    where $ \tilde{Q}_{ \mathbf{X}_{t:t+1}, \mathbf{Y}_{t:t+1} } ^{\mathrm{par}}$ is defined as follows:
    \begin{align}
        \tilde{Q}_{ \mathbf{X}_{t:t+1}, \mathbf{Y}_{t:t+1} } ^{\mathrm{par}} (\mathbf{x}_{t:t+1}, \mathbf{y}_{t:t+1})
        \coloneq& P_{ \mathbf{X}_{t+1} | \mathbf{Y}_{t+1}, \mathbf{X}_{t}, \mathbf{Y}_{t} } ( \mathbf{x}_{t} | \mathbf{y}_{t+1}, \mathbf{x}_{t+1}, \mathbf{y}_{t} ) P_{ \mathbf{X}_{t}, \mathbf{Y}_{t+1}, \mathbf{Y}_{t} } (\mathbf{x}_{t+1}, \mathbf{y}_{t+1}, \mathbf{y}_{t}) \\
        =& P_{ \mathbf{X}_{t+1}, \mathbf{Y}_{t+1}, \mathbf{X}_{t}, \mathbf{Y}_{t} } ( \mathbf{x}_{t}, \mathbf{y}_{t+1}, \mathbf{x}_{t+1}, \mathbf{y}_{t} )
    \end{align}
	${Q}_{ \mathbf{X}_{t:t+1}, \mathbf{Y}_{t:t+1} }^\mathrm{par}$ is a probability distribution obtained by swapping only the values of subsystem X in the forward distribution to the opposite temporal direction. In the following, we call this a partial time-reversed process of X. We also call the distribution of a partial time-reversed process a partial time-reversed distribution\footnote{In a partial time-reversed distribution, only the values taken by subsystem X in the probability distribution of the forward process are swapped in the opposite temporal direction. On the other hand, in the partial backward distribution, the values taken by both subsystems X and Y are swapped around in the temporal direction \cite{itoUnifiedFrameworkEntropy2020, auconiInformationThermodynamicsTime2019,crooksMarginalConditionalSecond2019}. In a non-bipartite time evolution, due to the dependency of X and Y at the same time, it is appropriate to swap the values taken by both X and Y as for a partial backward distribution, but even in a partial time-reversed distribution, swapping the values of both X and Y yields $\tilde{Q}_{ \mathbf{X}_{t:t+1}, \mathbf{Y}_{t:t+1} } ^\mathrm{par}=\tilde{Q}_{ \mathbf{X}_{t:t+1}, \mathbf{Y}_{t:t+1} }$, which results in the total dissipation function and the partial dissipation function being equivalent. Here, to avoid this problem, for the partial entropy production and partial dissipation function of X, we have used different definitions for the substitution of values taken by Y. }.
\end{dfn}
For these forms of partial entropy production and dissipation function, the non-negativity property holds.
\begin{result}[Non-Negativity of Partial Entropy Production]
For partial entropy production $\Sigma_{\mathrm{X}} ^{\mathrm{par}}$ and partial dissipation function $\tilde{\Sigma}_{\mathrm{X}} ^{\mathrm{par}}$,
\begin{align}
    \Sigma_{\mathrm{X}} ^{\mathrm{par}} &\geq 0, \\
    \tilde{\Sigma}_{\mathrm{X}} ^{\mathrm{par}} &\geq 0.
\end{align}
\end{result}
\begin{proof}
    This follows directly from the non-negativity of KLD. Based on the conditions under which KLD is zero, this relationship becomes an equality in the case of entropy production when the following condition holds for all $\mathbf{x}_{t},\mathbf{y}_{t},\mathbf{x}_{t+1},\mathbf{y}_{t+1}$:
    \begin{equation}
        P_{ \mathbf{X}_{t:t+1}, \mathbf{Y}_{t:t+1} } ( \mathbf{x}_{t:t+1}. \mathbf{y}_{t:t+1})= Q_{ \mathbf{X}_{t:t+1}, \mathbf{Y}_{t:t+1} } ^\mathrm{par}( \mathbf{x}_{t:t+1}, \mathbf{y}_{t:t+1}).
    \end{equation}
    Partial entropy production expresses the difference between the forward distribution and the partial backward distribution.

    For dissipation function, the following relationship holds for all $\mathbf{x}_{t},\mathbf{y}_{t},\mathbf{x}_{t+1},\mathbf{y}_{t+1}$:
    \begin{equation}
        P_{ \mathbf{X}_{t:t+1}. \mathbf{Y}_{t:t+1} }( \mathbf{x}_{t:t+1}, \mathbf{y}_{t:t+1}) = \tilde{Q}_{ \mathbf{X}_{t:t+1}, \mathbf{Y}_{t:t+1} } ^{\mathrm{par}}( \mathbf{x}_{t:t+1}, \mathbf{y}_{t:t+1})
    \end{equation}
    Partial dissipation function expresses the difference between the forward distribution and the partial time-reversed distribution.
\end{proof}
\subsubsection{Marginal Entropy Production}
Marginal entropy production can be defined using the marginal distribution of X obtained by summing the joint distributions of a composite system XY for X (By exchanging the random variables X and Y, the marginal entropy production of Y can also be defined in the same way).
\begin{dfn}[Marginal Entropy Production]
    \begin{align}
        \Sigma_{\mathrm{X}} ^{\mathrm{mar}} \coloneq \KL{ P_{ \mathbf{X}_{t:t+1} } }{ Q^{\mathrm{mar}}_{ \mathbf{X}_{t:t+1} } }
    \end{align}
    where $P_{ \mathbf{X}_{t:t+1} } (\mathbf{x}_{t:t+1})$ is obtained by marginalizing $P_{ \mathbf{X}_{t:t+1}, \mathbf{Y}_{t:t+1} } (\mathbf{x}_{t:t+1}, \mathbf{y}_{t:t+1})$ with respect to Y.
    \begin{align}
        P_{ \mathbf{X}_{t:t+1} } (\mathbf{x}_{t:t+1}) =& \dsum_{\mathbf{y}_{t:t+1}} P_{ \mathbf{X}_{t:t+1}, \mathbf{Y}_{t:t+1} } (\mathbf{x}_{t:t+1}, \mathbf{y}_{t:t+1})
    \end{align}
    Unlike the marginalization of $Q_{ \mathbf{X}_{t:t+1}, \mathbf{Y}_{t:t+1} } (\mathbf{x}_{t:t+1}, \mathbf{y}_{t:t+1})$ with respect to Y, $Q^{\mathrm{mar}}_{ \mathbf{X}_{t:t+1} }$ is defined separately as follows:
    \begin{align}
        Q^{\mathrm{mar}}_{ \mathbf{X}_{t:t+1} }(\mathbf{x}_{t:t+1}) =&P_{\mathbf{X}_{t+1}|\mathbf{X}_{t}} (\mathbf{x}_{t}|\mathbf{x}_{t+1}) P_{\mathbf{X}_{t+1}} (\mathbf{x}_{t+1}) \\
        \neq & \dsum_{\mathbf{y}_{t:t+1}} Q_{ \mathbf{X}_{t:t+1}, \mathbf{Y}_{t:t+1} } (\mathbf{x}_{t:t+1}, \mathbf{y}_{t:t+1})
    \end{align}
    $Q^{\mathrm{mar}}_{ \mathbf{X}_{t:t+1} }$ is mathematically equivalent to the backward distribution when regarding X as a single system. We call the backward process of X in a composite system XY a marginal backward process, and we call a distribution of marginal backward process a marginal backward distribution.
\end{dfn}
\begin{dfn}[Marginal Dissipation Function]
    \begin{align}
        \tilde{\Sigma}_{\mathrm{X}} ^{\mathrm{mar}} \coloneq \KL{ P_{ \mathbf{X}_{t:t+1} } }{ \tilde{Q}^{\mathrm{mar}}_{ \mathbf{X}_{t:t+1} } }
    \end{align}
    where $P_{ \mathbf{X}_{t:t+1} } (\mathbf{x}_{t:t+1})$ is obtained by marginalizing $P_{ \mathbf{X}_{t:t+1}, \mathbf{Y}_{t:t+1} } (\mathbf{x}_{t:t+1}, \mathbf{y}_{t:t+1})$ with respect to Y, and $\tilde{Q}^{\mathrm{mar}}_{ \mathbf{X}_{t:t+1} } (\mathbf{x}_{t:t+1})$ is also obtained by marginalizing $\tilde{Q}_{ \mathbf{X}_{t:t+1}, \mathbf{Y}_{t:t+1} } (\mathbf{x}_{t:t+1}, \mathbf{y}_{t:t+1})$ with respect to Y.
    \begin{align}
        P_{ \mathbf{X}_{t:t+1} } (\mathbf{x}_{t:t+1}) =& \dsum_{\mathbf{y}_{t:t+1}} P_{ \mathbf{X}_{t:t+1}, \mathbf{Y}_{t:t+1} } (\mathbf{x}_{t:t+1}, \mathbf{y}_{t:t+1}) \\
        \tilde{Q}^{\mathrm{mar}}_{ \mathbf{X}_{t:t+1} } (\mathbf{x}_{t:t+1})
        =&\dsum_{\mathbf{y}_{t:t+1}} \tilde{Q}_{ \mathbf{X}_{t:t+1}, \mathbf{Y}_{t:t+1} } (\mathbf{x}_{t:t+1}, \mathbf{y}_{t:t+1}) \\
        =& \tilde{Q}_{ \mathbf{X}_{t:t+1} } (\mathbf{x}_{t:t+1}) \\
        =& P_{\mathbf{X}_{t+1}, \mathbf{X}_{t}} (\mathbf{x}_{t}, \mathbf{x}_{t+1})
    \end{align}
    $\tilde{Q}^{\mathrm{mar}}_{ \mathbf{X}_{t:t+1} }$ is mathematically equivalent to the time-reversed distribution when regarding X as a single system. We call the time-reversed process of X in a composite system XY a marginal time-reversed process, and we call a distribution of marginal time-reversed process a marginal time-reversed distribution.
\end{dfn}
For these forms of marginal entropy production and dissipation function, the non-negativity property holds.
\begin{result}[Non-Negativity of Marginal Entropy Production]
For marginal entropy production $\Sigma_{\mathrm{X}} ^{\mathrm{mar}}$ and marginal dissipation function $\tilde{\Sigma}_{\mathrm{X}} ^{\mathrm{mar}}$,
\begin{align}
    \Sigma_{\mathrm{X}} ^{\mathrm{mar}} &\geq 0, \\
    \tilde{\Sigma}_{\mathrm{X}} ^{\mathrm{mar}} &\geq 0.
\end{align}
\end{result}
\begin{proof}
    This follows directly from the non-negativity of KLD. Based on the conditions under which KLD is zero, this relationship becomes an equality in the case of entropy production when the following condition holds for all $\mathbf{x}_{t},\mathbf{x}_{t+1}$.
    \begin{align}
        P_{ \mathbf{X}_{t:t+1} } (\mathbf{x}_{t}, \mathbf{x}_{t+1}) = Q_{ \mathbf{X}_{t:t+1}} (\mathbf{x}_{t}, \mathbf{x}_{t+1})
    \end{align}
    In dissipation function, the following condition holds for all $\mathbf{x}_{t},\mathbf{x}_{t+1}$.
    \begin{align}
        P_{ \mathbf{X}_{t:t+1} } (\mathbf{x}_{t}, \mathbf{x}_{t+1}) = \tilde{Q}_{ \mathbf{X}_{t:t+1} } (\mathbf{x}_{t}, \mathbf{x}_{t+1})
    \end{align} This means that
    \begin{align}
        P_{\mathbf{X}_{t+1}, |\mathbf{X}_{t}, } (\mathbf{x}_{t+1}|\mathbf{x}_{t}) P_{\mathbf{X}_{t}} (\mathbf{x}_{t}) =& P_{\mathbf{X}_{t+1}|\mathbf{X}_{t}} (\mathbf{x}_{t}|\mathbf{x}_{t+1}) P_{\mathbf{X}_{t}} (\mathbf{x}_{t+1})
    \end{align}
    which is zero in a reversible process where the detailed balance condition holds for the marginalized process of X.
\end{proof}
\subsubsection{Conditional Entropy Production}
When only Y is observed in a composite system XY, it is possible to define the conditional entropy production of X based on the difference between the total entropy production and the marginal entropy production of Y as follows \cite{crooksMarginalConditionalSecond2019,kawaguchiFluctuationTheoremHidden2013}:
\begin{dfn}[Conditional Entropy Production]
    \begin{align}
        \Sigma_{\mathrm{X|Y}} ^{\mathrm{con}} \coloneq& \Sigma_{\mathrm{XY}} ^{\mathrm{tot}} - \Sigma_{\mathrm{Y}} ^{\mathrm{mar}}
    \end{align}
\end{dfn}
\begin{dfn}[Conditional Dissipation Function]
    \begin{align}
        \tilde{\Sigma}_{\mathrm{X|Y}} ^{\mathrm{con}}\coloneq& \tilde{\Sigma}_{\mathrm{XY}} ^{\mathrm{tot}} - \tilde{\Sigma}_{\mathrm{Y}} ^{\mathrm{mar}} \label{def_condf}\\
        =& \KL{ P_{ \mathbf{X}_{t:t+1} | \mathbf{Y}_{t:t+1} } }{ \tilde{Q}_{ \mathbf{X}_{t:t+1} | \mathbf{Y}_{t:t+1} } }
    \end{align}
\end{dfn}
Here, the conditional entropy production cannot be expressed in KLD, but the conditional dissipation function can. In other words, the non-negativity of conditional entropy production does not hold, but that conditional dissipation-function does hold. This is also discussed in chapter \ref{IneEntPro} as the inequalities between total entropy production and marginal entropy production.

\subsubsection{Previous Studies}
Previous studies focused on either entropy production or dissipation function in discrete-time composite systems. We showed some cases in Table \ref{table:com_EP_table}. When a continuous-time stochastic process is discretized with a finite time interval, its definition is taken to be the one in discrete-time as Table \ref{table:EP_table}.
\begin{table}[H]
\caption{Definition types of entropy production in composite systems and previous studies.}
 \label{table:com_EP_table}
 \centering
  \begin{tabular}{cc}
   \toprule
   Types &Previous studies \\
   \midrule
    $\Sigma_{\mathrm{XY}}^{\mathrm{tot}}$
    & Crooks \textit{et al.}~(2019), Ito \textit{et al.}~(2020) \\
    $\tilde{\Sigma}_{\mathrm{XY}}^{\mathrm{tot}}$
    & Aunconi \textit{et al.}~(2019) \\
    $\Sigma_{\mathrm{X}}^{\mathrm{par}}$
    & Ito \textit{et al.}~(2020) \\
    $\tilde{\Sigma}_{\mathrm{X}}^{\mathrm{par}}$
    & Aunconi \textit{et al.}~(2019) \\
    $\Sigma_{\mathrm{X}}^{\mathrm{mar}}$
    & \\
    $\tilde{\Sigma}_{\mathrm{X}}^{\mathrm{mar}}$
    & Aunconi \textit{et al.}~(2019) \\
    $\Sigma_{\mathrm{X|Y}}^{\mathrm{con}}$
    & \\
    $\tilde{\Sigma}_{\mathrm{X|Y}}^{\mathrm{con}}$
    & Aunconi \textit{et al.}~(2019) \\
   \midrule
  \end{tabular}
\end{table}
\section{Decomposition of Entropy Production}
\label{DecEntPro}
This chapter describes the decomposition of the entropy production defined in chapter \ref{EntPro}, the derivation of inequalities from the non-negativity of entropy production, and their implications. Section \ref{sec:dec_thermo} discusses decomposition related to the inequality known as the second law of thermodynamics and we call this way of decomposition thermodynamic decomposition. Section \ref{sec:dec_infothermo} discusses decomposition related to the inequality known as the second law of information thermodynamics and we call that information-thermodynamic decomposition. Section \ref{sec:dec_info} focuses on the information-theoretic property of entropy production and discusses decomposition based on information-theoretic quantity and we call that way information-theoretic decomposition.

\subsection{Thermodynamic Decomposition}
\label{sec:dec_thermo}
In this section, we decompose the entropy production of a single system, the total entropy production of a composite system, and marginal entropy production into summated terms, entropy change and entropy flow, and we describe their physical interpretation based on an assumption called local detailed balance. We also derive the inequality known as the second law of thermodynamics from this decomposition and the non-negativity of KLD.

\subsubsection{Entropy Production of a Single System} \label{sig_2nd}

Consider a physical system whose state is represented by a random variable $\mathbf{S}_{t}$ in contact with a heat bath that has an inverse temperature $\beta$. In this system, the total entropy change that occurs in the temporal evolution from one time $t$ to the next time $t+1$ is expressed as the sum of the entropy change of the system and the entropy change of the heat bath that results from interaction between the system and the heat bath.

\subsubsection*{Entropy Production}
Taking the sum of the stochastic entropy of a system at  time $t$, $-\log P_{\mathbf{S}_{t}} (\mathbf{s}_{t})$, and the stochastic entropy change of the heat bath, $\sigma_{\mathrm{bath}}$, then the change in total entropy is given by
\begin{align}
    &\sigma_{\mathrm{sys}} + \sigma_{\mathrm{bath}}, \\
    \sigma_{\mathrm{sys}} \coloneq-&\log P_{\mathbf{S}_{t+1}} (\mathbf{s}_{t+1}) + \log P_{\mathbf{S}_{t}} (\mathbf{s}_{t})
\end{align}
and the expected value is given by
\begin{align}
    &\Sigma_{\mathrm{sys}} + \Sigma_{\mathrm{bath}}, \\
    \Sigma_{\mathrm{sys}} \coloneq& \expec{S}{\sigma_{\mathrm{sys}}} = H[\mathbf{S}_{t+1}] - H[\mathbf{S}_{t}], \\
    \Sigma_{\mathrm{bath}} \coloneq& \expec{S}{\sigma_{\mathrm{bath}}}.
\end{align}
Here, the following assumptions, so-called local detailed balance, are made.
\begin{asmp}[Local Detailed Balance for Entropy Production]
The entropy change of the heat bath is expressed in terms of transition probabilities as follows.
    \begin{align} \label{ldb_bep}
        \sigma_{\mathrm{bath}}=
        \log \dfrac{P_{\mathbf{S}_{t+1}|\mathbf{S}_{t}} (\mathbf{s}_{t+1}|\mathbf{s}_{t})}{P_{\mathbf{S}_{t+1}|\mathbf{S}_{t}} (\mathbf{s}_{t}|\mathbf{s}_{t+1})}
    \end{align}
\end{asmp}
By definition, entropy production $\Sigma$ can be decomposed as follows.
\begin{align}
    \Sigma =& \expec{S_{t:t+1}}{
    \log \dfrac{P_{\mathbf{S}_{t+1}|\mathbf{S}_{t}} (\mathbf{s}_{t+1}|\mathbf{s}_{t}) P_{\mathbf{S}_{t}} (\mathbf{s}_{t})}{P_{\mathbf{S}_{t+1}|\mathbf{S}_{t}} (\mathbf{s}_{t}|\mathbf{s}_{t+1}) P_{\mathbf{S}_{t+1}} (\mathbf{s}_{t+1})}
    } \\
    =& H[\mathbf{S}_{t+1}] - H[\mathbf{S}_{t}] + \expec{S_{t:t+1}}{
    \log \dfrac{P_{\mathbf{S}_{t+1}|\mathbf{S}_{t}} (\mathbf{s}_{t+1}|\mathbf{s}_{t})}{P_{\mathbf{S}_{t+1}|\mathbf{S}_{t}} (\mathbf{s}_{t}|\mathbf{s}_{t+1})}
    } \label{dec_singleEP}
\end{align}
Therefore, based on the local detailed balance \eqref{ldb_bep}, the entropy production can be decomposed as
\begin{align}
    \Sigma = \Sigma_{\mathrm{sys}} + \Sigma_{\mathrm{bath}}
\end{align}
and it can be seen that this is a physical quantity expressing the total entropy change. The non-negativity of entropy production can also be interpreted as the second law of thermodynamics, which states that total entropy is non-decreasing.
\begin{result}[Second Law of Thermodynamics]
    Based on the local detailed balance \eqref{ldb_bep}, the following relationship holds.
        \begin{align}
            \label{thr_2nd}
            \Sigma_{\mathrm{sys}} + \Sigma_{\mathrm{bath}} &\geq 0
        \end{align}
\end{result}
The equality holds when the forward distribution and the backward distribution are equal based on the condition that the entropy production becomes zero.

\subsubsection*{Dissipation Function}
A similar decomposition can be performed for dissipation function \cite{gaspardTimeReversedDynamicalEntropy2004}. The stochastic entropy changes in the system and heat bath are defined in the same way as for entropy production.
\begin{align}
    &\tilde{\sigma}_{\mathrm{sys}} + \tilde{\sigma}_{\mathrm{bath}}, \\
    \tilde{\sigma}_{\mathrm{sys}} \coloneq- &\log P_{\mathbf{S}_{t+1}} (\mathbf{s}_{t+1}) + \log P_{\mathbf{S}_{t}} (\mathbf{s}_{t}).
\end{align}
The expected value is given by
\begin{align}
    &\tilde{\Sigma}_{\mathrm{sys}} + \tilde{\Sigma}_{\mathrm{bath}}, \\
    \tilde{\Sigma}_{\mathrm{sys}} \coloneq& \expec{S}{\tilde{\sigma}_{\mathrm{sys}}} = H[\mathbf{S}_{t+1}] - H[\mathbf{S}_{t}], \\
    \tilde{\Sigma}_{\mathrm{bath}} \coloneq& \expec{S}{\tilde{\sigma}_{\mathrm{bath}}}.
\end{align}
Under the same assumptions as in case of entropy production, the entropy change of the heat bath is expressed as follows.
\begin{align}
	\label{ldb_trep} \tilde{\sigma}_{\mathrm{bath}}=
	\log \dfrac{P_{\mathbf{S}_{t+1}|\mathbf{S}_{t}} (\mathbf{s}_{t+1}|\mathbf{s}_{t}) P_{\mathbf{S}_{t+1}} (\mathbf{s}_{t+1})}{P_{\mathbf{S}_{t+1}|\mathbf{S}_{t}} (\mathbf{s}_{t}|\mathbf{s}_{t+1}) P_{\mathbf{S}_{t}} (\mathbf{s}_{t+1})}
\end{align}
Then, dissipation function $\tilde{\Sigma}$ can be decomposed as follows.
\begin{align}
    \tilde{\Sigma} =& \expec{S_{t:t+1}}{
    \log \dfrac{P_{\mathbf{S}_{t+1}|\mathbf{S}_{t}} (\mathbf{s}_{t+1}|\mathbf{s}_{t}) P_{\mathbf{S}_{t}} (\mathbf{s}_{t})}{P_{\mathbf{S}_{t+1}|\mathbf{S}_{t}} (\mathbf{s}_{t}|\mathbf{s}_{t+1}) P_{\mathbf{S}_{t}} (\mathbf{s}_{t+1})}
    } \\
    =& H[\mathbf{S}_{t+1}] - H[\mathbf{S}_{t}] + \expec{S_{t:t+1}}{
    \log \dfrac{P_{\mathbf{S}_{t+1}|\mathbf{S}_{t}} (\mathbf{s}_{t+1}|\mathbf{s}_{t}) P_{\mathbf{S}_{t+1}} (\mathbf{s}_{t+1})}{P_{\mathbf{S}_{t+1}|\mathbf{S}_{t}} (\mathbf{s}_{t}|\mathbf{s}_{t+1}) P_{\mathbf{S}_{t}} (\mathbf{s}_{t+1})}
    } \\
	=& \tilde{\Sigma}_{\mathrm{sys}} + \tilde{\Sigma}_{\mathrm{bath}}
\end{align}
The non-negativity of dissipation function can also be interpreted as the second law of thermodynamics.
\begin{align}
	\tilde{\Sigma}_{\mathrm{sys}} + \tilde{\Sigma}_{\mathrm{bath}} &\geq 0.
\end{align}
The equality holds when the system's forward process is time-reversed process based on the condition that the dissipation function becomes zero.

\subsubsection{Total Entropy Production in a Composite System}
The decomposition and inequality of total entropy production in a composite system can be discussed in the same way by setting $\mathbf{S}_{t} = \brc{\mathbf{X}_{t}, \mathbf{Y}_{t}}$ in the single-system.

\subsubsection*{Total Entropy Production}
Total entropy production can be decomposed as follows.
\begin{align}
    \Sigma_{\mathrm{XY}}^{\mathrm{tot}} = \Sigma_{\mathrm{sys}} ^{\mathrm{tot}} + \Sigma_{\mathrm{bath}}^{\mathrm{tot}} \geq 0
\end{align}
Here,
\begin{align}
    &\Sigma_{\mathrm{sys}} ^{\mathrm{tot}} \coloneq H[\mathbf{X}_{t+1},\mathbf{Y}_{t+1}] - H[\mathbf{X}_{t},\mathbf{Y}_{t}], \\
    &\Sigma_{\mathrm{bath}} ^{\mathrm{tot}} = \expec{ \mathbf{X}_{t:t+1}, \mathbf{Y}_{t:t+1} }{ \log \frac{ P_{ \mathbf{X}_{t+1}, \mathbf{Y}_{t+1} | \mathbf{X}_{t}, \mathbf{Y}_{t} } (\mathbf{x}_{t+1}, \mathbf{y}_{t+1} | \mathbf{x}_{t}, \mathbf{y}_{t}) }{ P_{ \mathbf{X}_{t+1}, \mathbf{Y}_{t+1} | \mathbf{X}_{t}, \mathbf{Y}_{t} } (\mathbf{x}_{t}, \mathbf{y}_{t} | \mathbf{x}_{t+1}, \mathbf{y}_{t+1}) } }.
\end{align}
The non-negativity of total entropy production can be interpreted as an expression of the second law of thermodynamics whereby the entropy of a composite system XY is non-decreasing.

\subsubsection*{Total Dissipation Function}
Total dissipation function can be decomposed as follows.
\begin{align}
    \tilde{\Sigma}_{\mathrm{XY}} ^{\mathrm{tot}} =& \tilde{\Sigma}_{\mathrm{XY}}^{\mathrm{tot}} + \tilde{\Sigma}_{\mathrm{bath}}^{\mathrm{tot}} \geq 0
    \label{dec_trep_tot}
\end{align}
Here, 
\begin{align}
    &\tilde{\Sigma}_{\mathrm{XY}}^{\mathrm{tot}} \coloneq H[\mathbf{X}_{t+1},\mathbf{Y}_{t+1}] - H[\mathbf{X}_{t},\mathbf{Y}_{t}], \\
    &\tilde{\Sigma}_{\mathrm{bath}}^{\mathrm{tot}} = \expec{ \mathbf{X}_{t:t+1}, \mathbf{Y}_{t:t+1} }{ \log \frac{ P_{ \mathbf{X}_{t+1}, \mathbf{Y}_{t+1} | \mathbf{X}_{t}, \mathbf{Y}_{t} } (\mathbf{x}_{t+1}, \mathbf{y}_{t+1} | \mathbf{x}_{t}, \mathbf{y}_{t}) P_{ \mathbf{X}_{t+1}, \mathbf{Y}_{t+1} } (\mathbf{x}_{t+1}, \mathbf{y}_{t+1}) }{ P_{ \mathbf{X}_{t+1}, \mathbf{Y}_{t+1} | \mathbf{X}_{t}, \mathbf{Y}_{t} } (\mathbf{x}_{t}, \mathbf{y}_{t} | \mathbf{x}_{t+1}, \mathbf{y}_{t+1}) P_{ \mathbf{X}_{t}, \mathbf{Y}_{t} } (\mathbf{x}_{t+1}, \mathbf{y}_{t+1}) } }.
\end{align}
The non-negativity of total dissipation function can also be interpreted as an expression of the second law of thermodynamics.

\subsubsection{Marginal Entropy Production in a Composite System}
The decomposition and inequality of marginal entropy production in a composite system can also be discussed in the same way as the case of single system\cite{crooksMarginalConditionalSecond2019}. 

\subsubsection*{Marginal Entropy Production}
\begin{gather}
    \Sigma_{\mathrm{X}} ^{\mathrm{mar}}
    = \Sigma_{\mathrm{sys}} ^{\mathrm{mar}} + \Sigma_{\mathrm{bath}} ^{\mathrm{mar}} \geq 0 \\
	\Sigma_{\mathrm{sys}} ^{\mathrm{mar}}=H[\mathbf{X}_{t+1}] - H[\mathbf{X}_{t}]\\
    \Sigma_{\mathrm{bath}} ^{\mathrm{mar}}= \expec{ \mathbf{X}_{t:t+1} }{ \log \frac{ P_{ \mathbf{X}_{t+1} | \mathbf{X}_{t} } (\mathbf{x}_{t+1} | \mathbf{x}_{t}) }{ P_{ \mathbf{X}_{t+1} | \mathbf{X}_{t} } (\mathbf{x}_{t} | \mathbf{x}_{t+1}) } }
    \label{dec_bep_mar}
\end{gather}
\subsubsection*{Marginal Dissipation Function}
\begin{gather}
    \tilde{\Sigma}_{\mathrm{X}} ^{\mathrm{mar}}
    = \tilde{\Sigma}_{\mathrm{sys}} ^{\mathrm{mar}} + \tilde{\Sigma}_{\mathrm{bath}} ^{\mathrm{mar}} \geq 0 \\
	\tilde{\Sigma}_{\mathrm{sys}} ^{\mathrm{mar}} =H[\mathbf{X}_{t+1}] - H[\mathbf{X}_{t}]\\
    \tilde{\Sigma}_{\mathrm{bath}} ^{\mathrm{mar}}= \expec{ \mathbf{X}_{t:t+1} }{ \log \frac{ P_{ \mathbf{X}_{t+1} | \mathbf{X}_{t} } (\mathbf{x}_{t+1} | \mathbf{x}_{t}) P_{ \mathbf{X}_{t+1} } (\mathbf{x}_{t+1}) }{ P_{ \mathbf{X}_{t+1} | \mathbf{X}_{t} } (\mathbf{x}_{t} | \mathbf{x}_{t+1})P_{ \mathbf{X}_{t} } (\mathbf{x}_{t+1}) } } \label{dec_trep_mar}
\end{gather}
The non-negativity of marginal entropy production/dissipation function can be interpreted as an expression of the second law of thermodynamics whereby the entropy of X is non-decreasing when only X is observable in a composite system XY.

\subsection{Information-Thermodynamic Decomposition}
\label{sec:dec_infothermo}
In this section, we decompose the partial entropy production, marginal entropy production and conditional entropy production in a composite system, and we describe their physical interpretation under the local detailed balance. We also derive the inequality known as the second law of information thermodynamics from this decomposition and the non-negativity of KLD.

\subsubsection{Partial Entropy Production}
Partial entropy production of a subsystem X in a composite system XY can be decomposed into the Shannon entropy change of X, the entropy change of the heat bath, and the flow of information from X to Y \cite{itoBackwardTransferEntropy2016,auconiInformationThermodynamicsTime2019}.

\subsubsection*{Partial Entropy Production}
The partial entropy production in X can be decomposed as follows by using the Shannon entropy, the conditional Shannon entropy, and the definition of mutual information \cite{coverElementsInformationTheory2012}.
\begin{align}
    \Sigma_{\mathrm{X}} ^{\mathrm{par}}
    =& H[\mathbf{X}_{t+1}] - H[\mathbf{X}_{t}]
    + \expec{ \mathbf{X}_{t:t+1}, \mathbf{Y}_{t:t+1} }{ \log \frac{ P_{ \mathbf{X}_{t+1} | \mathbf{X}_{t}, \mathbf{Y}_{t+1}, \mathbf{Y}_{t} } (\mathbf{x}_{t+1} | \mathbf{x}_{t}, \mathbf{y}_{t+1}, \mathbf{y}_{t}) }{ P_{ \mathbf{X}_{t+1} | \mathbf{X}_{t}, \mathbf{Y}_{t+1}, \mathbf{Y}_{t} } (\mathbf{x}_{t} | \mathbf{x}_{t+1}, \mathbf{y}_{t}, \mathbf{y}_{t+1}) } } \\
    & - I[\mathbf{X}_{t+1} ; \mathbf{Y}_{t:t+1}] + I[\mathbf{X}_{t} ; \mathbf{Y}_{t:t+1}] \\
    =& H[\mathbf{X}_{t+1}] - H[\mathbf{X}_{t}]
    + \expec{ \mathbf{X}_{t:t+1}, \mathbf{Y}_{t:t+1} }{ \log \frac{ P_{ \mathbf{X}_{t+1} | \mathbf{X}_{t}, \mathbf{Y}_{t+1}, \mathbf{Y}_{t} } (\mathbf{x}_{t+1} | \mathbf{x}_{t}, \mathbf{y}_{t+1}, \mathbf{y}_{t}) }{ P_{ \mathbf{X}_{t+1} | \mathbf{X}_{t}, \mathbf{Y}_{t+1}, \mathbf{Y}_{t} } (\mathbf{x}_{t} | \mathbf{x}_{t+1}, \mathbf{y}_{t}, \mathbf{y}_{t+1}) } } \label{parepdec} \\
    & - I[\mathbf{X}_{t+1} ; \mathbf{Y}_{t+1}] + I[\mathbf{X}_{t} ; \mathbf{Y}_{t}] - T^{\dagger}_{\mathrm{X} \rightarrow \mathrm{Y}} + T_{\mathrm{X} \rightarrow \mathrm{Y}}
\end{align}
Here, using the chain rule of mutual information\cite{coverElementsInformationTheory2012}, $T_{\mathrm{X} \rightarrow \mathrm{Y}}$ and $T^{\dagger}_{\mathrm{X} \rightarrow \mathrm{Y}}$ are quantities called the transfer entropy \cite{schreiberMeasuringInformationTransfer2000} and backward transfer entropy \cite{itoBackwardTransferEntropy2016} from X to Y respectively, which are defined in terms of conditional mutual information \cite{coverElementsInformationTheory2012} as follows.
\begin{align}
    T_{\mathrm{X} \rightarrow \mathrm{Y}} \coloneq I[\mathbf{X}_{t};\mathbf{Y}_{t+1}|\mathbf{Y}_{t}] \\
    T^{\dagger}_{\mathrm{X} \rightarrow \mathrm{Y}} \coloneq I[\mathbf{X}_{t+1};\mathbf{Y}_{t}|\mathbf{Y}_{t+1}]
\end{align}
$T_{\mathrm{X} \rightarrow \mathrm{Y}}$ represents the dependence of the current value of X and the next value of Y given the current value of Y. $T^{\dagger}_{\mathrm{X} \rightarrow \mathrm{Y}}$ represents the dependence of the next value of X and the current value of Y given the next value of Y.

Under the local detailed balance for the partial entropy production, third term of \eqref{parepdec} is equal to the entropy change of the heat bath.
\begin{align}
	\Sigma_{\mathrm{X}} ^{\mathrm{par}}=\expec{ \mathbf{X}_{t:t+1}, \mathbf{Y}_{t:t+1} }{ \log \frac{ P_{ \mathbf{X}_{t+1} | \mathbf{X}_{t}, \mathbf{Y}_{t+1}, \mathbf{Y}_{t} } (\mathbf{x}_{t+1} | \mathbf{x}_{t}, \mathbf{y}_{t+1}, \mathbf{y}_{t}) }{ P_{ \mathbf{X}_{t+1} | \mathbf{X}_{t}, \mathbf{Y}_{t+1}, \mathbf{Y}_{t} } (\mathbf{x}_{t} | \mathbf{x}_{t+1}, \mathbf{y}_{t}, \mathbf{y}_{t+1}) } }
\end{align}
From the above, the partial entropy production of X can be decomposed into the sum of the change of the marginal entropy of X, the entropy change of the heat bath, the change of the mutual information of X and Y, the transfer entropy, and the backward transfer entropy.
\begin{align}
    \Sigma_{\mathrm{X}} ^{\mathrm{par}}
    =& H[\mathbf{X}_{t+1}] - H[\mathbf{X}_{t}] + \Sigma_{\mathrm{bath}} ^{\mathrm{par}} - I[\mathbf{X}_{t+1} ; \mathbf{Y}_{t+1}] + I[\mathbf{X}_{t} ; \mathbf{Y}_{t}] - T^{\dagger}_{\mathrm{X} \rightarrow \mathrm{Y}} + T_{\mathrm{X} \rightarrow \mathrm{Y}} \label{dec_parEP_inf}
\end{align}
The non-negativity of partial entropy production can be interpreted as the second law of information thermodynamics for a subsystem X.
\begin{result}[Second Law of Information Thermodynamics]
    \begin{gather}
        H[\mathbf{X}_{t+1}] - H[\mathbf{X}_{t}] + \Sigma_{\mathrm{bath}} ^{\mathrm{par}} \geq \Theta_{\mathrm{X} \rightarrow \mathrm{Y}},\\
        \Theta_{\mathrm{X} \rightarrow \mathrm{Y}} \coloneq I[\mathbf{X}_{t+1} ; \mathbf{Y}_{t+1}] - I[\mathbf{X}_{t} ; \mathbf{Y}_{t}] + T^{\dagger}_{\mathrm{X} \rightarrow \mathrm{Y}} - T_{\mathrm{X} \rightarrow \mathrm{Y}}. \label{ddefdif}
    \end{gather}
    That is, the sum of the entropy change of system X and the entropy change of the heat bath $H[\mathbf{X}_{t+1}] - H[\mathbf{X}_{t}] + \Sigma_{\mathrm{bath}} ^{\mathrm{par}}$ is the thermodynamic cost\footnote{The quantity $H[\mathbf{X}_{t+1}] - H[\mathbf{X}_{t}] + \Sigma_{\mathrm{bath}} ^{\mathrm{par}}$ defined here is sometimes called the partial entropy of X \cite{hartichSensoryCapacityInformation2016}.}, $\Theta_{\mathrm{X} \rightarrow \mathrm{Y}}$ is called the dynamic information flow, and since this signifies the flow of information from X to Y, it expresses how the thermodynamic cost required for the dynamics of subsystem X is bounded from below by the flow of information from X to Y \cite{itoUnifiedFrameworkEntropy2020}.
\end{result}

\subsubsection*{Partial Dissipation Function}
A similar decomposition can be performed for partial dissipation function as follows.
\begin{align}
    \tilde{\Sigma}_{\mathrm{X}} ^{\mathrm{par}}
    =& H[\mathbf{X}_{t+1}] - H[\mathbf{X}_{t}]  \\
    &+ \expec{ \mathbf{X}_{t:t+1}, \mathbf{Y}_{t:t+1} }{ \log \frac{ P_{ \mathbf{X}_{t+1} | \mathbf{X}_{t}, \mathbf{Y}_{t+1}, \mathbf{Y}_{t} } (\mathbf{x}_{t+1} | \mathbf{x}_{t}, \mathbf{y}_{t+1}, \mathbf{y}_{t}) P_{ \mathbf{X}_{t+1}, \mathbf{Y}_{t+1}, \mathbf{Y}_{t} } (\mathbf{x}_{t}, \mathbf{y}_{t+1}, \mathbf{y}_{t}) }{ P_{ \mathbf{X}_{t+1} | \mathbf{X}_{t}, \mathbf{Y}_{t+1}, \mathbf{Y}_{t} } (\mathbf{x}_{t} | \mathbf{x}_{t+1}, \mathbf{y}_{t+1}, \mathbf{y}_{t}) P_{ \mathbf{X}_{t}, \mathbf{Y}_{t+1}, \mathbf{Y}_{t} } (\mathbf{x}_{t+1}, \mathbf{y}_{t+1}, \mathbf{y}_{t}) } } \\
    & - I[\mathbf{X}_{t+1} ; \mathbf{Y}_{t:t+1}] + I[\mathbf{X}_{t} ; \mathbf{Y}_{t:t+1}] \\
    =& H[\mathbf{X}_{t+1}] - H[\mathbf{X}_{t}]  \\
    &+ \expec{ \mathbf{X}_{t:t+1}, \mathbf{Y}_{t:t+1} }{ \log \frac{ P_{ \mathbf{X}_{t+1} | \mathbf{X}_{t}, \mathbf{Y}_{t+1}, \mathbf{Y}_{t} } (\mathbf{x}_{t+1} | \mathbf{x}_{t}, \mathbf{y}_{t+1}, \mathbf{y}_{t}) P_{ \mathbf{X}_{t+1}, \mathbf{Y}_{t+1}, \mathbf{Y}_{t} } (\mathbf{x}_{t}, \mathbf{y}_{t+1}, \mathbf{y}_{t}) }{ P_{ \mathbf{X}_{t+1} | \mathbf{X}_{t}, \mathbf{Y}_{t+1}, \mathbf{Y}_{t} } (\mathbf{x}_{t} | \mathbf{x}_{t+1}, \mathbf{y}_{t+1}, \mathbf{y}_{t}) P_{ \mathbf{X}_{t}, \mathbf{Y}_{t+1}, \mathbf{Y}_{t} } (\mathbf{x}_{t+1}, \mathbf{y}_{t+1}, \mathbf{y}_{t}) } } \\
    & - I[\mathbf{X}_{t+1} ; \mathbf{Y}_{t+1}] + I[\mathbf{X}_{t} ; \mathbf{Y}_{t}] - T^{\dagger}_{\mathrm{X} \rightarrow \mathrm{Y}} + T_{\mathrm{X} \rightarrow \mathrm{Y}}
\end{align}
Under the same assumption as in case of partial entropy production, the entropy change of the heat bath is
\begin{align}
	\tilde{\Sigma}_{\mathrm{X}} ^{\mathrm{par}}=& \expec{ \mathbf{X}_{t:t+1}, \mathbf{Y}_{t:t+1} }{ \log \frac{ P_{ \mathbf{X}_{t+1} | \mathbf{X}_{t}, \mathbf{Y}_{t+1}, \mathbf{Y}_{t} } (\mathbf{x}_{t+1} | \mathbf{x}_{t}, \mathbf{y}_{t+1}, \mathbf{y}_{t}) P_{ \mathbf{X}_{t+1}, \mathbf{Y}_{t+1}, \mathbf{Y}_{t} } (\mathbf{x}_{t}, \mathbf{y}_{t+1}, \mathbf{y}_{t}) }{ P_{ \mathbf{X}_{t+1} | \mathbf{X}_{t}, \mathbf{Y}_{t+1}, \mathbf{Y}_{t} } (\mathbf{x}_{t} | \mathbf{x}_{t+1}, \mathbf{y}_{t+1}, \mathbf{y}_{t}) P_{ \mathbf{X}_{t}, \mathbf{Y}_{t+1}, \mathbf{Y}_{t} } (\mathbf{x}_{t+1}, \mathbf{y}_{t+1}, \mathbf{y}_{t}) } }
\end{align}
Then, partial dissipation function $\tilde{\Sigma}_{\mathrm{X}} ^{\mathrm{par}}$ is expressed as follows.
\begin{align}
    \tilde{\Sigma}_{\mathrm{X}} ^{\mathrm{par}}
    =& H[\mathbf{X}_{t+1}] - H[\mathbf{X}_{t}] + \tilde{\Sigma}_{\mathrm{bath}} ^{\mathrm{par}} - I[\mathbf{X}_{t+1} ; \mathbf{Y}_{t+1}] + I[\mathbf{X}_{t} ; \mathbf{Y}_{t}] - T^{\dagger}_{\mathrm{X} \rightarrow \mathrm{Y}} + T_{\mathrm{X} \rightarrow \mathrm{Y}}
\end{align}
The non-negativity of partial dissipation function can also be interpreted as an expression of the second law of information thermodynamics for a subsystem X.
\begin{align}
	H[\mathbf{X}_{t+1}] - H[\mathbf{X}_{t}] + \tilde{\Sigma}_{\mathrm{bath}} \geq \Theta_{\mathrm{X} \rightarrow \mathrm{Y}}
\end{align}

\subsubsection{Relationship Between Marginal Entropy Production and Partial Entropy Production}
Marginal entropy production can be expressed as the sum of partial entropy production and the information-theoretic quantities\cite{crooksMarginalConditionalSecond2019}. Here, we discuss this relationship for both entropy production and dissipation function.

\subsubsection*{Marginal Entropy Production}
Marginal entropy production can be transformed as follows:
\begin{align}
    \Sigma_{\mathrm{X}} ^{\mathrm{mar}}
    =& H[\mathbf{X}_{t+1}] - H[\mathbf{X}_{t}]
    + \expec{ \mathbf{X}_{t:t+1} }{ \log \frac{ P_{ \mathbf{X}_{t+1} | \mathbf{X}_{t} } (\mathbf{x}_{t+1} | \mathbf{x}_{t}) }{ P_{ \mathbf{X}_{t+1} | \mathbf{X}_{t} } (\mathbf{x}_{t} | \mathbf{x}_{t+1}) } } \\
    & + \expec{ \mathbf{X}_{t:t+1},\mathbf{Y}_{t:t+1} }{ \log \frac{ P_{ \mathbf{X}_{t+1} | \mathbf{X}_{t}, \mathbf{Y}_{t+1}, \mathbf{Y}_{t} } (\mathbf{x}_{t+1} | \mathbf{x}_{t}, \mathbf{y}_{t+1}, \mathbf{y}_{t}) }{ P_{ \mathbf{X}_{t+1} | \mathbf{X}_{t}, \mathbf{Y}_{t+1}, \mathbf{Y}_{t} } (\mathbf{x}_{t} | \mathbf{x}_{t+1}, \mathbf{y}_{t}, \mathbf{y}_{t+1}) } } \\
    & - \expec{ \mathbf{X}_{t:t+1},\mathbf{Y}_{t:t+1} }{ \log \frac{ P_{ \mathbf{X}_{t+1} | \mathbf{X}_{t}, \mathbf{Y}_{t+1}, \mathbf{Y}_{t} } (\mathbf{x}_{t+1} | \mathbf{x}_{t}, \mathbf{y}_{t+1}, \mathbf{y}_{t}) }{ P_{ \mathbf{X}_{t+1} | \mathbf{X}_{t}, \mathbf{Y}_{t+1}, \mathbf{Y}_{t} } (\mathbf{x}_{t} | \mathbf{x}_{t+1}, \mathbf{y}_{t}, \mathbf{y}_{t+1}) } } \\
    =& H[\mathbf{X}_{t+1}] - H[\mathbf{X}_{t}]
    + \expec{ \mathbf{X}_{t:t+1},\mathbf{Y}_{t:t+1} }{ \log \frac{ P_{ \mathbf{X}_{t+1} | \mathbf{X}_{t}, \mathbf{Y}_{t+1}, \mathbf{Y}_{t} } (\mathbf{x}_{t} | \mathbf{x}_{t+1}, \mathbf{y}_{t+1}, \mathbf{y}_{t}) }{ P_{ \mathbf{X}_{t+1} | \mathbf{X}_{t}, \mathbf{Y}_{t+1}, \mathbf{Y}_{t} } (\mathbf{x}_{t+1} | \mathbf{x}_{t}, \mathbf{y}_{t}, \mathbf{y}_{t+1}) } }  \\
    & - \expec{ \mathbf{X}_{t:t+1},\mathbf{Y}_{t:t+1} }{ \log \frac{ P_{ \mathbf{X}_{t+1} | \mathbf{X}_{t}, \mathbf{Y}_{t+1}, \mathbf{Y}_{t} } (\mathbf{x}_{t+1} | \mathbf{x}_{t}, \mathbf{y}_{t+1}, \mathbf{y}_{t}) }{ P_{ \mathbf{X}_{t+1} | \mathbf{X}_{t} } (\mathbf{x}_{t+1} | \mathbf{x}_{t}) } } \\
    & + \expec{ \mathbf{X}_{t:t+1},\mathbf{Y}_{t:t+1} }{ \log \frac{  P_{ \mathbf{X}_{t+1} | \mathbf{X}_{t}, \mathbf{Y}_{t+1}, \mathbf{Y}_{t} } (\mathbf{x}_{t} | \mathbf{x}_{t+1}, \mathbf{y}_{t}, \mathbf{y}_{t+1}) }{ P_{ \mathbf{X}_{t+1} | \mathbf{X}_{t} } (\mathbf{x}_{t} | \mathbf{x}_{t+1}) } }\\
    =& H[\mathbf{X}_{t+1}] - H[\mathbf{X}_{t}]  + \Sigma_{\mathrm{bath}} ^{\mathrm{par}} \\
    & - \expec{ \mathbf{X}_{t:t+1},\mathbf{Y}_{t:t+1} }{ \log \frac{ P_{ \mathbf{X}_{t+1} | \mathbf{X}_{t}, \mathbf{Y}_{t+1}, \mathbf{Y}_{t} } (\mathbf{x}_{t+1} | \mathbf{x}_{t}, \mathbf{y}_{t+1}, \mathbf{y}_{t}) }{ P_{ \mathbf{X}_{t+1} | \mathbf{X}_{t} } (\mathbf{x}_{t+1} | \mathbf{x}_{t}) } } \\
    & + \expec{ \mathbf{X}_{t:t+1},\mathbf{Y}_{t:t+1} }{ \log \frac{  P_{ \mathbf{X}_{t+1} | \mathbf{X}_{t}, \mathbf{Y}_{t+1}, \mathbf{Y}_{t} } (\mathbf{x}_{t} | \mathbf{x}_{t+1}, \mathbf{y}_{t}, \mathbf{y}_{t+1}) }{ P_{ \mathbf{X}_{t+1} | \mathbf{X}_{t} } (\mathbf{x}_{t} | \mathbf{x}_{t+1}) } }
\end{align}
Here, assuming bipartiteness \eqref{def_bip},
\begin{align}
    \Sigma_{\mathrm{X}} ^{\mathrm{mar}}
    =& H[\mathbf{X}_{t+1}] - H[\mathbf{X}_{t}]  + \Sigma_{\mathrm{bath}} ^{\mathrm{par}} \\
    & - \expec{ \mathbf{X}_{t:t+1},\mathbf{Y}_{t:t+1} }{ \log \frac{ P_{ \mathbf{X}_{t+1} | \mathbf{X}_{t}, \mathbf{Y}_{t} } (\mathbf{x}_{t+1} | \mathbf{x}_{t}, \mathbf{y}_{t}) }{ P_{ \mathbf{X}_{t+1} | \mathbf{X}_{t} } (\mathbf{x}_{t+1} | \mathbf{x}_{t}) } } \\
    & + \expec{ \mathbf{X}_{t:t+1},\mathbf{Y}_{t:t+1} }{ \log \frac{  P_{ \mathbf{X}_{t+1} | \mathbf{X}_{t}, \mathbf{Y}_{t} } (\mathbf{x}_{t} | \mathbf{x}_{t+1}, \mathbf{y}_{t+1}) }{ P_{ \mathbf{X}_{t+1} | \mathbf{X}_{t} } (\mathbf{x}_{t} | \mathbf{x}_{t+1}) } }
\end{align}
The second term in the above equation is the transfer entropy $T_{\mathrm{Y} \rightarrow \mathrm{X}}$ from Y to X, and the third term is the quantity where both the values of X and Y in the second term are swapped. We represent this as $T^{\mathrm{R}}_{\mathrm{Y} \rightarrow \mathrm{X}}$ and the difference between the second and third terms is called the transferred dissipation\cite{crooksMarginalConditionalSecond2019}. Then, between the partial entropy production and marginal entropy production, the following relationship holds.
\begin{align}
    \Sigma_{\mathrm{X}} ^{\mathrm{mar}}
    =& H[\mathbf{X}_{t+1}] - H[\mathbf{X}_{t}] + \Sigma_{\mathrm{bath}} ^{\mathrm{par}} - T_{_{\mathrm{Y} \rightarrow \mathrm{X}}} +  T^{\mathrm{R}}_{\mathrm{Y} \rightarrow \mathrm{X}}\\
    =& \Sigma_{\mathrm{X}}^{\mathrm{par}} + \Theta_{\mathrm{X} \rightarrow \mathrm{Y}} -  T_{_{\mathrm{Y} \rightarrow \mathrm{X}}} +  T^{\mathrm{R}}_{\mathrm{Y} \rightarrow \mathrm{X}}
\end{align}
\subsubsection*{Marginal Dissipation Function}
Marginal dissipation function can also be transformed as follows.
\begin{align}
    \tilde{\Sigma}_{\mathrm{X}} ^{\mathrm{mar}}
    =& H[\mathbf{X}_{t+1}] - H[\mathbf{X}_{t}]
    + \expec{ \mathbf{X}_{t:t+1} }{ \log \frac{ P_{ \mathbf{X}_{t+1} | \mathbf{X}_{t} } (\mathbf{x}_{t+1} | \mathbf{x}_{t})
    P_{ \mathbf{X}_{t+1} } (\mathbf{x}_{t+1}) }{ P_{ \mathbf{X}_{t+1} | \mathbf{X}_{t} } (\mathbf{x}_{t} | \mathbf{x}_{t+1})
    P_{ \mathbf{X}_{t} } (\mathbf{x}_{t+1}) } } \\
    & + \expec{ \mathbf{X}_{t:t+1}, \mathbf{Y}_{t:t+1} }{ \log \frac{ P_{ \mathbf{X}_{t+1}| \mathbf{Y}_{t+1}, \mathbf{X}_{t}, \mathbf{Y}_{t} } (\mathbf{x}_{t+1}| \mathbf{y}_{t+1}, \mathbf{x}_{t}, \mathbf{y}_{t}) P_{ \mathbf{Y}_{t+1}, \mathbf{X}_{t+1}, \mathbf{Y}_{t} } (\mathbf{y}_{t+1}, \mathbf{x}_{t+1}, \mathbf{y}_{t}) }{ P_{ \mathbf{X}_{t+1}| \mathbf{Y}_{t+1}, \mathbf{X}_{t}, \mathbf{Y}_{t} } (\mathbf{x}_{t}| \mathbf{y}_{t+1}, \mathbf{x}_{t+1}, \mathbf{y}_{t})
    P_{ \mathbf{Y}_{t+1}, \mathbf{X}_{t}, \mathbf{Y}_{t} } (\mathbf{y}_{t+1}, \mathbf{x}_{t+1}, \mathbf{y}_{t})} }   \\
    & - \expec{ \mathbf{X}_{t:t+1}, \mathbf{Y}_{t:t+1} }{ \log \frac{ P_{ \mathbf{X}_{t+1}| \mathbf{Y}_{t+1}, \mathbf{X}_{t}, \mathbf{Y}_{t} } (\mathbf{x}_{t+1}| \mathbf{y}_{t+1}, \mathbf{x}_{t}, \mathbf{y}_{t}) P_{ \mathbf{Y}_{t+1}, \mathbf{X}_{t+1}, \mathbf{Y}_{t} } (\mathbf{y}_{t+1}, \mathbf{x}_{t+1}, \mathbf{y}_{t}) }{ P_{ \mathbf{X}_{t+1}| \mathbf{Y}_{t+1}, \mathbf{X}_{t}, \mathbf{Y}_{t} } (\mathbf{x}_{t}| \mathbf{y}_{t+1}, \mathbf{x}_{t+1}, \mathbf{y}_{t})
    P_{ \mathbf{Y}_{t+1}, \mathbf{X}_{t}, \mathbf{Y}_{t} } (\mathbf{y}_{t+1}, \mathbf{x}_{t+1}, \mathbf{y}_{t})} }  \\
    =& H[\mathbf{X}_{t+1}] - H[\mathbf{X}_{t}] \\
    &+ \expec{ \mathbf{X}_{t:t+1}, \mathbf{Y}_{t:t+1} }{ \log \frac{ P_{ \mathbf{X}_{t+1}| \mathbf{Y}_{t+1}, \mathbf{X}_{t}, \mathbf{Y}_{t} } (\mathbf{x}_{t+1}| \mathbf{y}_{t+1}, \mathbf{x}_{t}, \mathbf{y}_{t}) P_{ \mathbf{Y}_{t+1}, \mathbf{X}_{t+1}, \mathbf{Y}_{t} } (\mathbf{y}_{t+1}, \mathbf{x}_{t+1}, \mathbf{y}_{t}) }{ P_{ \mathbf{X}_{t+1}| \mathbf{Y}_{t+1}, \mathbf{X}_{t}, \mathbf{Y}_{t} } (\mathbf{x}_{t}| \mathbf{y}_{t+1}, \mathbf{x}_{t+1}, \mathbf{y}_{t})
    P_{ \mathbf{Y}_{t+1}, \mathbf{X}_{t}, \mathbf{Y}_{t} } (\mathbf{y}_{t+1}, \mathbf{x}_{t+1}, \mathbf{y}_{t})} }  \\
    & - \expec{ \mathbf{X}_{t:t+1},\mathbf{Y}_{t:t+1} }{ \log \frac{ P_{ \mathbf{X}_{t+1}| \mathbf{Y}_{t+1}, \mathbf{X}_{t}, \mathbf{Y}_{t} } (\mathbf{x}_{t+1}| \mathbf{y}_{t+1}, \mathbf{x}_{t}, \mathbf{y}_{t})
    P_{ \mathbf{Y}_{t+1}, \mathbf{X}_{t+1}, \mathbf{Y}_{t} } (\mathbf{y}_{t+1}, \mathbf{x}_{t+1}, \mathbf{y}_{t}) }{ P_{ \mathbf{X}_{t+1} | \mathbf{X}_{t} } (\mathbf{x}_{t+1} | \mathbf{x}_{t})
    P_{ \mathbf{X}_{t+1} } (\mathbf{x}_{t+1}) } } \\
    & + \expec{ \mathbf{X}_{t:t+1},\mathbf{Y}_{t:t+1} }{ \log \frac{  P_{ \mathbf{X}_{t+1}| \mathbf{Y}_{t+1}, \mathbf{X}_{t}, \mathbf{Y}_{t} } (\mathbf{x}_{t}| \mathbf{y}_{t+1}, \mathbf{x}_{t+1}, \mathbf{y}_{t})
    P_{ \mathbf{Y}_{t+1}, \mathbf{X}_{t}, \mathbf{Y}_{t} } (\mathbf{y}_{t+1}, \mathbf{x}_{t+1}, \mathbf{y}_{t}) }{ P_{ \mathbf{X}_{t+1} | \mathbf{X}_{t} } (\mathbf{x}_{t} | \mathbf{x}_{t+1})
    P_{ \mathbf{X}_{t} } (\mathbf{x}_{t+1}) } }
\end{align}
Here, assuming bipartiteness,
\begin{align}
    \tilde{\Sigma}_{\mathrm{X}} ^{\mathrm{mar}}
    =& H[\mathbf{X}_{t+1}] - H[\mathbf{X}_{t}] + \tilde{\Sigma}_{\mathrm{bath}} ^{\mathrm{par}} \\
    & - \expec{ \mathbf{X}_{t:t+1},\mathbf{Y}_{t:t+1} }{ \log \frac{ P_{ \mathbf{X}_{t+1}| \mathbf{X}_{t}, \mathbf{Y}_{t} } (\mathbf{x}_{t+1}| \mathbf{x}_{t}, \mathbf{y}_{t})
    }{ P_{ \mathbf{X}_{t+1} | \mathbf{X}_{t} } (\mathbf{x}_{t+1} | \mathbf{x}_{t})
    } } \\
    & + \expec{ \mathbf{X}_{t:t+1},\mathbf{Y}_{t:t+1} }{ \log \frac{  P_{ \mathbf{X}_{t+1}| \mathbf{X}_{t}, \mathbf{Y}_{t} } (\mathbf{x}_{t}| \mathbf{x}_{t+1}, \mathbf{y}_{t})
    }{ P_{ \mathbf{X}_{t+1} | \mathbf{X}_{t} } (\mathbf{x}_{t} | \mathbf{x}_{t+1})
    } } \label{trinmdf} \\
    & - \KL{P_{\mathbf{Y}_{t+1}, \mathbf{Y}_{t} | \mathbf{X}_{t+1}}}{P_{\mathbf{Y}_{t+1}, \mathbf{Y}_{t} | \mathbf{X}_{t}}}
\end{align}
We represent \eqref{trinmdf} as $\tilde{T}^{\mathrm{R}}_{\mathrm{Y} \rightarrow \mathrm{X}}$. This is different from $T^{\mathrm{R}}_{\mathrm{Y} \rightarrow \mathrm{X}}$ and the difference between them is also different from the transferred dissipation, but can be considered as a similar quantity. Therefore, the following relationship holds between partial dissipation function and marginal dissipation function.
\begin{align}
    \tilde{\Sigma}_{\mathrm{X}} ^{\mathrm{mar}}
    =& H[\mathbf{X}_{t+1}] - H[\mathbf{X}_{t}] + \tilde{\Sigma}_{\mathrm{bath}} ^{\mathrm{par}} - T_{\mathrm{Y} \rightarrow \mathrm{X}} +  \tilde{T}^{\mathrm{R}}_{\mathrm{Y} \rightarrow \mathrm{X}} - \KL{P_{\mathbf{Y}_{t+1}, \mathbf{Y}_{t} | \mathbf{X}_{t+1}}}{P_{\mathbf{Y}_{t+1}, \mathbf{Y}_{t} | \mathbf{X}_{t}}} \\
    =& \tilde{\Sigma}_{\mathrm{X}}^{\mathrm{par}} + \Theta_{\mathrm{X} \rightarrow \mathrm{Y}} -  T_{\mathrm{Y} \rightarrow \mathrm{X}} +  \tilde{T}^{\mathrm{R}}_{\mathrm{Y} \rightarrow \mathrm{X}} - \KL{P_{\mathbf{Y}_{t+1}, \mathbf{Y}_{t} | \mathbf{X}_{t+1}}}{P_{\mathbf{Y}_{t+1}, \mathbf{Y}_{t} | \mathbf{X}_{t}}}
\end{align}

\subsubsection{Relationship Between Conditional Entropy Production and Partial Entropy Production}
Conditional entropy production can be expressed as the sum of partial entropy production and the information-theoretic quantities\cite{crooksMarginalConditionalSecond2019}.

By definition, conditional entropy production can be transformed as follows.
\begin{align}
    \Sigma_{\mathrm{X|Y}} ^{\mathrm{con}}
    =& \Sigma_{\mathrm{XY}} ^{\mathrm{tot}} - \Sigma_{\mathrm{Y}} ^{\mathrm{mar}} \\
    =& H[\mathbf{X}_{t+1}, \mathbf{Y}_{t+1}] - H[\mathbf{X}_{t}, \mathbf{Y}_{t}]
    + \expec{ \mathbf{X}_{t:t+1}, \mathbf{Y}_{t:t+1} }{ \log \frac{ P_{ \mathbf{X}_{t+1}, \mathbf{Y}_{t+1} | \mathbf{X}_{t}, \mathbf{Y}_{t} } (\mathbf{x}_{t+1}, \mathbf{y}_{t+1} | \mathbf{x}_{t}, \mathbf{y}_{t}) }{ P_{ \mathbf{X}_{t+1}, \mathbf{Y}_{t+1} | \mathbf{X}_{t}, \mathbf{Y}_{t} } (\mathbf{x}_{t}, \mathbf{y}_{t} | \mathbf{x}_{t+1}, \mathbf{y}_{t+1}) } } \\
    &- H[\mathbf{Y}_{t+1}] + H[\mathbf{Y}_{t}]
    - \expec{ \mathbf{Y}_{t:t+1} }{ \log \frac{ P_{ \mathbf{Y}_{t+1} | \mathbf{Y}_{t} } (\mathbf{y}_{t+1} | \mathbf{y}_{t}) }{ P_{ \mathbf{Y}_{t+1} | \mathbf{Y}_{t} } (\mathbf{y}_{t} | \mathbf{y}_{t+1}) } } \\
    =& H[\mathbf{X}_{t+1} | \mathbf{Y}_{t+1}] - H[\mathbf{X}_{t} | \mathbf{Y}_{t}] + \expec{ \mathbf{X}_{t:t+1}, \mathbf{Y}_{t:t+1} }{ \log \frac{ P_{ \mathbf{X}_{t+1}| \mathbf{Y}_{t+1}, \mathbf{X}_{t}, \mathbf{Y}_{t} } (\mathbf{x}_{t+1} | \mathbf{y}_{t+1}, \mathbf{x}_{t}, \mathbf{y}_{t}) }{ P_{ \mathbf{X}_{t+1} | \mathbf{Y}_{t+1}, \mathbf{X}_{t}, \mathbf{Y}_{t} } (\mathbf{x}_{t}|\mathbf{y}_{t}, \mathbf{x}_{t+1}, \mathbf{y}_{t+1}) } }\\
    & + \expec{ \mathbf{X}_{t:t+1}, \mathbf{Y}_{t:t+1} }{ \log \frac{ P_{ \mathbf{Y}_{t+1}| \mathbf{X}_{t}, \mathbf{Y}_{t} } (\mathbf{y}_{t+1} | \mathbf{x}_{t}, \mathbf{y}_{t}) }{ P_{\mathbf{Y}_{t+1} | \mathbf{X}_{t}, \mathbf{Y}_{t} } (\mathbf{y}_{t} | \mathbf{x}_{t+1}, \mathbf{y}_{t+1}) } }- \expec{ \mathbf{Y}_{t:t+1} }{ \log \frac{ P_{ \mathbf{Y}_{t+1} | \mathbf{Y}_{t} } (\mathbf{y}_{t+1} | \mathbf{y}_{t}) }{ P_{ \mathbf{Y}_{t+1} | \mathbf{Y}_{t} } (\mathbf{y}_{t} | \mathbf{y}_{t+1}) } }
\end{align}
Here, using the relationship between the conditional Shannon entropy and mutual information \cite{coverElementsInformationTheory2012}, we have
\begin{align}
    &H[\mathbf{X}_{t+1}] - H[\mathbf{X}_{t}] = H[\mathbf{X}_{t+1} | \mathbf{Y}_{t+1}] - H[\mathbf{X}_{t} | \mathbf{Y}_{t}] + I[\mathbf{X}_{t+1};\mathbf{Y}_{t+1}] - I[\mathbf{X}_{t};\mathbf{Y}_{t}]
\end{align}
and therefore the relationship between conditional entropy production and partial entropy production is as follows.
\begin{align}
    \Sigma_{\mathrm{X|Y}} ^{\mathrm{con}} =& H[\mathbf{X}_{t+1}] - H[\mathbf{X}_{t}] - I[\mathbf{X}_{t+1};\mathbf{Y}_{t+1}] + I[\mathbf{X}_{t};\mathbf{Y}_{t}] + \Sigma_{\mathrm{bath}} ^{\mathrm{par}} + T_{\mathrm{X} \rightarrow \mathrm{Y}} - T^{\mathrm{R}}_{\mathrm{X} \rightarrow \mathrm{Y}} \\
    =& \Sigma_{\mathrm{X}} ^{\mathrm{par}} + \Theta_{\mathrm{X} \rightarrow \mathrm{Y}} - I[\mathbf{X}_{t+1};\mathbf{Y}_{t+1}] + I[\mathbf{X}_{t};\mathbf{Y}_{t}] + T_{\mathrm{X} \rightarrow \mathrm{Y}} - T^{\mathrm{R}}_{\mathrm{X} \rightarrow \mathrm{Y}} \\
    =& \Sigma_{\mathrm{X}} ^{\mathrm{par}} + T^{\dagger}_{\mathrm{X} \rightarrow \mathrm{Y}} - T_{\mathrm{X} \rightarrow \mathrm{Y}} + T_{\mathrm{X} \rightarrow \mathrm{Y}} - T^{\mathrm{R}}_{\mathrm{X} \rightarrow \mathrm{Y}} \label{cconpar} \\
    =& \Sigma_{\mathrm{X}} ^{\mathrm{par}} + T^{\dagger}_{\mathrm{X} \rightarrow \mathrm{Y}} - T^{\mathrm{R}}_{\mathrm{X} \rightarrow \mathrm{Y}}
\end{align}
Here, \eqref{cconpar} was obtained using \eqref{ddefdif}. In other words, the difference between conditional entropy production and partial entropy production is equal to the difference between $T^{\dagger}_{\mathrm{X} \rightarrow \mathrm{Y}}$ and $T^{\mathrm{R}}_{\mathrm{X} \rightarrow \mathrm{Y}}$.

Note that this relation cannot be derived for conditional dissipation function.

\subsection{Information-Theoretic Decomposition}
\label{sec:dec_info}
In this section, we discuss decomposition of the total entropy production using only information-theoretic quantities such as the Shannon entropy and mutual information.

\subsubsection{Decomposition Based on Shannon Entropy} \label{dec_ent}
The total entropy production can be decomposed into the difference between the Shannon entropy of the forward process and the cross entropy of the backward/time-reversed process. Although this sort of decomposition has been discussed for stationary Markov processes \cite{gaspardTimeReversedDynamicalEntropy2004,dianaMutualEntropyProduction2014}, it can be shown that both total entropy production and total dissipation function can be decomosed for non-stationary Markov processes.
\subsubsection*{Total Entropy Production}
\begin{align}
    \Sigma_{\mathrm{XY}} ^{\mathrm{tot}} =& \KL{ P_{ \mathbf{X}_{t:t+1}, \mathbf{Y}_{t:t+1} } }{ Q_{ \mathbf{X}_{t:t+1}, \mathbf{Y}_{t:t+1} } } \\
    =& \expec{\mathbf{X}_{t:t+1}, \mathbf{Y}_{t:t+1}}{
    \log P_{ \mathbf{X}_{t:t+1}, \mathbf{Y}_{t:t+1} }
    }
    - \expec{\mathbf{X}_{t:t+1}, \mathbf{Y}_{t:t+1}}{
    \log Q_{ \mathbf{X}_{t:t+1}, \mathbf{Y}_{t:t+1} }
    } \\
    =& H_{\mathrm{R}} [\mathbf{X}_{t:t+1}, \mathbf{Y}_{t:t+1}] - H[\mathbf{X}_{t:t+1}, \mathbf{Y}_{t:t+1}]
\end{align}
Here, the cross entropy of the backward distribution is defined as:
\begin{align}
    H_{\mathrm{R}} [\mathbf{X}_{t:t+1}, \mathbf{Y}_{t:t+1}]
    \coloneq - \expec{\mathbf{X}_{t:t+1}, \mathbf{Y}_{t:t+1}}{
    \log Q_{ \mathbf{X}_{t:t+1}, \mathbf{Y}_{t:t+1} }
    }
\end{align}
\subsubsection*{Total Dissipation Function}
For dissipation function,
\begin{align}
    \tilde{\Sigma}_{\mathrm{XY}} ^{\mathrm{tot}} =& \KL{ P_{ \mathbf{X}_{t:t+1}, \mathbf{Y}_{t:t+1} } }{ \tilde{Q}_{ \mathbf{X}_{t:t+1}, \mathbf{Y}_{t:t+1} } } \\
    =& \expec{\mathbf{X}_{t:t+1}, \mathbf{Y}_{t:t+1}}{
    \log P_{ \mathbf{X}_{t:t+1}, \mathbf{Y}_{t:t+1} }
    }
    - \expec{\mathbf{X}_{t:t+1}, \mathbf{Y}_{t:t+1}}{
    \log \tilde{Q}_{ \mathbf{X}_{t:t+1}, \mathbf{Y}_{t:t+1} }
    } \\
    =& \tilde{H}_{\mathrm{R}} [\mathbf{X}_{t:t+1}, \mathbf{Y}_{t:t+1}] - H[\mathbf{X}_{t:t+1}, \mathbf{Y}_{t:t+1}]
\end{align}
Here, the cross entropy of the time-reversed distribution is defined as:
\begin{align}
    \tilde{H}_{\mathrm{R}} [\mathbf{X}_{t:t+1}, \mathbf{Y}_{t:t+1}]
    \coloneq - \expec{\mathbf{X}_{t:t+1}, \mathbf{Y}_{t:t+1}}{
    \log \tilde{Q}_{ \mathbf{X}_{t:t+1}, \mathbf{Y}_{t:t+1} }
    }
\end{align}
A similar decomposition can be performed for marginal entropy production/dissipation function and entropy production/dissipation for single system.

\subsubsection{Decomposition Based on Mutual Information}
Total entropy production can be decomposed into the difference between the mutual information of X and Y at time $t$ and X and Y at time $t+1$, and the mutual information in the time-reversed process. A similar decomposition for another physical quantity has been discussed in \cite{dianaMutualEntropyProduction2014}\footnote{Diana and Esposito introduces mutual entropy production, which is defined by the difference between the total entropy production and marginal entropy production for each subsystems.}, but this decomposition is discussed here for the first time.
\subsubsection*{Total Entropy Production}
From the results of section \ref{dec_ent}, entropy production can be decomposed as follows.
\begin{align}
    \Sigma_{\mathrm{XY}} ^{\mathrm{tot}} =& H_{\mathrm{R}} [\mathbf{X}_{t:t+1}, \mathbf{Y}_{t:t+1}] - H[\mathbf{X}_{t:t+1}, \mathbf{Y}_{t:t+1}] \\
    =& \pr{
    H[\mathbf{X}_{t+1}, \mathbf{Y}_{t+1}]
    + H[\mathbf{X}_{t}, \mathbf{Y}_{t}]
	- H[\mathbf{X}_{t:t+1}, \mathbf{Y}_{t:t+1}]
    } \\
    &- \pr{
    H[\mathbf{X}_{t+1}, \mathbf{Y}_{t+1}]
    + H[\mathbf{X}_{t}, \mathbf{Y}_{t}]
	- H_{\mathrm{R}} [\mathbf{X}_{t:t+1}, \mathbf{Y}_{t:t+1}]
    } \\
    =& I[\mathbf{X}_{t+1}, \mathbf{Y}_{t+1} ; \mathbf{X}_{t}, \mathbf{Y}_{t}]
	- I_{\mathrm{R}}[\mathbf{X}_{t+1}, \mathbf{Y}_{t+1} ; \mathbf{X}_{t}, \mathbf{Y}_{t}]
\end{align}
Here, the mutual information of the backward process is defined as follows.
\begin{align}
    I_{\mathrm{R}}[\mathbf{X}_{t+1}, \mathbf{Y}_{t+1} ; \mathbf{X}_{t}, \mathbf{Y}_{t}]
    \coloneq H[\mathbf{X}_{t+1}, \mathbf{Y}_{t+1}]
    + H[\mathbf{X}_{t}, \mathbf{Y}_{t}]
	- H_{\mathrm{R}} [\mathbf{X}_{t:t+1}, \mathbf{Y}_{t:t+1}]
\end{align}
\subsubsection*{Total Dissipation Function}
Similarly, dissipation function can be decomposed as follows.
\begin{align}
    \tilde{\Sigma}_{\mathrm{XY}} ^{\mathrm{tot}} =& \tilde{H}_{\mathrm{R}} [\mathbf{X}_{t:t+1}, \mathbf{Y}_{t:t+1}] - H[\mathbf{X}_{t:t+1}, \mathbf{Y}_{t:t+1}] \\
    =& 
    \pr{
	H[\mathbf{X}_{t+1}, \mathbf{Y}_{t+1}]
	+ H[\mathbf{X}_{t}, \mathbf{Y}_{t}]
	- H[\mathbf{X}_{t:t+1}, \mathbf{Y}_{t:t+1}]
	} \\
	&- \pr{
	H[\mathbf{X}_{t+1}, \mathbf{Y}_{t+1}]
	+ H[\mathbf{X}_{t}, \mathbf{Y}_{t}]
	- \tilde{H}_{\mathrm{R}} [\mathbf{X}_{t:t+1}, \mathbf{Y}_{t:t+1}]
	} \\
    =& I[\mathbf{X}_{t+1}, \mathbf{Y}_{t+1} ; \mathbf{X}_{t}, \mathbf{Y}_{t}]
    - \tilde{I}_{\mathrm{R}}[\mathbf{X}_{t+1}, \mathbf{Y}_{t+1} ; \mathbf{X}_{t}, \mathbf{Y}_{t}]
\end{align}
Here, the time-reversed mutual information is defined as follows.
\begin{align}
    \tilde{I}_{\mathrm{R}}[\mathbf{X}_{t+1}, \mathbf{Y}_{t+1} ; \mathbf{X}_{t}, \mathbf{Y}_{t}]
    \coloneq H[\mathbf{X}_{t+1}, \mathbf{Y}_{t+1}]
    + H[\mathbf{X}_{t}, \mathbf{Y}_{t}]
	- \tilde{H}_{\mathrm{R}} [\mathbf{X}_{t:t+1}, \mathbf{Y}_{t:t+1}]
\end{align}
In this equation, $I_{\mathrm{R}}[\mathbf{X}_{t+1}, \mathbf{Y}_{t+1} ; \mathbf{X}_{t}, \mathbf{Y}_{t}]$ and $\tilde{I}_{\mathrm{R}}[\mathbf{X}_{t+1}, \mathbf{Y}_{t+1} ; \mathbf{X}_{t}, \mathbf{Y}_{t}]$ are not always non-negative, unlike the ordinary mutual information.

A similar decomposition can be performed for marginal entropy production/dissipation function and entropy production/dissipation for single system.

\section{Inequalities for Entropy Production in Composite Systems}
\label{IneEntPro}
In this chapter, we discuss comparison of entropy production in composite systems. In section \ref{sec:tot_mar}, we derive the inequality between total entropy production and marginal entropy production, and in section \ref{sec:par_mar}, we derive the inequality between partial entropy production and marginal entropy production.

\subsection{Inequalities for Total Entropy Production and Marginal Entropy Production}
\label{sec:tot_mar}
\begin{figure}[bt]
    \centering
    \includegraphics[width=70mm]{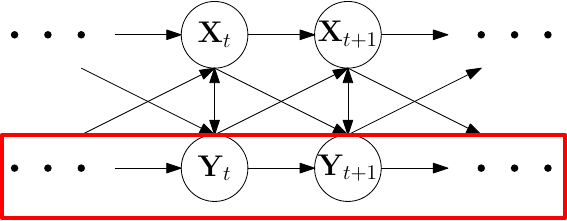}
    \caption{Schematic diagram of a discrete-time Markov process in a composite system (where only Y is observable).}
    \label{fig:mar-pro}
\end{figure}

In the composite system XY in section \ref{sec:composite_system}, consider the case where X is not observable but Y is (Fig.~\ref{fig:mar-pro}). Under such conditions, it has been shown that total entropy production bounded below by marginal entropy production in a continuous-time system\cite{kawaiDissipationPhaseSpacePerspective2007,kawaguchiFluctuationTheoremHidden2013}. In other words, since entropy production that can be calculated only from the data of observable system is an underestimate of the total entropy production including unobservable system, it is useful for estimating the total entropy production. Here, we consider whether the same inequality holds in discrete-time systems.

\subsubsection{Dissipation Function}
The inequalities between total dissipation function and marginal dissipation function is shown below.
\begin{result}[Inequality for Total Dissipation Function and Marginal Dissipation Function]
    \begin{align}
        \tilde{\Sigma}_{\mathrm{XY}} ^{\mathrm{tot}} \geq \tilde{\Sigma}_{\mathrm{Y}} ^{\mathrm{mar}}
        \label{ineq_trep_tot_mar}
    \end{align}
\end{result}
\begin{proof}
	We use the log sum inequality:
	\begin{equation}
	\sum_{i} a_i \log \frac{a_i}{b_i} \geq \sum_{i} a_i \log \frac{\sum_{i} a_i}{\sum_{i} b_i}
	\end{equation}
    \eqref{ineq_trep_tot_mar} is obtained by setting
    \begin{align}
        i &\rightarrow \pr{\mathbf{x}_{t}, \mathbf{x}_{t+1}} \\
        a_i &\rightarrow P_{ \mathbf{X}_{t+1}, \mathbf{Y}_{t+1}, \mathbf{X}_{t}, \mathbf{Y}_{t} } ( \mathbf{x}_{t+1}, \mathbf{y}_{t+1}, \mathbf{x}_{t}, \mathbf{y}_{t} ) \\
        b_i &\rightarrow \tilde{Q}_{ \mathbf{X}_{t+1}, \mathbf{Y}_{t+1}, \mathbf{X}_{t}, \mathbf{Y}_{t} } ( \mathbf{x}_{t+1}, \mathbf{y}_{t+1}, \mathbf{x}_{t}, \mathbf{y}_{t} )
    \end{align}
    in the log sum inequality. The equality holds when the ratio of $\tilde{Q}_{ \mathbf{X}_{t+1}, \mathbf{Y}_{t+1}, \mathbf{X}_{t}, \mathbf{Y}_{t} }$ to $P_{ \mathbf{X}_{t+1}, \mathbf{Y}_{t+1}, \mathbf{X}_{t}, \mathbf{Y}_{t} }$ depends only on $\mathbf{y}_{t}$ and $\mathbf{y}_{t+1}$, so that
    \begin{equation}
        \dfrac{P_{ \mathbf{X}_{t+1}, \mathbf{Y}_{t+1}, \mathbf{X}_{t}, \mathbf{Y}_{t} } ( \mathbf{x}_{t+1}, \mathbf{y}_{t+1}, \mathbf{x}_{t}, \mathbf{y}_{t} )}{\tilde{Q}_{ \mathbf{X}_{t+1}, \mathbf{Y}_{t+1}, \mathbf{X}_{t}, \mathbf{Y}_{t} } ( \mathbf{x}_{t+1}, \mathbf{y}_{t+1}, \mathbf{x}_{t}, \mathbf{y}_{t} )} = f(\mathbf{y}_{t},\mathbf{y}_{t+1})
    \end{equation}
    Here, $f$ is any function. This means that the ratio of the time-reversed distribution to the forward distribution does not depend on the dynamics of X.
\end{proof}
\eqref{ineq_trep_tot_mar} means that the correct total dissipation function is bounded from below by the marginal dissipation function calculated from the data of partially observable system in multiple interacting systems.

Furthermore, from the above inequalities, it can be seen that the conditional dissipation function \eqref{def_condf} is always non-negative.
\begin{result}[Non-Negativity of Conditional Dissipation Function]
\begin{equation}
    \tilde{\Sigma}_{\mathrm{X|Y}} ^{\mathrm{con}} \geq 0.
\end{equation}
\end{result}

\subsubsection{Entropy Production}
Using the inequalities of the total dissipation function and marginal dissipation function, it is possible to derive a similar inrequality for entropy production.
\begin{result}[Inequality for Total Entropy Production and Marginal Entropy Production]
    \begin{equation}
        \Sigma_{\mathrm{XY}} ^{\mathrm{tot}} + \KL{P_{\mathbf{X}_{t+1} | \mathbf{Y}_{t+1}} }{P_{\mathbf{X}_{t} | \mathbf{Y}_{t}}}
        \geq \Sigma_{\mathrm{Y}} ^{\mathrm{mar}}
        \label{ineq_bep_tot_mar}
    \end{equation}
\end{result}
\begin{proof}
    From the definitions of dissipation function and entropy production,
    \begin{align}
        \tilde{\Sigma}_{\mathrm{XY}} ^\mathrm{tot} - \Sigma_{\mathrm{XY}} ^\mathrm{tot}
        =& \expec{\mathrm{XY}}{
        \log \dfrac{
        P_{ \mathbf{X}_{t+1}, \mathbf{Y}_{t+1} | \mathbf{X}_{t}, \mathbf{Y}_{t} } ( \mathbf{x}_{t}, \mathbf{y}_{t} | \mathbf{x}_{t+1}, \mathbf{y}_{t+1} ) P_{ \mathbf{X}_{t+1}, \mathbf{Y}_{t+1} } (\mathbf{x}_{t+1}, \mathbf{y}_{t+1})
        }{
        P_{ \mathbf{X}_{t+1}, \mathbf{Y}_{t+1} | \mathbf{X}_{t}, \mathbf{Y}_{t} } ( \mathbf{x}_{t}, \mathbf{y}_{t} | \mathbf{x}_{t+1}, \mathbf{y}_{t+1} )
        P_{ \mathbf{X}_{t}, \mathbf{Y}_{t} } (\mathbf{x}_{t+1}, \mathbf{y}_{t+1})
        } }\\
        =& \KL{P_{ \mathbf{X}_{t+1}, \mathbf{Y}_{t+1} }}{P_{ \mathbf{X}_{t}, \mathbf{Y}_{t} }} \\
        \tilde{\Sigma}_{\mathrm{Y}} ^\mathrm{mar} - \Sigma_{\mathrm{Y}} ^\mathrm{mar}
        =& \expec{\mathrm{XY}}{
        \log \dfrac{
        P_{ \mathbf{Y}_{t+1} | \mathbf{Y}_{t} } ( \mathbf{y}_{t} | \mathbf{y}_{t+1} ) P_{ \mathbf{Y}_{t+1} } (\mathbf{y}_{t+1})
        }{
        P_{ \mathbf{Y}_{t+1} | \mathbf{Y}_{t} } ( \mathbf{y}_{t} | \mathbf{y}_{t+1} )
        P_{ \mathbf{Y}_{t} } (\mathbf{y}_{t+1})
        } }\\
        =& \KL{P_{\mathbf{Y}_{t+1} }}{P_{ \mathbf{Y}_{t} }}
    \end{align}
    Based on this relationship and the inequality $\tilde{\Sigma}_{\mathrm{XY}} ^\mathrm{tot} \geq \tilde{\Sigma}_{\mathrm{Y}} ^\mathrm{mar}$,
    \begin{align}
        \Sigma_{\mathrm{XY}} ^\mathrm{tot} + \KL{P_{ \mathbf{X}_{t+1}, \mathbf{Y}_{t+1} }}{P_{ \mathbf{X}_{t}, \mathbf{Y}_{t} }} &\geq \Sigma_{\mathrm{Y}} ^\mathrm{mar} + \KL{P_{\mathbf{Y}_{t+1} }}{P_{ \mathbf{Y}_{t} }} \\
        \Sigma_{\mathrm{XY}} ^{\mathrm{tot}} + \KL{P_{\mathbf{X}_{t+1} | \mathbf{Y}_{t+1}} }{P_{\mathbf{X}_{t} | \mathbf{Y}_{t}}}
        &\geq \Sigma_{\mathrm{Y}} ^{\mathrm{mar}}
    \end{align}
\end{proof}
Unlike the case of dissipation function, \eqref{ineq_bep_tot_mar} means that the correct total entropy production is not bounded from below by the marginal entropy production calculated from the data of partially observable system in multiple interacting systems.

In addition, the inequality indicates that the conditional entropy production is not always non-negative.

\subsection{Inequalities for Partial Entropy Production and Marginal Entropy Production}
\label{sec:par_mar}
In the composite system XY in section \ref{sec:composite_system}, consider the case where X is not observable but Y is, and the case where Y is observed given the observation of X. As in the case of total entropy production and marginal entropy production, if a certain inequality between partial entropy production and marginal entropy production holds, it can be useful for estimating the partial entropy production from observable data.

\subsubsection{Dissipation Function}
The inequality between partial dissipation function and marginal dissipation function is shown below.
\begin{result}[Inequality for Partial Dissipation Function and Marginal Dissipation Function]
    \begin{equation}
        \tilde{\Sigma}_{\mathrm{Y}} ^{\mathrm{par}} \geq \tilde{\Sigma}_{\mathrm{Y}} ^{\mathrm{mar}}
        \label{ineq_trep_par_mar}
    \end{equation}
\end{result}
\begin{proof}
    \eqref{ineq_trep_par_mar} is obtained by setting
    \begin{align}
        i &\rightarrow \pr{\mathbf{x}_{t}, \mathbf{x}_{t+1}} \\
        a_i &\rightarrow P_{ \mathbf{X}_{t+1}, \mathbf{Y}_{t+1}, \mathbf{X}_{t}, \mathbf{Y}_{t} } ( \mathbf{x}_{t+1}, \mathbf{y}_{t+1}, \mathbf{x}_{t}, \mathbf{y}_{t} ) \\
        b_i &\rightarrow \tilde{Q}^{\mathrm{par}}_{ \mathbf{X}_{t+1}, \mathbf{Y}_{t+1}, \mathbf{X}_{t}, \mathbf{Y}_{t} } ( \mathbf{x}_{t+1}, \mathbf{y}_{t+1}, \mathbf{x}_{t}, \mathbf{y}_{t} )
    \end{align}
	in the log sum inequality.
    The equality holds when the ratio of $\tilde{Q}^{\mathrm{par}}_{ \mathbf{X}_{t+1}, \mathbf{Y}_{t+1}, \mathbf{X}_{t}, \mathbf{Y}_{t} }$ to $P_{ \mathbf{X}_{t+1}, \mathbf{Y}_{t+1}, \mathbf{X}_{t}, \mathbf{Y}_{t} }$ depends only on $\mathbf{y}_{t}$ and $\mathbf{y}_{t+1}$, so that
    \begin{equation}
        \dfrac{P_{ \mathbf{X}_{t+1}, \mathbf{Y}_{t+1}, \mathbf{X}_{t}, \mathbf{Y}_{t} } ( \mathbf{x}_{t+1}, \mathbf{y}_{t+1}, \mathbf{x}_{t}, \mathbf{y}_{t} )}{\tilde{Q}^{\mathrm{par}}_{ \mathbf{X}_{t+1}, \mathbf{Y}_{t+1}, \mathbf{X}_{t}, \mathbf{Y}_{t} } ( \mathbf{x}_{t+1}, \mathbf{y}_{t+1}, \mathbf{x}_{t}, \mathbf{y}_{t} )} = f(\mathbf{y}_{t},\mathbf{y}_{t+1})
    \end{equation}
    Here, $f$ is any function. This means that the ratio of the partial time-reversed distribution to the forward distribution does not depend on the dynamics of X.
\end{proof}
\eqref{ineq_trep_par_mar} means that the correct partial dissipation function is bounded from below by the marginal dissipation function calculated from the data of partially observable system in multiple interacting systems.

\subsubsection{Entropy Production}
Using the inequalities of the partial dissipation function and marginal dissipation function, it is possible to derive a similar relationship for entropy production.
\begin{result}[Inequality for Partial Entropy Production and Marginal Entropy Production]
    \begin{equation}
        \Sigma_{\mathrm{Y}} ^{\mathrm{par}} + \KL{P_{\mathbf{X}_{t+1},\mathbf{X}_{t} | \mathbf{Y}_{t+1}} }{P_{\mathbf{X}_{t+1},\mathbf{X}_{t} | \mathbf{Y}_{t}}}
        + \expec{\mathrm{XY}}{
        \log \dfrac{
        P_{\mathbf{Y}_{t+1} | \mathbf{Y}_{t}, \mathbf{X}_{t+1}, \mathbf{X}_{t}} (\mathbf{y}_{t} | \mathbf{y}_{t+1}, \mathbf{x}_{t}, \mathbf{x}_{t+1})
        }{
        P_{\mathbf{Y}_{t+1} | \mathbf{Y}_{t}, \mathbf{X}_{t+1}, \mathbf{X}_{t}} (\mathbf{y}_{t} | \mathbf{y}_{t+1}, \mathbf{x}_{t+1}, \mathbf{x}_{t})
        } }
        \geq \Sigma_{\mathrm{Y}} ^{\mathrm{mar}}
        \label{ineq_bep_par_mar}
    \end{equation}
\end{result}
\begin{proof}
    From the relationship between dissipation function and entropy production,
    \begin{align}
        \tilde{\Sigma}_{\mathrm{Y}} ^\mathrm{par} - \Sigma_{\mathrm{Y}} ^\mathrm{par}
        =& \expec{\mathrm{XY}}{
        \log \dfrac{
        P_{\mathbf{Y}_{t+1} | \mathbf{Y}_{t}, \mathbf{X}_{t+1}, \mathbf{X}_{t}} (\mathbf{y}_{t} | \mathbf{y}_{t+1}, \mathbf{x}_{t+1}, \mathbf{x}_{t}) P_{\mathbf{Y}_{t+1}, \mathbf{X}_{t+1}, \mathbf{X}_{t}} (\mathbf{y}_{t+1}, \mathbf{x}_{t+1}, \mathbf{x}_{t})
        }{
        P_{\mathbf{Y}_{t+1} | \mathbf{Y}_{t}, \mathbf{X}_{t+1}, \mathbf{X}_{t}} (\mathbf{y}_{t} | \mathbf{y}_{t+1}, \mathbf{x}_{t+1}, \mathbf{x}_{t}) P_{\mathbf{Y}_{t}, \mathbf{X}_{t+1}, \mathbf{X}_{t}} (\mathbf{y}_{t+1}, \mathbf{x}_{t+1}, \mathbf{x}_{t})
        } }\\
        & + \expec{\mathrm{XY}}{
        \log \dfrac{
        P_{\mathbf{Y}_{t+1} | \mathbf{Y}_{t}, \mathbf{X}_{t+1}, \mathbf{X}_{t}} (\mathbf{y}_{t} | \mathbf{y}_{t+1}, \mathbf{x}_{t}, \mathbf{x}_{t+1})
        }{
        P_{\mathbf{Y}_{t+1} | \mathbf{Y}_{t}, \mathbf{X}_{t+1}, \mathbf{X}_{t}} (\mathbf{y}_{t} | \mathbf{y}_{t+1}, \mathbf{x}_{t+1}, \mathbf{x}_{t})
        } } \\
        =& \KL{P_{\mathbf{Y}_{t+1}, \mathbf{X}_{t+1}, \mathbf{X}_{t}}}{P_{\mathbf{Y}_{t}, \mathbf{X}_{t+1}, \mathbf{X}_{t}}}
        + \expec{\mathrm{XY}}{
        \log \dfrac{
        P_{\mathbf{Y}_{t+1} | \mathbf{Y}_{t}, \mathbf{X}_{t+1}, \mathbf{X}_{t}} (\mathbf{y}_{t} | \mathbf{y}_{t+1}, \mathbf{x}_{t}, \mathbf{x}_{t+1})
        }{
        P_{\mathbf{Y}_{t+1} | \mathbf{Y}_{t}, \mathbf{X}_{t+1}, \mathbf{X}_{t}} (\mathbf{y}_{t} | \mathbf{y}_{t+1}, \mathbf{x}_{t+1}, \mathbf{x}_{t})
        } } \\
        \tilde{\Sigma}_{\mathrm{Y}} ^\mathrm{mar} - \Sigma_{\mathrm{Y}} ^\mathrm{mar}
        =& \expec{\mathrm{XY}}{
        \log \dfrac{
        P_{ \mathbf{Y}_{t+1} | \mathbf{Y}_{t} } ( \mathbf{y}_{t} | \mathbf{y}_{t+1} ) P_{ \mathbf{Y}_{t+1} } (\mathbf{y}_{t+1})
        }{
        P_{ \mathbf{Y}_{t+1} | \mathbf{Y}_{t} } ( \mathbf{y}_{t} | \mathbf{y}_{t+1} )
        P_{ \mathbf{Y}_{t} } (\mathbf{y}_{t+1})
        } }
         \\
        =& \KL{P_{\mathbf{Y}_{t+1} }}{P_{ \mathbf{Y}_{t} }}
    \end{align}
    Based on this relationship and the inequality $\tilde{\Sigma}_{\mathrm{Y}} ^\mathrm{par} \geq \tilde{\Sigma}_{\mathrm{Y}} ^\mathrm{mar}$,
    \begin{gather}
        \Sigma_{\mathrm{Y}} ^\mathrm{par} + \KL{P_{\mathbf{Y}_{t+1}, \mathbf{X}_{t+1}, \mathbf{X}_{t}}}{P_{\mathbf{Y}_{t}, \mathbf{X}_{t+1}, \mathbf{X}_{t}}}
        + \expec{\mathrm{XY}}{
        \log \dfrac{
        P_{\mathbf{Y}_{t+1} | \mathbf{Y}_{t}, \mathbf{X}_{t+1}, \mathbf{X}_{t}} (\mathbf{y}_{t} | \mathbf{y}_{t+1}, \mathbf{x}_{t}, \mathbf{x}_{t+1})
        }{
        P_{\mathbf{Y}_{t+1} | \mathbf{Y}_{t}, \mathbf{X}_{t+1}, \mathbf{X}_{t}} (\mathbf{y}_{t} | \mathbf{y}_{t+1}, \mathbf{x}_{t+1}, \mathbf{x}_{t})
        } } \nonumber \\
        \geq \Sigma_{\mathrm{Y}} ^\mathrm{mar} + \KL{P_{\mathbf{Y}_{t+1} }}{P_{ \mathbf{Y}_{t} }} \\
        \Sigma_{\mathrm{Y}} ^{\mathrm{par}} + \KL{P_{\mathbf{X}_{t+1},\mathbf{X}_{t} | \mathbf{Y}_{t+1}} }{P_{\mathbf{X}_{t+1},\mathbf{X}_{t} | \mathbf{Y}_{t}}}
        + \expec{\mathrm{XY}}{
        \log \dfrac{
        P_{\mathbf{Y}_{t+1} | \mathbf{Y}_{t}, \mathbf{X}_{t+1}, \mathbf{X}_{t}} (\mathbf{y}_{t} | \mathbf{y}_{t+1}, \mathbf{x}_{t}, \mathbf{x}_{t+1})
        }{
        P_{\mathbf{Y}_{t+1} | \mathbf{Y}_{t}, \mathbf{X}_{t+1}, \mathbf{X}_{t}} (\mathbf{y}_{t} | \mathbf{y}_{t+1}, \mathbf{x}_{t+1}, \mathbf{x}_{t})
        } }
        \geq \Sigma_{\mathrm{Y}} ^{\mathrm{mar}}
    \end{gather}
\end{proof}
Unlike the case of dissipation function, \eqref{ineq_bep_par_mar} means that the correct partial entropy production is not bounded from below by the marginal entropy production calculated from the data of partially observable system in multiple interacting systems.

\section{Example: Gaussian process}
\label{Ex}
In this chapter, we derive the entropy production for Gaussian process. Section \ref{sec:def_langevin} summarize the settings for Gaussian process. In sections \ref{sec:tot_langevin}, \ref{sec:par_langevin} and \ref{sec:mar_langevin}, we calculate the total entropy production, partial entropy production and marginal entropy production for Gaussian process. In section \ref{sec:numerical}, we show the numerical results based on the analytical calculation.

Note that the partial entropy production for composite Gaussian process in continuous-time as discussed in references \cite{hartichSensoryCapacityInformation2016,matsumotoRoleSufficientStatistics2018} among others.

\subsection{Setup}
\label{sec:def_langevin}
Gaussian process is a stochastic process where probability distribution at an arbitrary time is Gaussian distribution. As an application example, Kalman filter, which is widely used in time-series data analysis, is known\cite{wellingKalmanFilter}.

Here, we first define the distributions of system X and Y at times $t$ and $t+1$, and then derive the distribution at each time, the conditional distribution, the difference equation, and the marginal distribution of X.
\subsubsection*{Joint Distribution at Times $t$ and $t+1$}
When the systems X and Y evolve over time from $t$ to $t+1$ while interacting, the joint distribution $\mathbf{X}_{t:t+1}, \mathbf{Y}_{t:t+1}$ is given by the following multidimensional Gaussian distribution:
\begin{align}
    &P_{ \mathbf{X}_{t:t+1}, \mathbf{Y}_{t:t+1} } (\mathbf{x}_{t:t+1}, \mathbf{y}_{t:t+1})
    = \mathcal{N}_{\mathbf{x}_{t:t+1}, \mathbf{y}_{t:t+1}} (\bm{\mu},\mathbf{K}) \\
    &\bm{\mu} =
    \tra{
    \begin{pmatrix}
        \bm{\mu}_{\mathbf{X}_{t}} &
        \bm{\mu}_{\mathbf{Y}_{t}} &
        \bm{\mu}_{\mathbf{X}_{t+1}} &
        \bm{\mu}_{\mathbf{Y}_{t+1}}
    \end{pmatrix}
    } \\
    &\mathbf{K} =
    \begin{pmatrix}
    \mathbf{K}_{\mathbf{x}_{t} \mathbf{x}_{t}}
    & \mathbf{K}_{\mathbf{x}_{t} \mathbf{y}_{t}}
    & \mathbf{K}_{\mathbf{x}_{t} \mathbf{x}_{t+1}}
    & \mathbf{K}_{\mathbf{x}_{t} \mathbf{y}_{t+1}} \\
    \mathbf{K}_{\mathbf{y}_{t} \mathbf{x}_{t}}
    & \mathbf{K}_{\mathbf{y}_{t} \mathbf{y}_{t}}
    & \mathbf{K}_{\mathbf{y}_{t} \mathbf{x}_{t+1}}
    & \mathbf{K}_{\mathbf{y}_{t} \mathbf{y}_{t+1}} \\
    \mathbf{K}_{\mathbf{x}_{t+1} \mathbf{x}_{t}}
    & \mathbf{K}_{\mathbf{x}_{t+1} \mathbf{y}_{t}}
    & \mathbf{K}_{\mathbf{x}_{t+1} \mathbf{x}_{t+1}}
    & \mathbf{K}_{\mathbf{x}_{t+1} \mathbf{y}_{t+1}} \\
    \mathbf{K}_{\mathbf{y}_{t+1} \mathbf{x}_{t}}
    & \mathbf{K}_{\mathbf{y}_{t+1} \mathbf{y}_{t}}
    & \mathbf{K}_{\mathbf{y}_{t+1} \mathbf{x}_{t+1}}
    & \mathbf{K}_{\mathbf{y}_{t+1} \mathbf{y}_{t+1}}
\end{pmatrix}
\end{align}
where
\begin{align}
    \bm{\mu}_{\mathbf{z}_{i}} =& \expec{\mathbf{Z}_i}{\mathbf{Z}_i} \\
    \mathbf{K}_{\mathbf{z}_{i} \mathbf{z}'_{j}} =& \text{Cov} \br{
    \mathbf{Z}_{i}, \mathbf{Z}'_{j}
    } \\
    ( \mathbf{Z}, \mathbf{Z}' = \mathbf{X}, \mathbf{Y} , \quad & i,j = t,t+1)
\end{align}
Since $\mathbf{K}_{\mathbf{z}'_{j} \mathbf{z}_{i}} = \tra{\mathbf{K}}_{\mathbf{z}_{i} \mathbf{z}'_{j}}$ by definition, this is equivalent to
\begin{align}
    \mathbf{K} =
    \begin{pmatrix}
    \mathbf{K}_{\mathbf{x}_{t} \mathbf{x}_{t}}
    & \mathbf{K}_{\mathbf{x}_{t} \mathbf{y}_{t}}
    & \mathbf{K}_{\mathbf{x}_{t} \mathbf{x}_{t+1}}
    & \mathbf{K}_{\mathbf{x}_{t} \mathbf{y}_{t+1}} \\
    \tra{\mathbf{K}}_{\mathbf{x}_{t} \mathbf{y}_{t}}
    & \mathbf{K}_{\mathbf{y}_{t} \mathbf{y}_{t}}
    & \tra{\mathbf{K}}_{\mathbf{x}_{t} \mathbf{y}_{t+1}}
    & \mathbf{K}_{\mathbf{y}_{t} \mathbf{y}_{t+1}} \\
    \tra{\mathbf{K}}_{\mathbf{x}_{t} \mathbf{x}_{t+1}}
    & \tra{\mathbf{K}}_{\mathbf{x}_{t} \mathbf{y}_{t+1}}
    & \mathbf{K}_{\mathbf{x}_{t+1} \mathbf{x}_{t+1}}
    & \mathbf{K}_{\mathbf{x}_{t+1} \mathbf{y}_{t+1}} \\
    \mathbf{K}_{\mathbf{x}_{t} \mathbf{y}_{t+1}}
    &\tra{\mathbf{K}}_{\mathbf{y}_{t} \mathbf{y}_{t+1}}
    & \tra{\mathbf{K}}_{\mathbf{x}_{t+1} \mathbf{y}_{t+1}}
    & \mathbf{K}_{\mathbf{y}_{t+1} \mathbf{y}_{t+1}}
\end{pmatrix}
\end{align}
Furthermore, $\mathbf{X}_{t}$, $\mathbf{X}_{t+1}$, $\mathbf{Y}_{t}$ and $\mathbf{Y}_{t+1}$ are all assumed to be $d$-dimensional vectors for simplicity.

\subsubsection*{Joint Distribution at Each Time}
The joint distribution at each time is given by
\begin{align}
    P_{ \mathbf{X}_{t}, \mathbf{Y}_{t} }
    (\mathbf{x}_{t}, \mathbf{y}_{t}) =& \mathcal{N}_{\mathbf{x}_{t}, \mathbf{y}_{t}} (\bm{\mu}_{t},\mathbf{K}_{t,t}) \\
    P_{ \mathbf{X}_{t+1}, \mathbf{Y}_{t+1} } (\mathbf{x}_{t+1}, \mathbf{y}_{t+1})
    =& \mathcal{N}_{\mathbf{x}_{t+1}, \mathbf{y}_{t+1}} (\bm{\mu}_{t+1},\mathbf{K}_{t+1,t+1})
\end{align}
where
\begin{align}
    &\bm{\mu}_i =
    \tra{
    \begin{pmatrix}
        \bm{\mu}_{\mathbf{x}_{i}} &
        \bm{\mu}_{\mathbf{y}_{i}}
    \end{pmatrix}
    } \\
    &\mathbf{K}_{i,j} =
    \begin{pmatrix}
        \mathbf{K}_{\mathbf{x}_{i} \mathbf{x}_{j}}
        &\mathbf{K}_{\mathbf{x}_{i} \mathbf{y}_{j}} \\
        \tra{\mathbf{K}}_{\mathbf{x}_{i} \mathbf{y}_{j}}
        &\mathbf{K}_{\mathbf{y}_{i} \mathbf{y}_{j}}
    \end{pmatrix} \\
    &( i,j = t, t+1 )
\end{align}
\subsubsection*{Conditional Distribution}
The conditional distribution of X and Y at time $t+1$, based on the conditions of X and Y at time $t$ is given by
\begin{equation}
    P_{ \mathbf{X}_{t+1}, \mathbf{Y}_{t+1} | \mathbf{X}_{t}, \mathbf{Y}_{t} } (\mathbf{x}_{t+1}, \mathbf{y}_{t+1} | \mathbf{x}_{t}, \mathbf{y}_{t})
    = \mathcal{N}_{\mathbf{x}_{t+1}, \mathbf{y}_{t+1} | \mathbf{x}_{t}, \mathbf{y}_{t}} (\bm{\mu}_{t+1|t},\mathbf{K}_{t+1|t})
\end{equation}
where
\begin{align}
    \bm{\mu}_{t+1|t} =& \bm{\mu}_{t+1} + \tra{\mathbf{K}}_{t,t+1} \mathbf{K}_{t,t} ^{-1} \pr{\mathbf{s}_t - \bm{\mu}_{t}} \\
    \mathbf{K}_{t+1|t} =& \mathbf{K}_{t+1,t+1} - \tra{\mathbf{K}}_{t,t+1} \mathbf{K}_{t,t} ^{-1} \mathbf{K}_{t,t+1}
\end{align}
\subsubsection*{Difference Equation}
In the conditional distribution, $\bm{\mu}_{t+1|t}$ can also be expressed as
\begin{align}
    \bm{\mu}_{t+1|t} =& \mathbf{A} \mathbf{s}_t + \mathbf{b} \\
    \mathbf{s}_t =& \tra{
    \begin{pmatrix}
        \mathbf{x}_t & \mathbf{y}_t
    \end{pmatrix}
    } \\
    \mathbf{A} \coloneq \tra{\mathbf{K}}_{t,t+1} \mathbf{K}_{t,t} ^{-1}, \quad
    &\mathbf{b} \coloneq \bm{\mu}_{t+1} - \tra{\mathbf{K}}_{t,t+1} \mathbf{K}_{t,t} ^{-1} \bm{\mu}_{t}
\end{align}
and thus the difference equation representing the time evolution of systems X and Y can be expressed as follows.
\begin{align}
    \label{dif_equ}
    \mathbf{s}_{t+1} =& \mathbf{A} \mathbf{s}_t + \mathbf{b} + \bm{\xi}_t \\
    \mathbf{s}_{t+1} =& \tra{
    \begin{pmatrix}
        \mathbf{x}_{t+1} & \mathbf{y}_{t+1}
    \end{pmatrix}
    } \\
    \bm{\xi}_t \coloneq&
    \tra{
    \begin{pmatrix}
        \bm{\xi}_{\mathbf{X}_{t}} & \bm{\xi}_{\mathbf{Y}_{t}}
    \end{pmatrix}
    }
\end{align}
Here, $\mathbf{A}$ represents the interaction of systems X and Y, and $\mathbf{b}$ represents a constant shift. Suppose $\mathbf{A}$ and $\mathbf{b}$ are expressed as follows.
\begin{gather}
    \mathbf{A} =
    \begin{pmatrix}
        \mathbf{A_{xx}} & \mathbf{A_{xy}} \\
        \mathbf{A_{yx}} & \mathbf{A_{yy}}
    \end{pmatrix}
    =
    \begin{pmatrix}
        \mathbf{A_x} \\
        \mathbf{A_y}
    \end{pmatrix}
    \\
    \mathbf{A_x} =
    \begin{pmatrix}
        \mathbf{A_{xx}} & \mathbf{A_{xy}}
    \end{pmatrix}
    \quad
    \mathbf{A_y} =
    \begin{pmatrix}
        \mathbf{A_{yx}} & \mathbf{A_{yy}}
    \end{pmatrix} \\
    \mathbf{b} = \tra{
    \begin{pmatrix}
        \mathbf{b_x} & \mathbf{b_y}
    \end{pmatrix}
    }
\end{gather}
Furthermore, $\bm{\xi}_t$ represents Gaussian noise with a mean of zero and a covariance matrix of $\mathbf{K}_{t+1|t}$. In the following, we assume that the Gaussian noise $\bm{\xi}_t$ is uniform and independent of systems X and Y:
\begin{equation}
    \mathbf{K}_{t+1|t} = \lambda^{-1} \mathbf{I}_{2d}
\end{equation}
where $\lambda$ denotes the accuracy of the noise\footnote{For a physical system in contact with a heat bath, $\lambda$ corresponds to the inverse temperature of the heat bath \cite{hartichSensoryCapacityInformation2016,matsumotoRoleSufficientStatistics2018}.}, and $\mathbf{I}_{2d}$ is a $2d$-dimensional unit matrix.

Also, by using $\mathbf{K}_{t+1|t}$, the following relation holds between the covariance matrices of the joint distributions of X and Y at each time $t$ and $t+1$.
\begin{align}
    \mathbf{K}_{t+1,t+1}
    =& \tra{\mathbf{K}}_{t,t+1} \mathbf{K}_{t,t} ^{-1} \mathbf{K}_{t,t+1} + \mathbf{K}_{t+1|t} \\
    =& \tra{\mathbf{K}}_{t,t+1} \mathbf{K}_{t,t} ^{-1} \mathbf{K}_{t,t} \mathbf{K}_{t,t} ^{-1} \mathbf{K}_{t,t+1} + \mathbf{K}_{t+1|t} \\
    =& \tra{\mathbf{K}}_{t,t+1} \mathbf{K}_{t,t} ^{-1} \mathbf{K}_{t,t} \tra{(\tra{\mathbf{K}}_{t,t+1} \mathbf{K}_{t,t} ^{-1})} + \mathbf{K}_{t+1|t} \\
    =& \mathbf{A} \mathbf{K}_{t,t} \tra{\mathbf{A}} + \mathbf{K}_{t+1|t}
\end{align}
\subsubsection*{Marginal Distribution}
The marginal distribution of X is given by
\begin{equation}
    P_{ \mathbf{X}_{t:t+1} } (\mathbf{x}_{t:t+1})
    = \mathcal{N}_{\mathbf{x}_{t:t+1}} (\bm{\mu}_{\mathbf{x}},\mathbf{K}_{\mathbf{x} \mathbf{x}})
\end{equation}
where
\begin{align}
    &\bm{\mu}_{\mathbf{x}} =
    \tra{
    \begin{pmatrix}
        \bm{\mu}_{\mathbf{x}_t} & \bm{\mu}_{\mathbf{x}_{t+1}}
    \end{pmatrix}
    } \\
    &\mathbf{K}_{\mathbf{x} \mathbf{x}} =
    \begin{pmatrix}
        \mathbf{K}_{\mathbf{x}_{t} \mathbf{x}_{t}}
        &\mathbf{K}_{\mathbf{x}_{t} \mathbf{x}_{t+1}} \\
        \tra{\mathbf{K}}_{\mathbf{x}_{t} \mathbf{x}_{t+1}}
        &\mathbf{K}_{\mathbf{x}_{t+1} \mathbf{x}_{t+1}}
    \end{pmatrix}
\end{align}

\subsection{Total Entropy Production}
\label{sec:tot_langevin}
\subsubsection*{Total Dissipation Function}
In the total dissipation function
\begin{equation}
    \tilde{\Sigma}_{\mathrm{XY}} ^{\mathrm{tot}} = \KL{ P_{ \mathbf{X}_{t:t+1}, \mathbf{Y}_{t:t+1} } }{ \tilde{Q}_{ \mathbf{X}_{t:t+1}, \mathbf{Y}_{t:t+1} } },
\end{equation}
we assume the following:
\begin{align}
    P_{ \mathbf{X}_{t:t+1}, \mathbf{Y}_{t:t+1} }
    (\mathbf{x}_{t:t+1}, \mathbf{y}_{t:t+1})
    =& \mathcal{N}_{\mathbf{s}} (\bm{\mu},\mathbf{K})\\
    =& \frac{1}{\sqrt{|2 \pi \mathbf{K}|}} \exp \br{ -\frac{1}{2}
    \tra{(\mathbf{s} - \bm{\mu})} \mathbf{K}^{-1} (\mathbf{s} - \bm{\mu})
    } \\
    \tilde{Q}_{ \mathbf{X}_{t:t+1}, \mathbf{Y}_{t:t+1} }
    (\mathbf{x}_{t:t+1}, \mathbf{y}_{t:t+1})
    =& \frac{1}{\sqrt{|2 \pi \mathbf{K}|}} \exp \br{ -\frac{1}{2}
    \tra{(\tilde{\mathbf{s}} - \bm{\mu})} \mathbf{K}^{-1} (\tilde{\mathbf{s}} - \bm{\mu})
    }
\end{align}
where
\begin{equation}
    \mathbf{s} = \tra{
    \begin{pmatrix}
        \mathbf{s}_{t} & \mathbf{s}_{t+1}
    \end{pmatrix}
    }, \quad
    \tilde{\mathbf{s}} = \tra{
    \begin{pmatrix}
        \mathbf{s}_{t+1} & \mathbf{s}_{t}
    \end{pmatrix}
    }
\end{equation}
Consider rewriting the variable of the time-reversed distribution with $\mathbf{s}$. From the inverse matrix of the block matrix\cite{wellingKalmanFilter,petersenMatrixCookbook}, we obtain
\begin{align}
    \mathbf{K}^{-1} =&
    \begin{pmatrix}
        \mathbf{K}_{t|t+1}^{-1}
        &-\mathbf{K}_{t,t}^{-1} \mathbf{K}_{t,t+1} \mathbf{K}_{t+1|t}^{-1} \\
        -\mathbf{K}_{t+1|t}^{-1} \tra{\mathbf{K}}_{t,t+1} \mathbf{K}_{t,t}^{-1}
        &\mathbf{K}_{t+1|t}^{-1}
    \end{pmatrix} \\
    =&
    \begin{pmatrix}
        \mathbf{K}_{t|t+1}^{-1}
        &-\tra{\mathbf{A}} \mathbf{K}_{t+1|t}^{-1} \\
        -\mathbf{K}_{t+1|t}^{-1} \mathbf{A}
        &\mathbf{K}_{t+1|t}^{-1}
    \end{pmatrix}
\end{align}
where it is assumed that
\begin{equation}
    \mathbf{K}_{t|t+1} \coloneq \mathbf{K}_{t,t} - \mathbf{K}_{t,t+1} \mathbf{K}_{t+1,t+1}^{-1} \tra{\mathbf{K}}_{t,t+1}
\end{equation}
Therefore, the second term in the exponent of the time-reversed distribution $\tilde{Q}_{ \mathbf{X}_{t:t+1}, \mathbf{Y}_{t:t+1} }$ becomes
\begin{align}
    \tra{(\tilde{\mathbf{s}} - \bm{\mu})} \mathbf{K}^{-1} (\tilde{\mathbf{s}} - \bm{\mu})
    =& \tra{(\mathbf{s}_{t+1} - \bm{\mu}_{t})} \mathbf{K}_{t|t+1}^{-1} (\mathbf{s}_{t+1} - \bm{\mu}_{t})
    + \tra{(\mathbf{s}_{t+1} - \bm{\mu}_{t})} \pr{-\tra{\mathbf{A}} \mathbf{K}_{t+1|t}^{-1}} (\mathbf{s}_{t} - \bm{\mu}_{t+1}) \\
    &+ \tra{(\mathbf{s}_{t} - \bm{\mu}_{t+1})} \pr{-\mathbf{K}_{t+1|t}^{-1} \mathbf{A}} (\mathbf{s}_{t+1} - \bm{\mu}_{t})
    + \tra{(\mathbf{s}_{t} - \bm{\mu}_{t+1})} \mathbf{K}_{t+1|t}^{-1} (\mathbf{s}_{t} - \bm{\mu}_{t+1}) \\
    =& \tra{(\mathbf{s} - \tilde{\bm{\mu}})} \tilde{\mathbf{K}}^{-1} (\mathbf{s} - \tilde{\bm{\mu}})
\end{align}
where it is assumed that
\begin{align}
    \tilde{\bm{\mu}} \coloneq&
    \tra{
    \begin{pmatrix}
        \bm{\mu}_{t+1} & \bm{\mu}_{t}
    \end{pmatrix}
    } \\
    \tilde{\mathbf{K}}^{-1} \coloneq&
    \begin{pmatrix}
        \mathbf{K}_{t+1|t}^{-1}
        &-\mathbf{K}_{t+1|t}^{-1} \mathbf{A} \\
        -\tra{\mathbf{A}} \mathbf{K}_{t+1|t}^{-1}
        &\mathbf{K}_{t|t+1}^{-1}
    \end{pmatrix}
\end{align}
Also, from the determinant of the block matrix \cite{wellingKalmanFilter,petersenMatrixCookbook},
\begin{align}
    |\mathbf{K}^{-1}| =& \left|
    \mathbf{K}_{t|t+1}^{-1}
    \right |
    \left|
    \mathbf{K}_{t+1|t}^{-1} - \mathbf{K}_{t+1|t}^{-1} \mathbf{A} \mathbf{K}_{t|t+1} \tra{\mathbf{A}} \mathbf{K}_{t+1|t}^{-1}
    \right| \\
    |\tilde{\mathbf{K}}^{-1}| =& \left |
    \mathbf{K}_{t|t+1}^{-1}
    \right |
    \left |
    \mathbf{K}_{t+1|t}^{-1} - \mathbf{K}_{t+1|t}^{-1} \mathbf{A} \mathbf{K}_{t|t+1} \tra{\mathbf{A}} \mathbf{K}_{t+1|t}^{-1}
    \right | \\
    =& |\mathbf{K}^{-1}|
\end{align}
which yields the following expression:
\begin{align}
    \tilde{Q}_{ \mathbf{X}_{t:t+1}, \mathbf{Y}_{t:t+1} }
    (\mathbf{x}_{t:t+1}, \mathbf{y}_{t:t+1})
    =& \frac{1}{\sqrt{|2 \pi \tilde{\mathbf{K}}|}} \exp \br{ - \frac{1}{2}
    \tra{(\mathbf{s} - \tilde{\bm{\mu}})} \tilde{\mathbf{K}}^{-1} (\mathbf{s} - \tilde{\bm{\mu}})
    } \\
    =& \mathcal{N}_{\mathbf{s}} (\tilde{\bm{\mu}}, \tilde{\mathbf{K}})
\end{align}
Based on the above and the KLD of the multidimensional Gaussian distribution, the total dissipation function is as follows:
\begin{align}
    \tilde{\Sigma}_{\mathrm{XY}} ^{\mathrm{tot}} =&
    \frac{1}{2} \brc{
    \trace{}{\tilde{\mathbf{K}}^{-1} \mathbf{K} } - 4d
    + \tra{( \tilde{\bm{\mu}} - \bm{\mu})} \tilde{\mathbf{K}}^{-1} ( \tilde{\bm{\mu}} - \bm{\mu})
    + \log \frac{|\tilde{\mathbf{K}}|}{|\mathbf{K}|}
    } \\
    =&
    \frac{1}{2} \brc{
    \trace{}{\tilde{\mathbf{K}}^{-1} \mathbf{K} } - 4d
    + \tra{( \tilde{\bm{\mu}} - \bm{\mu})} \tilde{\mathbf{K}}^{-1} ( \tilde{\bm{\mu}} - \bm{\mu})
    }
    \label{trep_tot_gau}
\end{align}
Here, we used the relationship $|\mathbf{K}| = |\tilde{\mathbf{K}}| $ since $|\mathbf{K}^{-1}| = |\tilde{\mathbf{K}}^{-1}|$.

In the total dissipation function \eqref{trep_tot_gau}, the first term is
\begin{align}
    \tilde{\mathbf{K}}^{-1} \mathbf{K}
    =& \begin{pmatrix}
        \mathbf{K}_{t+1|t}^{-1}
        &-\mathbf{K}_{t+1|t}^{-1} \mathbf{A} \\
        -\tra{\mathbf{A}} \mathbf{K}_{t+1|t}^{-1}
        &\mathbf{K}_{t|t+1}^{-1}
    \end{pmatrix}
    \begin{pmatrix}
        \mathbf{K}_{t,t}
        &\mathbf{K}_{t,t+1} \\
        \tra{\mathbf{K}}_{t,t+1}
        &\mathbf{K}_{t+1,t+1}
    \end{pmatrix}  \\
    =&
    \begin{pmatrix}
        \mathbf{K}_{t+1|t}^{-1}\mathbf{K}_{t,t} -\mathbf{K}_{t+1|t}^{-1} \mathbf{A} \tra{\mathbf{K}}_{t,t+1}
        & \mathbf{K}_{t+1|t}^{-1} \mathbf{K}_{t,t+1} -\mathbf{K}_{t+1|t}^{-1} \mathbf{A} \mathbf{K}_{t+1,t+1} \\
        -\tra{\mathbf{A}} \mathbf{K}_{t+1|t}^{-1}\mathbf{K}_{t,t} + \mathbf{K}_{t|t+1}^{-1}\tra{\mathbf{K}}_{t,t+1}
        & -\tra{\mathbf{A}} \mathbf{K}_{t+1|t}^{-1}\mathbf{K}_{t,t+1} + \mathbf{K}_{t|t+1}^{-1}\mathbf{K}_{t+1,t+1}
    \end{pmatrix}
\end{align}
and so
\begin{align}
    \trace{}{\tilde{\mathbf{K}}^{-1} \mathbf{K}} =&
    \trace{}{\mathbf{K}_{t+1|t}^{-1}\mathbf{K}_{t,t}
    -\mathbf{K}_{t+1|t}^{-1} \mathbf{A} \tra{\mathbf{K}}_{t,t+1}
    -\tra{\mathbf{A}} \mathbf{K}_{t+1|t}^{-1}\mathbf{K}_{t,t+1}
    + \mathbf{K}_{t|t+1}^{-1}\mathbf{K}_{t+1,t+1}
    } \\
    =& \lambda \trace{}{\mathbf{K}_{t,t}
    - \mathbf{A} \tra{\mathbf{K}}_{t,t+1}
    -\tra{\mathbf{A}} \mathbf{K}_{t,t+1}
    }
    + \trace{}{\mathbf{K}_{t|t+1}^{-1}\mathbf{K}_{t+1,t+1}}
\end{align}
According to the Woodbury identity\cite{wellingKalmanFilter,petersenMatrixCookbook}, the last term in the above formula can be expanded as shown below.
\begin{align}
    \trace{}{\mathbf{K}_{t|t+1}^{-1}\mathbf{K}_{t+1,t+1}}
    =& \trace{}{
    \pr{\mathbf{K}_{t,t}^{-1} + \tra{\mathbf{A}} \mathbf{K}_{t+1|t}^{-1} \mathbf{A}} \mathbf{K}_{t+1,t+1}
    } \\
    =& \trace{}{\mathbf{K}_{t,t}^{-1}\mathbf{K}_{t+1,t+1}}
    + \lambda \trace{}{\tra{\mathbf{A}} \mathbf{A}\mathbf{K}_{t+1,t+1}}
\end{align}
Therefore,
\begin{align}
    \trace{}{\tilde{\mathbf{K}}^{-1} \mathbf{K}}
    =& \lambda \trace{}{\mathbf{K}_{t,t}
    - \mathbf{A} \tra{\mathbf{K}}_{t,t+1}
    - \tra{\mathbf{A}} \mathbf{K}_{t,t+1}
    + \tra{\mathbf{A}} \mathbf{A}\mathbf{K}_{t+1,t+1}
    }
    + \trace{}{\mathbf{K}_{t,t}^{-1}\mathbf{K}_{t+1,t+1}} \\
    =& \lambda \trace{}{\mathbf{K}_{t,t}
    - \mathbf{A}^2 \mathbf{K}_{t,t}
    - \tra{\mathbf{A}} \mathbf{K}_{t,t} \tra{\mathbf{A}}
    + \tra{\mathbf{A}} \mathbf{A} \pr{\mathbf{A} \mathbf{K}_{t,t} \tra{\mathbf{A}} + \mathbf{K}_{t+1|t}}
    }
    + \trace{}{\mathbf{K}_{t,t}^{-1}\mathbf{K}_{t+1,t+1}} \\
    =& \lambda \trace{}{\mathbf{K}_{t,t}
    - \mathbf{A}^2 \mathbf{K}_{t,t}
    - (\tra{\mathbf{A}})^2 \mathbf{K}_{t,t}
    + \tra{\mathbf{A}} \mathbf{A} \pr{\mathbf{A} \mathbf{K}_{t,t} \tra{\mathbf{A}} + \mathbf{K}_{t+1|t}}
    }
    + \trace{}{\mathbf{K}_{t,t}^{-1}\mathbf{K}_{t+1,t+1}} \\
    =& \lambda \trace{}{\mathbf{K}_{t,t}
    - \mathbf{A}^2 \mathbf{K}_{t,t}
    - (\tra{\mathbf{A}})^2 \mathbf{K}_{t,t}
    + (\tra{\mathbf{A}})^2 \mathbf{A}^2 \mathbf{K}_{t,t} + \tra{\mathbf{A}} \mathbf{A} \mathbf{K}_{t+1|t}
    }
    + \trace{}{\mathbf{K}_{t,t}^{-1}\mathbf{K}_{t+1,t+1}} \\
    =& \lambda \trace{}{ \pr{ \mathbf{I}_{2d} - ( \tra{\mathbf{A}} )^2 }\pr{ \mathbf{I}_{2d} - \mathbf{A}^2 }\mathbf{K}_{t,t}
    }
    + \lambda \trace{}{ \tra{\mathbf{A}} \mathbf{A} \mathbf{K}_{t+1|t}
    }
    + \trace{}{\mathbf{K}_{t,t}^{-1}\mathbf{K}_{t+1,t+1}} \\
    =& \lambda \trace{}{ \pr{ \mathbf{I}_{2d} - \tra{\mathbf{A}} } \pr{ \mathbf{I}_{2d} + \tra{\mathbf{A}} } \pr{ \mathbf{I}_{2d} - \mathbf{A} } \pr{ \mathbf{I}_{2d} + \mathbf{A} } \mathbf{K}_{t,t}
    }
    + \trace{}{ \tra{\mathbf{A}} \mathbf{A}
    }
    + \trace{}{\mathbf{K}_{t,t}^{-1}\mathbf{K}_{t+1,t+1}}
\end{align}
The third term can be expanded as follows:
\begin{align}
    &\tra{( \tilde{\bm{\mu}} - \bm{\mu})} \tilde{\mathbf{K}}^{-1} ( \tilde{\bm{\mu}} - \bm{\mu}) \\
    =&
    \tra{(\bm{\mu}_{t+1} - \bm{\mu}_{t})} \mathbf{K}_{t+1|t}^{-1} (\bm{\mu}_{t+1} - \bm{\mu}_{t})
    + \tra{(\bm{\mu}_{t+1} - \bm{\mu}_{t})} (-\mathbf{K}_{t+1|t}^{-1} \mathbf{A}) (\bm{\mu}_{t} - \bm{\mu}_{t+1}) \\
    &+ \tra{(\bm{\mu}_{t} - \bm{\mu}_{t+1})} (-\tra{\mathbf{A}} \mathbf{K}_{t+1|t}^{-1}) (\bm{\mu}_{t+1} - \bm{\mu}_{t})
    + \tra{(\bm{\mu}_{t} - \bm{\mu}_{t+1})} \mathbf{K}_{t|t+1}^{-1} (\bm{\mu}_{t} - \bm{\mu}_{t+1}) \\
    =& \tra{\bm{\mu}}_{t+1} \pr{
    \mathbf{K}_{t+1|t}^{-1} + \mathbf{K}_{t+1|t}^{-1} \mathbf{A} + \tra{\mathbf{A}} \mathbf{K}_{t+1|t}^{-1} + \mathbf{K}_{t|t+1}^{-1}
    } \bm{\mu}_{t+1} \\
    &- \tra{\bm{\mu}}_{t+1} \pr{
     \mathbf{K}_{t+1|t}^{-1} + \mathbf{K}_{t+1|t}^{-1} \mathbf{A} +\tra{\mathbf{A}} \mathbf{K}_{t+1|t}^{-1} + \mathbf{K}_{t|t+1}^{-1}
    } \bm{\mu}_{t} \\
    &- \tra{\bm{\mu}}_{t} \pr{
    \mathbf{K}_{t+1|t}^{-1} + \mathbf{K}_{t+1|t}^{-1} \mathbf{A} +\tra{\mathbf{A}} \mathbf{K}_{t+1|t}^{-1} + \mathbf{K}_{t|t+1}^{-1}
    } \bm{\mu}_{t+1} \\
    &+ \tra{\bm{\mu}}_{t} \pr{
    \mathbf{K}_{t+1|t}^{-1} + \mathbf{K}_{t+1|t}^{-1} \mathbf{A} +\tra{\mathbf{A}} \mathbf{K}_{t+1|t}^{-1} + \mathbf{K}_{t|t+1}^{-1}
    } \bm{\mu}_{t} \\
    =& \tra{ (\bm{\mu}_{t+1} - \bm{\mu}_{t}) } \mathbf{W} (\bm{\mu}_{t+1} - \bm{\mu}_{t}) \\
    &\mathbf{W} \coloneq \mathbf{K}_{t+1|t}^{-1} + \mathbf{K}_{t+1|t}^{-1} \mathbf{A} +\tra{\mathbf{A}} \mathbf{K}_{t+1|t}^{-1} + \mathbf{K}_{t|t+1}^{-1}
\end{align}
Here, according to the Woodbury identity,
\begin{align}
    \mathbf{K}_{t|t+1}^{-1} =& \pr{
    \mathbf{K}_{t,t} - \mathbf{K}_{t,t+1} \mathbf{K}_{t+1,t+1} \tra{\mathbf{K}}_{t,t+1}
    }^{-1} \\
    =& \mathbf{K}_{t,t}^{-1} - \mathbf{K}_{t,t}^{-1}\mathbf{K}_{t,t+1} \pr{
    -\mathbf{K}_{t+1,t+1} + \tra{\mathbf{K}}_{t,t+1} \mathbf{K}_{t,t}^{-1} \mathbf{K}_{t,t+1}
    }^{-1} \tra{\mathbf{K}}_{t,t+1} \mathbf{K}_{t,t}^{-1}\\
    =& \mathbf{K}_{t,t}^{-1} - \tra{\mathbf{A}} \pr{
    -\mathbf{K}_{t+1,t+1} + \mathbf{A} \mathbf{K}_{t,t+1}
    }^{-1} \mathbf{A} \\
    =& \mathbf{K}_{t,t}^{-1} - \tra{\mathbf{A}} \pr{
    -\mathbf{K}_{t+1,t+1} + \mathbf{A} \mathbf{K}_{t,t} \tra{\mathbf{A}}
    }^{-1} \mathbf{A} \\
    =& \mathbf{K}_{t,t}^{-1} + \tra{\mathbf{A}} \mathbf{K}_{t+1|t}^{-1} \mathbf{A}
\end{align}
And so,
\begin{align}
    \mathbf{W} =& \mathbf{K}_{t+1|t}^{-1} + \mathbf{K}_{t+1|t}^{-1} \mathbf{A} +\tra{\mathbf{A}} \mathbf{K}_{t+1|t}^{-1} + \mathbf{K}_{t,t}^{-1} + \tra{\mathbf{A}} \mathbf{K}_{t+1|t}^{-1} \mathbf{A} \\
    =& \lambda \pr{\mathbf{I}_{2d} + \mathbf{A}} \pr{\mathbf{I}_{2d} + \tra{\mathbf{A}}} + \mathbf{K}_{t,t}^{-1}
\end{align}
Therefore,
\begin{align}
    &\tra{ (\bm{\mu}_{t+1} - \bm{\mu}_{t}) } \mathbf{W} (\bm{\mu}_{t+1} - \bm{\mu}_{t}) \\
    =& \lambda \tra{ (\bm{\mu}_{t+1} - \bm{\mu}_{t}) } \pr{\mathbf{I}_{2d} + \mathbf{A}} \pr{\mathbf{I}_{2d} + \tra{\mathbf{A}}} (\bm{\mu}_{t+1} - \bm{\mu}_{t})
    + \tra{ (\bm{\mu}_{t+1} - \bm{\mu}_{t}) } \mathbf{K}_{t,t}^{-1} (\bm{\mu}_{t+1} - \bm{\mu}_{t}) \\
    =& \lambda \tra{\bm{\mu}}_{t} \pr{ \tra{\mathbf{A}} - \mathbf{I}_{2d}} \pr{ \mathbf{A} + \mathbf{I}_{2d} } \pr{ \tra{\mathbf{A}} + \mathbf{I}_{2d}} \pr{ \mathbf{A} - \mathbf{I}_{2d}} \bm{\mu}_{t}
    + \tra{ (\bm{\mu}_{t+1} - \bm{\mu}_{t}) } \mathbf{K}_{t,t}^{-1} (\bm{\mu}_{t+1} - \bm{\mu}_{t}) \\
    =& \lambda | \pr{\tra{\mathbf{A}} \mathbf{A} - \mathbf{I}_{2d} - \tra{\mathbf{A}} + \mathbf{A}} \bm{\mu}_{t} |^2
    + \tra{ (\bm{\mu}_{t+1} - \bm{\mu}_{t}) } \mathbf{K}_{t,t}^{-1} (\bm{\mu}_{t+1} - \bm{\mu}_{t})
\end{align}
From the above, the total dissipation function is expressed as follows:
\begin{align}
    \tilde{\Sigma}_{\mathrm{XY}} ^{\mathrm{tot}}
    =& \frac{1}{2} \bigg\{ \lambda \trace{}{ \pr{ \mathbf{I}_{2d} - \tra{\mathbf{A}} } \pr{ \mathbf{I}_{2d} + \tra{\mathbf{A}} } \pr{ \mathbf{I}_{2d} - \mathbf{A} } \pr{ \mathbf{I}_{2d} + \mathbf{A} } \mathbf{K}_{t,t}
    }
    + \trace{}{ \tra{\mathbf{A}} \mathbf{A}
    + \mathbf{K}_{t,t}^{-1}\mathbf{K}_{t+1,t+1}}
    - 4d \nonumber \\
    &+ \lambda | \pr{\tra{\mathbf{A}} \mathbf{A} - \mathbf{I}_{2d} - \tra{\mathbf{A}} + \mathbf{A}} \bm{\mu}_{t} | ^2
    + \tra{ (\bm{\mu}_{t+1} - \bm{\mu}_{t}) } \mathbf{K}_{t,t}^{-1} (\bm{\mu}_{t+1} - \bm{\mu}_{t})
    \bigg\} \label{trep_tot_gau2}
\end{align}

\subsubsection*{Total Entropy Production}
The total entropy production is related to the total dissipation function by the formula
\begin{equation}
    \Sigma_{\mathrm{XY}} ^{\mathrm{tot}} =
    \tilde{\Sigma}_{\mathrm{XY}} ^{\mathrm{tot}}
    - \KL{P_{ \mathbf{X}_{t+1}, \mathbf{Y}_{t+1} }}{P_{ \mathbf{X}_{t}, \mathbf{Y}_{t} }}
\end{equation}
where the second term is equal to
\begin{equation}
    \KL{P_{ \mathbf{X}_{t+1}, \mathbf{Y}_{t+1} }}{P_{ \mathbf{X}_{t}, \mathbf{Y}_{t} }}
    =
    \frac{1}{2} \brc{
    \trace{}{\mathbf{K}_{t,t}^{-1} \mathbf{K}_{t+1,t+1} } - 2d
    + \tra{( \bm{\mu}_{t} - \bm{\mu}_{t+1})} \mathbf{K}_{t,t}^{-1} ( \bm{\mu}_{t} - \bm{\mu}_{t+1})
    + \log \frac{|\mathbf{K}_{t,t}|}{|\mathbf{K}_{t+1,t+1}|}
    }
\end{equation}
Therefore, the total entropy production is as follows:
\begin{align}
    \Sigma_{\mathrm{XY}} ^{\mathrm{tot}}
     =& \frac{1}{2} \brc{
    \trace{}{\tilde{\mathbf{K}}^{-1} \mathbf{K} } - 4d
    + \tra{( \tilde{\bm{\mu}} - \bm{\mu})} \tilde{\mathbf{K}}^{-1} ( \tilde{\bm{\mu}} - \bm{\mu})
    } \\
    & - \frac{1}{2} \brc{
    \trace{}{\mathbf{K}_{t,t}^{-1} \mathbf{K}_{t+1,t+1} } - 2d
    + \tra{( \bm{\mu}_{t} - \bm{\mu}_{t+1})} \mathbf{K}_{t,t}^{-1} ( \bm{\mu}_{t} - \bm{\mu}_{t+1})
    + \log \frac{|\mathbf{K}_{t,t}|}{|\mathbf{K}_{t+1,t+1}|}
    } \\
    =& \frac{1}{2} \bigg\{
    \trace{}{\tilde{\mathbf{K}}^{-1} \mathbf{K} }
    - \trace{}{\mathbf{K}_{t,t}^{-1} \mathbf{K}_{t+1,t+1} }
    - 2d \\
    &+ \tra{( \tilde{\bm{\mu}} - \bm{\mu})} \tilde{\mathbf{K}}^{-1} ( \tilde{\bm{\mu}} - \bm{\mu})
    - \tra{( \bm{\mu}_{t} - \bm{\mu}_{t+1})} \mathbf{K}_{t,t}^{-1} ( \bm{\mu}_{t} - \bm{\mu}_{t+1})
    + \log \frac{|\mathbf{K}_{t+1,t+1}|}{|\mathbf{K}_{t,t}|}
    \bigg\} \\
    =& \frac{1}{2} \bigg\{ \lambda \trace{}{ \pr{ \mathbf{I}_{2d} - \tra{\mathbf{A}} } \pr{ \mathbf{I}_{2d} + \tra{\mathbf{A}} } \pr{ \mathbf{I}_{2d} - \mathbf{A} } \pr{ \mathbf{I}_{2d} + \mathbf{A} } \mathbf{K}_{t,t}
    }
    + \trace{}{ \tra{\mathbf{A}} \mathbf{A}
    }
    - 2d \nonumber \\
    &+ \lambda | \pr{\tra{\mathbf{A}} \mathbf{A} - \mathbf{I}_{2d} - \tra{\mathbf{A}} + \mathbf{A}} \bm{\mu}_{t} | ^2
    + \log \frac{|\mathbf{K}_{t+1,t+1}|}{|\mathbf{K}_{t,t}|}
    \bigg\}
    \label{bep_tot_gau}
\end{align}
\subsection{Partial Entropy Production}
\label{sec:par_langevin}
\subsubsection*{Partial Dissipation Function}
In the partial dissipation function
\begin{equation}
    \tilde{\Sigma}_{\mathrm{X}} ^{\mathrm{par}} = \KL{ P_{ \mathbf{X}_{t:t+1}, \mathbf{Y}_{t:t+1} } }{ \tilde{Q}_{ \mathbf{X}_{t:t+1}, \mathbf{Y}_{t:t+1} } ^{\mathrm{par}} },
\end{equation}
we assume the following:
\begin{align}
    P_{ \mathbf{X}_{t:t+1}, \mathbf{Y}_{t:t+1} }
    (\mathbf{x}_{t:t+1}, \mathbf{y}_{t:t+1})
    =& \mathcal{N}_{\mathbf{s}} (\bm{\mu},\mathbf{K})\\
    =& \frac{1}{\sqrt{|2 \pi \mathbf{K}|}} \exp \br{ -\frac{1}{2}
    \tra{(\mathbf{s} - \bm{\mu})} \mathbf{K}^{-1} (\mathbf{s} - \bm{\mu})
    } \\
    \tilde{Q}_{ \mathbf{X}_{t:t+1}, \mathbf{Y}_{t:t+1} } ^{\mathrm{par}}
    (\mathbf{x}_{t:t+1}, \mathbf{y}_{t:t+1})
    =& \frac{1}{\sqrt{|2 \pi \mathbf{K}|}} \exp \br{ -\frac{1}{2}
    \tra{(\tilde{\mathbf{s}}^{\mathrm{par}} - \bm{\mu})} \mathbf{K}^{-1} (\tilde{\mathbf{s}}^{\mathrm{par}} - \bm{\mu})
    }
\end{align}
where it is assumed that
\begin{equation}
    \tilde{\mathbf{s}}^{\mathrm{par}} = \tra{
    \begin{pmatrix}
        \mathbf{x}_{t+1} & \mathbf{y}_{t} & \mathbf{x}_{t} & \mathbf{y}_{t+1}
    \end{pmatrix}
    }
\end{equation}
Consider rewriting the variables of the partial time-reversed distribution with $\mathbf{s}$. In what follows, let the covariance matrix be
\begin{align}
    \mathbf{K}^{-1} =&
    \begin{pmatrix}
        \mathbf{P}^1
        &\mathbf{P}^2 \\
        \mathbf{P}^3
        &\mathbf{P}^4
    \end{pmatrix}  \\
    \mathbf{P}^1 = \mathbf{K}_{t|t+1}^{-1}, \quad
    \mathbf{P}^2 = -\tra{\mathbf{A}} \mathbf{K}_{t+1|t}^{-1},& \quad
    \mathbf{P}^3 = -\mathbf{K}_{t+1|t}^{-1} \mathbf{A}, \quad
    \mathbf{P}^4 = \mathbf{K}_{t+1|t}^{-1} \\
    \mathbf{P}^i =&
    \begin{pmatrix}
        \mathbf{P}^i _{\mathbf{xx}} & \mathbf{P}^i _{\mathbf{xy}} \\
        \mathbf{P}^i _{\mathbf{yx}} & \mathbf{P}^i _{\mathbf{yy}}
    \end{pmatrix} \quad
    \pr{i=1, 2, 3, 4}
\end{align}
then the second term in the exponent of $\tilde{Q}_{ \mathbf{X}_{t:t+1}, \mathbf{Y}_{t:t+1} } ^{\mathrm{par}}$ becomes
\begin{align}
    &\tra{(\tilde{\mathbf{s}}^{\mathrm{par}} - \bm{\mu})} \mathbf{K}^{-1} (\tilde{\mathbf{s}}^{\mathrm{par}} - \bm{\mu}) \\
    =& \tra{(\mathbf{s}_{t+1} - \bm{\mu}_{t})} \mathbf{P}^1 (\mathbf{s}_{t+1} - \bm{\mu}_{t})
    + \tra{(\mathbf{s}_{t+1} - \bm{\mu}_{t})} \mathbf{P}^2 (\mathbf{s}_{t} - \bm{\mu}_{t+1}) \\
    &+ \tra{(\mathbf{s}_{t} - \bm{\mu}_{t+1})} \mathbf{P}^3 (\mathbf{s}_{t+1} - \bm{\mu}_{t})
    + \tra{(\mathbf{s}_{t} - \bm{\mu}_{t+1})} \mathbf{P}^4 (\mathbf{s}_{t} - \bm{\mu}_{t+1}) \\
    =& \tra{(\mathbf{x}_{t+1} - \bm{\mu}_{\mathbf{x}_t})} \mathbf{P}^1 _{\mathbf{xx}} (\mathbf{x}_{t+1} - \bm{\mu}_{\mathbf{x}_t})
    + \tra{(\mathbf{x}_{t+1} - \bm{\mu}_{\mathbf{x}_t})} \mathbf{P}^1 _{\mathbf{xy}} (\mathbf{y}_{t} - \bm{\mu}_{\mathbf{y}_t}) \\
    &+ \tra{(\mathbf{y}_{t} - \bm{\mu}_{\mathbf{y}_t})} \mathbf{P}^1 _{\mathbf{yx}} (\mathbf{x}_{t+1} - \bm{\mu}_{\mathbf{x}_t})
    + \tra{(\mathbf{y}_{t} - \bm{\mu}_{\mathbf{y}_t})} \mathbf{P}^1 _{\mathbf{yy}} (\mathbf{y}_{t} - \bm{\mu}_{\mathbf{y}_t})\\
    &+\tra{(\mathbf{x}_{t+1} - \bm{\mu}_{\mathbf{x}_t})} \mathbf{P}^2 _{\mathbf{xx}} (\mathbf{x}_{t} - \bm{\mu}_{\mathbf{x}_{t+1}})
    + \tra{(\mathbf{x}_{t+1} - \bm{\mu}_{\mathbf{x}_t})} \mathbf{P}^2 _{\mathbf{xy}} (\mathbf{y}_{t+1} - \bm{\mu}_{\mathbf{y}_{t+1}}) \\
    &+ \tra{(\mathbf{y}_{t} - \bm{\mu}_{\mathbf{y}_t})} \mathbf{P}^2 _{\mathbf{yx}} (\mathbf{x}_{t} - \bm{\mu}_{\mathbf{x}_{t+1}})
    + \tra{(\mathbf{y}_{t} - \bm{\mu}_{\mathbf{y}_t})} \mathbf{P}^2 _{\mathbf{yy}} (\mathbf{y}_{t+1} - \bm{\mu}_{\mathbf{y}_{t+1}})\\
    &+\tra{(\mathbf{x}_{t} - \bm{\mu}_{\mathbf{x}_{t+1}})} \mathbf{P}^3 _{\mathbf{xx}} (\mathbf{x}_{t+1} - \bm{\mu}_{\mathbf{x}_t})
    + \tra{(\mathbf{x}_{t} - \bm{\mu}_{\mathbf{x}_{t+1}})} \mathbf{P}^3 _{\mathbf{xy}} (\mathbf{y}_{t} - \bm{\mu}_{\mathbf{y}_t}) \\
    &+ \tra{(\mathbf{y}_{t+1} - \bm{\mu}_{\mathbf{y}_{t+1}})} \mathbf{P}^3 _{\mathbf{yx}} (\mathbf{x}_{t+1} - \bm{\mu}_{\mathbf{x}_{t}})
    + \tra{(\mathbf{y}_{t+1} - \bm{\mu}_{\mathbf{y}_{t+1}})} \mathbf{P}^3 _{\mathbf{yy}} (\mathbf{y}_{t} - \bm{\mu}_{\mathbf{y}_t})\\
    &+\tra{(\mathbf{x}_{t} - \bm{\mu}_{\mathbf{x}_{t+1}})} \mathbf{P}^4 _{\mathbf{xx}} (\mathbf{x}_{t} - \bm{\mu}_{\mathbf{x}_{t+1}})
    + \tra{(\mathbf{x}_{t} - \bm{\mu}_{\mathbf{x}_{t+1}})} \mathbf{P}^4 _{\mathbf{xy}} (\mathbf{y}_{t+1} - \bm{\mu}_{\mathbf{y}_{t+1}}) \\
    &+ \tra{(\mathbf{y}_{t+1} - \bm{\mu}_{\mathbf{y}_{t+1}})} \mathbf{P}^4 _{\mathbf{yx}} (\mathbf{x}_{t} - \bm{\mu}_{\mathbf{x}_{t+1}})
    + \tra{(\mathbf{y}_{t+1} - \bm{\mu}_{\mathbf{y}_{t+1}})} \mathbf{P}^4 _{\mathbf{yy}} (\mathbf{y}_{t+1} - \bm{\mu}_{\mathbf{y}_{t+1}}) \\
    =& \tra{(\mathbf{s}_{t} - \bm{\mu}_{t+1})} \mathbf{P'}^1 (\mathbf{s}_{t} - \bm{\mu}_{t+1})
    + \tra{(\mathbf{s}_{t} - \bm{\mu}_{t+1})} \mathbf{P'}^2 (\mathbf{s}_{t+1} - \bm{\mu}_{t}) \\
    &+ \tra{(\mathbf{s}_{t+1} - \bm{\mu}_{t})} \mathbf{P'}^3 (\mathbf{s}_{t} - \bm{\mu}_{t+1})
    + \tra{(\mathbf{s}_{t+1} - \bm{\mu}_{t})} \mathbf{P'}^4 (\mathbf{s}_{t+1} - \bm{\mu}_{t}) \\
    =& \tra{(\mathbf{s}- \tilde{\bm{\mu}}^{\mathrm{par}})} \pr{\tilde{\mathbf{K}}^{\mathrm{par}}}^{-1} (\mathbf{s} - \tilde{\bm{\mu}}^{\mathrm{par}})
\end{align}
Here, it is assumed that
\begin{align}
    \tilde{\bm{\mu}}^{\mathrm{par}} \coloneq&
    \tra{
    \begin{pmatrix}
        \bm{\mu}_{\mathbf{x}_{t+1}} & \bm{\mu}_{\mathbf{y}_{t}} & \bm{\mu}_{\mathbf{x}_{t}} & \bm{\mu}_{\mathbf{y}_{t+1}}
    \end{pmatrix}
    } \\
    \pr{\tilde{\mathbf{K}}^{\mathrm{par}}}^{-1} \coloneq&
    \begin{pmatrix}
        \mathbf{P'}^1
        &\mathbf{P'}^2 \\
        \mathbf{P'}^3
        &\mathbf{P'}^4
    \end{pmatrix} \\
    \mathbf{P'}^1
    = \begin{pmatrix}
        \mathbf{P}^4 _{\mathbf{xx}} & \mathbf{P}^3 _{\mathbf{xy}} \\
        \mathbf{P}^2 _{\mathbf{yx}} & \mathbf{P}^1 _{\mathbf{yy}}
    \end{pmatrix}
    \mathbf{P'}^2
    = \begin{pmatrix}
        \mathbf{P}^3 _{\mathbf{xx}} & \mathbf{P}^4 _{\mathbf{xy}} \\
        \mathbf{P}^1 _{\mathbf{yx}} & \mathbf{P}^2 _{\mathbf{yy}}
    \end{pmatrix}
    &\mathbf{P'}^3
    = \begin{pmatrix}
        \mathbf{P}^2 _{\mathbf{xx}} & \mathbf{P}^1 _{\mathbf{xy}} \\
        \mathbf{P}^4 _{\mathbf{yx}} & \mathbf{P}^3 _{\mathbf{yy}}
    \end{pmatrix}
    \mathbf{P'}^4
    = \begin{pmatrix}
        \mathbf{P}^1 _{\mathbf{xx}} & \mathbf{P}^2 _{\mathbf{xy}} \\
        \mathbf{P}^3 _{\mathbf{yx}} & \mathbf{P}^4 _{\mathbf{yy}}
    \end{pmatrix}
\end{align}
Therefore it can be expressed as follows
\begin{align}
    \tilde{Q}_{ \mathbf{X}_{t:t+1}, \mathbf{Y}_{t:t+1} }^{\mathrm{par}}
    (\mathbf{x}_{t:t+1}, \mathbf{y}_{t:t+1})
    =& \frac{1}{\sqrt{|2 \pi \tilde{\mathbf{K}}^{\mathrm{par}}|}} \exp \br{ - \frac{1}{2}
    \tra{(\mathbf{s} - \tilde{\bm{\mu}}^{\mathrm{par}} )} \pr{\tilde{\mathbf{K}}^{\mathrm{par}}}^{-1} (\mathbf{s} - \tilde{\bm{\mu}}^{\mathrm{par}})
    } \\
    =&\mathcal{N}_{\mathbf{s}} (\tilde{\bm{\mu}}^{\mathrm{par}},\tilde{\mathbf{K}}^{\mathrm{par}})
\end{align}
by using the identity $|\tilde{\mathbf{K}}^{\mathrm{par}}|=|\mathbf{K}|$.

From the above, the partial dissipation function is expressed as follows:
\begin{equation}
    \tilde{\Sigma}_{\mathrm{X}} ^{\mathrm{par}} =
    \frac{1}{2} \brc{
    \trace{}{\pr{\tilde{\mathbf{K}}^{\mathrm{par}}}^{-1} \mathbf{K} } - 4d
    + \tra{( \tilde{\bm{\mu}}^{\mathrm{par}} - \bm{\mu})} \pr{\tilde{\mathbf{K}}^{\mathrm{par}}}^{-1} ( \tilde{\bm{\mu}}^{\mathrm{par}} - \bm{\mu})
    }
    \label{trep_par_gau}
\end{equation}
\subsubsection*{Partial Entropy Production}
By definition, the difference between the partial entropy production and total dissipation function is as follows.
\begin{align}
    \Sigma_{\mathrm{X}} ^{\mathrm{par}} - \tilde{\Sigma} _{\mathrm{XY}} ^{\mathrm{tot}}
    =& \expec{\mathbf{X}_{t:t+1}, \mathbf{Y}_{t:t+1}}{\log \frac{
    P_{ \mathbf{X}_{t+1} | \mathbf{Y}_{t+1}, \mathbf{X}_{t}, \mathbf{Y}_{t} } ( \mathbf{x}_{t} | \mathbf{y}_{t}, \mathbf{x}_{t+1}, \mathbf{y}_{t+1} )
    P_{ \mathbf{X}_{t}, \mathbf{Y}_{t+1}, \mathbf{Y}_{t} } (\mathbf{x}_{t+1}, \mathbf{y}_{t+1}, \mathbf{y}_{t})
    }{P_{ \mathbf{X}_{t+1} | \mathbf{Y}_{t+1}, \mathbf{X}_{t}, \mathbf{Y}_{t} } ( \mathbf{x}_{t} | \mathbf{y}_{t}, \mathbf{x}_{t+1}, \mathbf{y}_{t+1} )
    P_{ \mathbf{X}_{t+1}, \mathbf{Y}_{t+1}, \mathbf{Y}_{t} } (\mathbf{x}_{t+1}, \mathbf{y}_{t+1}, \mathbf{y}_{t})}} \\
    =& - \KL{P_{ \mathbf{X}_{t+1}, \mathbf{Y}_{t+1}, \mathbf{Y}_{t} }
    }{
    P_{ \mathbf{X}_{t}, \mathbf{Y}_{t+1}, \mathbf{Y}_{t} }
	} \label{dif_parep_totdf}
\end{align}
Here,
\begin{align}
    P_{\mathbf{X}_{t+1}, \mathbf{Y}_{t:t+1}} (\mathbf{x}_{t+1}, \mathbf{y}_{t:t+1})
    =& \mathcal{N}_{\mathbf{x}_{t+1}, \mathbf{y}_{t:t+1}} (\bm{\mu}_{\mathbf{x}_{t+1}, \mathbf{y}_{t:t+1}},\mathbf{K}_{\mathbf{x}_{t+1}, \mathbf{y}_{t:t+1}}) \\
    \bm{\mu}_{\mathbf{x}_{t+1}, \mathbf{y}_{t:t+1}} =&
    \tra{
    \begin{pmatrix}
        \bm{\mu}_{\mathbf{x}_{t+1}} &
        \bm{\mu}_{\mathbf{y}_{t}} &
        \bm{\mu}_{\mathbf{y}_{t+1}}
    \end{pmatrix}
    } \\
    \mathbf{K}_{\mathbf{x}_{t+1}, \mathbf{y}_{t:t+1}}
    =& \begin{pmatrix}
    \mathbf{K}_{\mathbf{x}_{t+1} \mathbf{x}_{t+1}}
    & \mathbf{K}_{\mathbf{x}_{t+1} \mathbf{y}_{t}}
    & \mathbf{K}_{\mathbf{x}_{t+1} \mathbf{y}_{t+1}} \\
    \tra{\mathbf{K}}_{\mathbf{x}_{t+1} \mathbf{y}_{t}}
    & \mathbf{K}_{\mathbf{y}_{t} \mathbf{y}_{t}}
    & \mathbf{K}_{\mathbf{y}_{t} \mathbf{y}_{t+1}} \\
    \tra{\mathbf{K}}_{\mathbf{x}_{t+1} \mathbf{y}_{t+1}}
    & \tra{\mathbf{K}}_{\mathbf{y}_{t} \mathbf{y}_{t+1}}
    & \mathbf{K}_{\mathbf{y}_{t+1} \mathbf{y}_{t+1}}
\end{pmatrix}
\end{align}
\begin{align}
    P_{\mathbf{X}_{t}, \mathbf{Y}_{t:t+1}} (\mathbf{x}_{t}, \mathbf{y}_{t:t+1})
    =& \mathcal{N}_{\mathbf{x}_{t}, \mathbf{y}_{t:t+1}} (\bm{\mu}_{\mathbf{x}_{t}, \mathbf{y}_{t:t+1}},\mathbf{K}_{\mathbf{x}_{t}, \mathbf{y}_{t:t+1}}) \\
    \bm{\mu}_{\mathbf{x}_{t}, \mathbf{y}_{t:t+1}} =&
    \tra{
    \begin{pmatrix}
        \bm{\mu}_{\mathbf{x}_{t}} &
        \bm{\mu}_{\mathbf{y}_{t}} &
        \bm{\mu}_{\mathbf{y}_{t+1}}
    \end{pmatrix}
    } \\
    \mathbf{K}_{\mathbf{x}_{t}, \mathbf{y}_{t:t+1}}
    =& \begin{pmatrix}
    \mathbf{K}_{\mathbf{x}_{t} \mathbf{x}_{t}}
    & \mathbf{K}_{\mathbf{x}_{t} \mathbf{y}_{t}}
    & \mathbf{K}_{\mathbf{x}_{t} \mathbf{y}_{t+1}} \\
    \tra{\mathbf{K}}_{\mathbf{x}_{t} \mathbf{y}_{t}}
    & \mathbf{K}_{\mathbf{y}_{t} \mathbf{y}_{t}}
    & \mathbf{K}_{\mathbf{y}_{t} \mathbf{y}_{t+1}} \\
    \tra{\mathbf{K}}_{\mathbf{x}_{t} \mathbf{y}_{t+1}}
    & \tra{\mathbf{K}}_{\mathbf{y}_{t} \mathbf{y}_{t+1}}
    & \mathbf{K}_{\mathbf{y}_{t+1} \mathbf{y}_{t+1}}
\end{pmatrix}
\end{align}
and so, from the KLD of the multidimensional Gaussian distribution, the right hand side of \eqref{dif_parep_totdf} is
\begin{align}
    &\KL{P_{ \mathbf{X}_{t+1}, \mathbf{Y}_{t+1}, \mathbf{Y}_{t} }
    }{
    P_{ \mathbf{X}_{t}, \mathbf{Y}_{t+1}, \mathbf{Y}_{t} }
    } \\
	=& \frac{1}{2} \bigg\{
    \trace{}{ \mathbf{K}_{\mathbf{x}_{t}, \mathbf{y}_{t:t+1}}^{-1}  \mathbf{K}_{\mathbf{x}_{t+1}, \mathbf{y}_{t:t+1}} } - 3d
    + \tra{( \bm{\mu}_{\mathbf{x}_{t}, \mathbf{y}_{t:t+1}} - \bm{\mu}_{\mathbf{x}_{t+1}, \mathbf{y}_{t:t+1}})} \mathbf{K}_{\mathbf{x}_{t}, \mathbf{y}_{t:t+1}}^{-1} ( \bm{\mu}_{\mathbf{x}_{t}, \mathbf{y}_{t:t+1}} - \bm{\mu}_{\mathbf{x}_{t+1}, \mathbf{y}_{t:t+1}})\\
    &+\log \frac{|\mathbf{K}_{\mathbf{x}_{t}, \mathbf{y}_{t:t+1}}|}{|\mathbf{K}_{\mathbf{x}_{t+1}, \mathbf{y}_{t:t+1}}|}
    \bigg\}
    \label{kl_xyy}
\end{align}
Furthermore, the covariance matrix and mean vector is expressed as follows:
\begin{align}
    \mathbf{K}_{\mathbf{x}_{t}, \mathbf{y}_{t:t+1}}
    =& \begin{pmatrix}
        \mathbf{K}_{\mathbf{x}_{t}, \mathbf{x}_{t}}
        & \mathbf{P}_t \\
        \tra{\mathbf{P}}_t
        & \mathbf{K}_{\mathbf{y} \mathbf{y}}
    \end{pmatrix} \\
    \mathbf{P}_t \coloneq& \begin{pmatrix}
        \mathbf{K}_{\mathbf{x}_{t}, \mathbf{y}_{t}} & \mathbf{K}_{\mathbf{x}_{t}, \mathbf{y}_{t+1}}
    \end{pmatrix} \\
    \mathbf{K}_{\mathbf{x}_{t+1}, \mathbf{y}_{t:t+1}}
    =& \begin{pmatrix}
        \mathbf{K}_{\mathbf{x}_{t+1}, \mathbf{x}_{t+1}}
        & \mathbf{P}_{t+1} \\
        \tra{\mathbf{P}}_{t+1}
        & \mathbf{K}_{\mathbf{y} \mathbf{y}}
    \end{pmatrix} \\
    \mathbf{P}_{t+1} \coloneq& \begin{pmatrix}
        \mathbf{K}_{\mathbf{x}_{t+1}, \mathbf{y}_{t}} & \mathbf{K}_{\mathbf{x}_{t+1}, \mathbf{y}_{t+1}}
    \end{pmatrix} \\
    \bm{\mu}_{\mathbf{x}_{t}, \mathbf{y}_{t:t+1}} - \bm{\mu}_{\mathbf{x}_{t+1}, \mathbf{y}_{t:t+1}}
    =& \tra{
    \begin{pmatrix}
        \bm{\mu}_{\mathbf{x}_{t}} - \bm{\mu}_{\mathbf{x}_{t+1}} & 0 & 0
    \end{pmatrix}
    }
\end{align}
Therefore the third term of \eqref{kl_xyy} becomes
\begin{equation}
    \tra{( \bm{\mu}_{\mathbf{x}_{t}, \mathbf{y}_{t:t+1}} - \bm{\mu}_{\mathbf{x}_{t+1}, \mathbf{y}_{t:t+1}})} \mathbf{K}_{\mathbf{x}_{t}, \mathbf{y}_{t:t+1}}^{-1} ( \bm{\mu}_{\mathbf{x}_{t}, \mathbf{y}_{t:t+1}} - \bm{\mu}_{\mathbf{x}_{t+1}, \mathbf{y}_{t:t+1}})
    = \tra{( \bm{\mu}_{\mathbf{x}_{t}} - \bm{\mu}_{\mathbf{x}_{t+1}})} \mathbf{K}_{\mathbf{x}_{t} \mathbf{x}_{t}}^{-1} ( \bm{\mu}_{\mathbf{x}_{t}} - \bm{\mu}_{\mathbf{x}_{t+1}})
\end{equation}
From the determinant of the block matrix, the fourth term is
\begin{align}
    |\mathbf{K}_{\mathbf{x}_{t}, \mathbf{y}_{t:t+1}}| =&
    |\mathbf{K}_{\mathbf{x}_{t} \mathbf{x}_{t}}| |\mathbf{K}_{\mathbf{y} \mathbf{y}} - \tra{\mathbf{P}}_{t} \mathbf{K}_{\mathbf{x}_{t+1} \mathbf{x}_{t+1}} ^{-1} \mathbf{P}_{t}| \\
    |\mathbf{K}_{\mathbf{x}_{t+1}, \mathbf{y}_{t:t+1}}| =&
    |\mathbf{K}_{\mathbf{x}_{t+1} \mathbf{x}_{t+1}}| |\mathbf{K}_{\mathbf{y} \mathbf{y}} - \tra{\mathbf{P}}_{t} \mathbf{K}_{\mathbf{x}_{t+1} \mathbf{x}_{t+1}} ^{-1} \mathbf{P}_{t}|
\end{align}
and thus
\begin{equation}
    \log \frac{|\mathbf{K}_{\mathbf{x}_{t}, \mathbf{y}_{t:t+1}}|}{|\mathbf{K}_{\mathbf{x}_{t+1}, \mathbf{y}_{t:t+1}}|}
    = \log \frac{|\mathbf{K}_{\mathbf{x}_{t} \mathbf{x}_{t}}|}{    |\mathbf{K}_{\mathbf{x}_{t+1} \mathbf{x}_{t+1}}|}
\end{equation}
Finally, tshe partial entropy production is as follows:
\begin{equation}
    \Sigma_{\mathrm{X}} ^{\mathrm{par}} =
     \tilde{\Sigma} _{\mathrm{XY}} ^{\mathrm{tot}}
     - \frac{1}{2} \bigg\{
    \trace{}{ \mathbf{K}_{\mathbf{x}_{t}, \mathbf{y}_{t:t+1}}^{-1}  \mathbf{K}_{\mathbf{x}_{t+1}, \mathbf{y}_{t:t+1}} } - 3d + \tra{( \bm{\mu}_{\mathbf{x}_{t}} - \bm{\mu}_{\mathbf{x}_{t+1}})} \mathbf{K}_{\mathbf{x}_{t} \mathbf{x}_{t}}^{-1} ( \bm{\mu}_{\mathbf{x}_{t}} - \bm{\mu}_{\mathbf{x}_{t+1}}) + \log \frac{|\mathbf{K}_{\mathbf{x}_{t} \mathbf{x}_{t}}|}{    |\mathbf{K}_{\mathbf{x}_{t+1} \mathbf{x}_{t+1}}|}
    \bigg\}
    \label{bep_par_gau}
\end{equation}
\subsection{Marginal Entropy Production}
\label{sec:mar_langevin}
When only system X is observable, the marginal entropy production can be obtained by substituting the total entropy production \eqref{trep_tot}:
\begin{align}
    \mathbf{K} \rightarrow& \mathbf{K_{xx}} \\
    \bm{\mu} \rightarrow& \bm{\mu}_{\mathbf{x}}
\end{align}
\subsubsection*{Marginal Dissipation Function}
The marginal dissipation function is given by
\begin{equation}
    \tilde{\Sigma}_{\mathrm{X}} ^{\mathrm{mar}}
    =
    \frac{1}{2} \brc{
    \trace{}{\tilde{\mathbf{K}}_{\mathbf{x} \mathbf{x}}^{-1} \mathbf{K}_{\mathbf{x} \mathbf{x}} } - 2d
    + \tra{( \tilde{\bm{\mu}}_{\mathbf{x}} - \bm{\mu}_{\mathbf{x}})} \tilde{\mathbf{K}}_{\mathbf{x} \mathbf{x}}^{-1} ( \tilde{\bm{\mu}}_{\mathbf{x}} - \bm{\mu}_{\mathbf{x}})
    }
    \label{trep_mar}
\end{equation}
where
\begin{align}
    \bm{\mu}_{\mathbf{x}} \coloneq&
    \tra{
    \begin{pmatrix}
        \bm{\mu}_{\mathbf{x}_{t}} & \bm{\mu}_{\mathbf{x}_{t+1}}
    \end{pmatrix}
    } \\
    \tilde{\bm{\mu}}_{\mathbf{x}} \coloneq&
    \tra{
    \begin{pmatrix}
        \bm{\mu}_{\mathbf{x}_{t+1}} & \bm{\mu}_{\mathbf{x}_{t}}
    \end{pmatrix}
    } \\
    \tilde{\mathbf{K}}_{\mathbf{x} \mathbf{x}}^{-1} \coloneq&
    \begin{pmatrix}
        \mathbf{K}_{\mathbf{x}_{t+1}|\mathbf{x}_{t}}^{-1}
        &-\mathbf{K}_{\mathbf{x}_{t+1}|\mathbf{x}_{t}}^{-1} \mathbf{C} \\
        -\tra{\mathbf{C}}  \mathbf{K}_{\mathbf{x}_{t+1}|\mathbf{x}_{t}}^{-1}
       &\mathbf{K}_{\mathbf{x}_{t}|\mathbf{x}_{t+1}}^{-1}
    \end{pmatrix} \\
    \mathbf{C} \coloneq& \tra{\mathbf{K}}_{\mathbf{x}_{t},\mathbf{x}_{t+1}} \mathbf{K}_{\mathbf{x}_{t},\mathbf{x}_{t}}^{-1} \\
    \mathbf{K}_{\mathbf{x}_{t+1}|\mathbf{x}_{t}} \coloneq& \mathbf{K}_{\mathbf{x}_{t+1},\mathbf{x}_{t+1}} - \tra{\mathbf{K}}_{\mathbf{x}_{t},\mathbf{x}_{t+1}} \mathbf{K}_{\mathbf{x}_{t},\mathbf{x}_{t}}^{-1} \mathbf{K}_{\mathbf{x}_{t},\mathbf{x}_{t+1}}  \\
    \mathbf{K}_{\mathbf{x}_{t}|\mathbf{x}_{t+1}} \coloneq& \mathbf{K}_{\mathbf{x}_{t},\mathbf{x}_{t}} - \mathbf{K}_{\mathbf{x}_{t},\mathbf{x}_{t+1}} \mathbf{K}_{\mathbf{x}_{t+1},\mathbf{x}_{t+1}}^{-1} \tra{\mathbf{K}}_{\mathbf{x}_{t},\mathbf{x}_{t+1}}
\end{align}
Also, by expanding the first and third terms, this can be expressed as follows:
\begin{align}
    \tilde{\Sigma}_{\mathrm{X}} ^{\mathrm{mar}}
    =& \frac{1}{2} \bigg\{
    \trace{}{\mathbf{K}_{\mathbf{x}_{t+1} | \mathbf{x}_{t} } ^{-1}\mathbf{K}_{\mathbf{x}_t,\mathbf{x}_t}
    -\mathbf{K}_{\mathbf{x}_{t+1} | \mathbf{x}_{t} } ^{-1} \mathbf{C} \tra{\mathbf{K}}_{\mathbf{x}_t,\mathbf{x}_{t+1}}
    -\tra{\mathbf{C}} \mathbf{K}_{\mathbf{x}_{t+1} | \mathbf{x}_{t} } ^{-1}\mathbf{K}_{\mathbf{x}_t,\mathbf{x}_{t+1}}
    + \mathbf{K}_{\mathbf{x}_{t} | \mathbf{x}_{t+1} } ^{-1}\mathbf{K}_{\mathbf{x}_{t+1},\mathbf{x}_{t+1}}
    }
    - 2d \nonumber \\
     &+ \tra{ (\bm{\mu}_{\mathbf{x}_{t+1}} - \bm{\mu}_{\mathbf{x}_t}) } \pr{\mathbf{K}_{\mathbf{x}_{t+1} | \mathbf{x}_{t} } ^{-1} + \mathbf{K}_{\mathbf{x}_{t+1} | \mathbf{x}_{t} } ^{-1} \mathbf{C} +\tra{\mathbf{C}} \mathbf{K}_{\mathbf{x}_{t+1} | \mathbf{x}_{t} } ^{-1} + \mathbf{K}_{\mathbf{x}_{t} | \mathbf{x}_{t+1} } ^{-1}} (\bm{\mu}_{\mathbf{x}_{t+1}} - \bm{\mu}_{\mathbf{x}_t})
    \bigg\} \\
    =& \frac{1}{2} \bigg\{
    \trace{}{\mathbf{K}_{\mathbf{x}_{t+1} | \mathbf{x}_{t} } ^{-1}\mathbf{K}_{\mathbf{x}_t,\mathbf{x}_t}
    -\mathbf{K}_{\mathbf{x}_{t+1} | \mathbf{x}_{t} } ^{-1} \mathbf{C} \tra{\mathbf{K}}_{\mathbf{x}_t,\mathbf{x}_{t+1}}
    -\tra{\mathbf{C}} \mathbf{K}_{\mathbf{x}_{t+1} | \mathbf{x}_{t} } ^{-1}\mathbf{K}_{\mathbf{x}_t,\mathbf{x}_{t+1}}
    + \tra{\mathbf{C}} \mathbf{K}_{\mathbf{x}_{t+1} | \mathbf{x}_{t} } ^{-1} \mathbf{C} \mathbf{K}_{\mathbf{x}_{t+1},\mathbf{x}_{t+1}}
    }
    - 2d \nonumber \\
    &+ \tra{ (\bm{\mu}_{\mathbf{x}_{t+1}} - \bm{\mu}_{\mathbf{x}_t}) } \pr{\mathbf{K}_{\mathbf{x}_{t+1} | \mathbf{x}_{t} } ^{-1} + \mathbf{K}_{\mathbf{x}_{t+1} | \mathbf{x}_{t} } ^{-1} \mathbf{C} +\tra{\mathbf{C}} \mathbf{K}_{\mathbf{x}_{t+1} | \mathbf{x}_{t} } ^{-1} + \tra{\mathbf{C}} \mathbf{K}_{\mathbf{x}_{t+1} | \mathbf{x}_{t} } ^{-1} \mathbf{C}} (\bm{\mu}_{\mathbf{x}_{t+1}} - \bm{\mu}_{\mathbf{x}_t}) \\
    &+ \trace{}{\mathbf{K}_{\mathbf{x}_t,\mathbf{x}_{t}} ^{-1} \mathbf{K}_{\mathbf{x}_{t+1},\mathbf{x}_{t+1}}} + \tra{ (\bm{\mu}_{\mathbf{x}_{t+1}} - \bm{\mu}_{\mathbf{x}_t}) }
    \mathbf{K}_{\mathbf{x}_t,\mathbf{x}_{t}} ^{-1}
    (\bm{\mu}_{\mathbf{x}_{t+1}} - \bm{\mu}_{\mathbf{x}_t})
    \bigg\} \\
    =& \frac{1}{2} \bigg\{
    \trace{}{\mathbf{K}_{\mathbf{x}_{t+1} | \mathbf{x}_{t} } ^{-1} \pr{\mathbf{K}_{\mathbf{x}_t,\mathbf{x}_t} - \mathbf{C} \tra{\mathbf{K}}_{\mathbf{x}_t,\mathbf{x}_{t+1}}  }
    -\tra{\mathbf{C}} \mathbf{K}_{\mathbf{x}_{t+1} | \mathbf{x}_{t} }^{-1} \pr{\mathbf{K}_{\mathbf{x}_t,\mathbf{x}_{t+1}} - \mathbf{C} \mathbf{K}_{\mathbf{x}_{t+1},\mathbf{x}_{t+1}} }
    }
    - 2d \nonumber \\
    &+ \tra{ (\bm{\mu}_{\mathbf{x}_{t+1}} - \bm{\mu}_{\mathbf{x}_t}) } \pr{\mathbf{K}_{\mathbf{x}_{t+1} | \mathbf{x}_{t} } ^{-1}
    \pr{\mathbf{I}_{d} +  \mathbf{C}}
    +\tra{\mathbf{C}} \mathbf{K}_{\mathbf{x}_{t+1} | \mathbf{x}_{t} } ^{-1}
    \pr{\mathbf{I}_{d} + \mathbf{C}}
    }
    (\bm{\mu}_{\mathbf{x}_{t+1}} - \bm{\mu}_{\mathbf{x}_t}) \\
    &+ \trace{}{\mathbf{K}_{\mathbf{x}_t,\mathbf{x}_{t}} ^{-1} \mathbf{K}_{\mathbf{x}_{t+1},\mathbf{x}_{t+1}}} + \tra{ (\bm{\mu}_{\mathbf{x}_{t+1}} - \bm{\mu}_{\mathbf{x}_t}) }
    \mathbf{K}_{\mathbf{x}_t,\mathbf{x}_{t}} ^{-1}
    (\bm{\mu}_{\mathbf{x}_{t+1}} - \bm{\mu}_{\mathbf{x}_t})
    \bigg\}
    \label{trep_mar_gau}
\end{align}
\subsubsection*{Marginal Entropy Production}
By definition, the difference between the marginal entropy production and the marginal dissipation function as follows:
\begin{align}
    \Sigma_{\mathrm{X}} ^{\mathrm{mar}}
    =& \tilde{\Sigma}_{\mathrm{X}} ^{\mathrm{mar}} - \KL{P_{\mathbf{X}_{t+1}}}{P_{\mathbf{X}_{t}}}
\end{align}
Here, the second term is equal to
\begin{equation}
    \KL{P_{\mathbf{X}_{t+1}}}{P_{\mathbf{X}_{t}}} = \frac{1}{2} \brc{
    \trace{}{\mathbf{K}_{\mathbf{x}_t,\mathbf{x}_{t}} ^{-1} \mathbf{K}_{\mathbf{x}_{t+1},\mathbf{x}_{t+1}}  } - d
    + \tra{( \bm{\mu}_{\mathbf{x}_{t}}  - \bm{\mu}_{\mathbf{x}_{t+1}} )} \mathbf{K}_{\mathbf{x}_t,\mathbf{x}_{t}} ^{-1} ( \bm{\mu}_{\mathbf{x}_{t}}  - \bm{\mu}_{\mathbf{x}_{t+1}} )
    + \log \frac{|\mathbf{K}_{\mathbf{x}_t,\mathbf{x}_{t}} |}{|\mathbf{K}_{\mathbf{x}_{t+1},\mathbf{x}_{t+1}} |}
    }
\end{equation}
Thus, the marginal entropy production is
\begin{align}
    \Sigma_{\mathrm{X}} ^{\mathrm{mar}}
    =& \frac{1}{2} \bigg\{
    \trace{}{\mathbf{K}_{\mathbf{x}_{t+1} | \mathbf{x}_{t} } ^{-1} \pr{\mathbf{K}_{\mathbf{x}_t,\mathbf{x}_t} - \mathbf{C} \tra{\mathbf{K}}_{\mathbf{x}_t,\mathbf{x}_{t+1}}  }
    -\tra{\mathbf{C}} \mathbf{K}_{\mathbf{x}_{t+1} | \mathbf{x}_{t} }^{-1} \pr{\mathbf{K}_{\mathbf{x}_t,\mathbf{x}_{t+1}} - \mathbf{C} \mathbf{K}_{\mathbf{x}_{t+1},\mathbf{x}_{t+1}} }
    }
    - 2d \nonumber \\
    &+ \tra{ (\bm{\mu}_{\mathbf{x}_{t+1}} - \bm{\mu}_{\mathbf{x}_t}) } \pr{\mathbf{K}_{\mathbf{x}_{t+1} | \mathbf{x}_{t} } ^{-1}
    \pr{\mathbf{I}_{d} +  \mathbf{C}}
    +\tra{\mathbf{C}} \mathbf{K}_{\mathbf{x}_{t+1} | \mathbf{x}_{t} } ^{-1}
    \pr{\mathbf{I}_{d} + \mathbf{C}}
    }
    (\bm{\mu}_{\mathbf{x}_{t+1}} - \bm{\mu}_{\mathbf{x}_t}) \\
    &+ \trace{}{\mathbf{K}_{\mathbf{x}_t,\mathbf{x}_{t}} ^{-1} \mathbf{K}_{\mathbf{x}_{t+1},\mathbf{x}_{t+1}}} + \tra{ (\bm{\mu}_{\mathbf{x}_{t+1}} - \bm{\mu}_{\mathbf{x}_t}) }
    \mathbf{K}_{\mathbf{x}_t,\mathbf{x}_{t}} ^{-1}
    (\bm{\mu}_{\mathbf{x}_{t+1}} - \bm{\mu}_{\mathbf{x}_t})
    \bigg\} \\
    &- \frac{1}{2} \brc{
    \trace{}{\mathbf{K}_{\mathbf{x}_t,\mathbf{x}_{t}} ^{-1} \mathbf{K}_{\mathbf{x}_{t+1},\mathbf{x}_{t+1}}  } - d
    + \tra{( \bm{\mu}_{\mathbf{x}_{t}}  - \bm{\mu}_{\mathbf{x}_{t+1}} )} \mathbf{K}_{\mathbf{x}_t,\mathbf{x}_{t}} ^{-1} ( \bm{\mu}_{\mathbf{x}_{t}}  - \bm{\mu}_{\mathbf{x}_{t+1}} )
    + \log \frac{|\mathbf{K}_{\mathbf{x}_t,\mathbf{x}_{t}} |}{|\mathbf{K}_{\mathbf{x}_{t+1},\mathbf{x}_{t+1}} |}
    } \\
    =& \frac{1}{2} \bigg\{
    \trace{}{\mathbf{K}_{\mathbf{x}_{t+1} | \mathbf{x}_{t} } ^{-1} \pr{\mathbf{K}_{\mathbf{x}_t,\mathbf{x}_t} - \mathbf{C} \tra{\mathbf{K}}_{\mathbf{x}_t,\mathbf{x}_{t+1}}  }
    -\tra{\mathbf{C}} \mathbf{K}_{\mathbf{x}_{t+1} | \mathbf{x}_{t} }^{-1} \pr{\mathbf{K}_{\mathbf{x}_t,\mathbf{x}_{t+1}} - \mathbf{C} \mathbf{K}_{\mathbf{x}_{t+1},\mathbf{x}_{t+1}} }
    }
    - d \nonumber \\
    &+ \tra{ (\bm{\mu}_{\mathbf{x}_{t+1}} - \bm{\mu}_{\mathbf{x}_t}) } \pr{\mathbf{K}_{\mathbf{x}_{t+1} | \mathbf{x}_{t} } ^{-1}
    \pr{\mathbf{I}_{d} +  \mathbf{C}}
    +\tra{\mathbf{C}} \mathbf{K}_{\mathbf{x}_{t+1} | \mathbf{x}_{t} } ^{-1}
    \pr{\mathbf{I}_{d} + \mathbf{C}}
    }
    (\bm{\mu}_{\mathbf{x}_{t+1}} - \bm{\mu}_{\mathbf{x}_t})
     + \log \frac{|\mathbf{K}_{\mathbf{x}_{t+1},\mathbf{x}_{t+1}} |}{|\mathbf{K}_{\mathbf{x}_t,\mathbf{x}_{t}} |}
    \bigg\}
    \label{bep_mar_gau}
\end{align}
\subsection{Numerical Illustration}
\label{sec:numerical}
Here, we illustrate the numerical results using analytical calculations of the total entropy production and marginal entropy production for Gaussian process \eqref{trep_tot_gau}, \eqref{bep_tot_gau}, \eqref{trep_mar_gau}, \eqref{bep_mar_gau}. In particular, we confirm the inequalities of total entropy production and marginal entropy production discussed in chapter \ref{IneEntPro}.

\subsubsection*{Settings}
The settings for numerical calculations are as follows.
\begin{itemize}
    \item Systems X and Y are both one-dimensional, and the composite system is two-dimensional. In other words, $d=1$.
    \item The mean vector and covariance matrix of the distribution at time $t$ are $\bm{\mu}_{t}=\tra{\begin{pmatrix}
        1 & 1
    \end{pmatrix}}$ and $\mathbf{K}_{t,t}=\mathbf{I}_2$, respectively.
    \item The covariance matrix of noise  is $\mathbf{K}_{t+1|t}=\mathbf{I}_2$. In other words, $\lambda=1$.
    \item The interaction matrix expressed as $\mathbf{A}=\begin{pmatrix}
        1 & a \\
        -a & 1
    \end{pmatrix}$ using the parameter $a \in [-1, 1]$. Here, the difference $2a$ of the off-diagonal elements represents the asymmetry of the interaction, and when $a=0$, $\mathbf{A}=\tra{\mathbf{A}}$.
\end{itemize}
Note that in the above settings, since $\mathbf{A}^{-1}=\tra{\mathbf{A}}$, i.e., since $\mathbf{A}$ is an orthogonal matrix, so the dependence of total entropy production on $a$ \eqref{trep_tot_gau2}, \eqref{bep_tot_gau} can be considered to represent only the asymmetry of the interaction.

\subsubsection*{Results}
\begin{figure}[H]
    \centering
    \includegraphics[width=70mm]{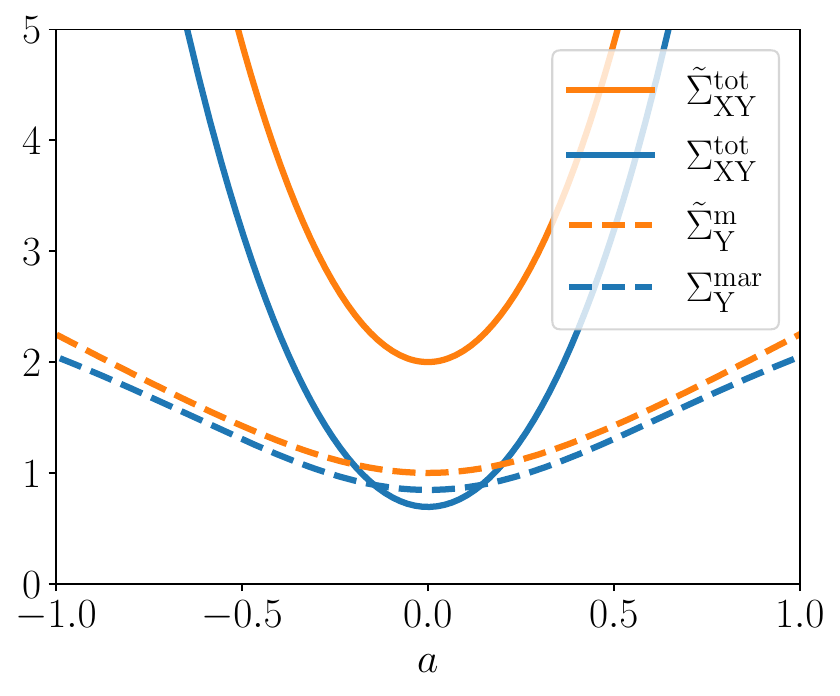}
    \caption{Relationship between total entropy production and marginal entropy production for Gaussian process.}
    \label{fig:tot-mar}
\end{figure}

Figure \ref{fig:tot-mar} shows the results of using the above settings to calculate the total entropy production and the marginal entropy production, respectively. The solid red line represents $\tilde{\Sigma}_{\mathrm{XY}} ^{\mathrm{tot}}$, the solid blue line represents $\Sigma_{\mathrm{XY}} ^{\mathrm{tot}}$, the dashed red line represents $\tilde{\Sigma}_{\mathrm{X}} ^{\mathrm{mar}}$, and the dashed blue line represents $\Sigma_{\mathrm{X}} ^{\mathrm{mar}}$. From this figure, it can be seen that
\begin{itemize}
    \item $\tilde{\Sigma}_{\mathrm{XY}} ^{\mathrm{tot}} \geq \tilde{\Sigma}_{\mathrm{X}} ^{\mathrm{mar}}$ always holds
    \item There are also regions where $\Sigma_{\mathrm{XY}}^{\mathrm{tot}} \geq \Sigma_{\mathrm{X}} ^{\mathrm{mar}} $ dose not hold
    \item $\Sigma_{\mathrm{XY}}^{\mathrm{tot}}, \tilde{\Sigma}_{\mathrm{XY}} ^{\mathrm{tot}}, \Sigma_{\mathrm{X}} ^{\mathrm{mar}},\tilde{\Sigma}_{\mathrm{X}} ^{\mathrm{mar}}$ is symmetric relative to $a=0$
    \item $\Sigma_{\mathrm{XY}}^{\mathrm{tot}}, \tilde{\Sigma}_{\mathrm{XY}} ^{\mathrm{tot}}, \Sigma_{\mathrm{X}} ^{\mathrm{mar}},\tilde{\Sigma}_{\mathrm{X}} ^{\mathrm{mar}}$ is smallest at $a=0$
\end{itemize}
This is consistent with the results of chapter \ref{IneEntPro} and the analytical calculation of total entropy production obtained in the previous section.

\section{Summary and Discussion}
\label{Sum}
\subsection{Summary}
In this paper, we discussed entropy production for discrete-time Markov processes. In previous studies of stochastic thermodynamics, the entropy production is often discussed based on a continuous-time master equation. However, considering applications to information-theoretic systems that have been actively studied in recent years, such as neural networks and time-series data analysis, it is necessary to derive formulas for the entropy production for use as a measure of irreversibility in discrete-time stochastic dynamics. However, there have not been many studies of discrete-time systems, these studies have used multiple definitions of entropy production without drawing any distinctions between them. Therefore, in this study, we have explicitly distinguished and discussed the multiple definitions of entropy production that have been used in previous studies of discrete-time Markov processes.

First, we defined entropy production and dissipation function that satisfy the condition of non-negativity in single systems, and we showed that they are equal in the steady state and in the continuous-time limit. We also discussed the implications of entropy production and dissipation function, using a two-dimensional Gaussian distribution as an example. We revisited the implications, which were also discussed in previous studies, and obtained different interpretations the entropy production represents the difference between the posterior distribution based on Bayesian inference and the transition probability of the backward process, and dissipation function represents the degree of violation of the detailed balance. Next, we defined total entropy production, total dissipation function, partial entropy production, partial dissipation function, marginal entropy production, marginal dissipation function, conditional entropy production, and conditional dissipation function that satisfy non-negativity in composite systems.

Previous studies have considered various decomposition of entropy production. In this paper, we we classify these decompositions as thermodynamic, information-thermodynamic and information-theoretic decomposition, and derive them for the entropy production and dissipation function. Thermodynamic and information-thermodynamic decomposition were used to derive inequalities known as the second law of thermodynamics and the second law of information thermodynamics, respectively.

In continuous-time systems, it is known that total entropy production is bounded from below by marginal entropy production. This means that when an experiment is conducted to measure entropy production in a composite system, the entropy production calculated only from the data of observable subsystem is an underestimate of the correct total entropy production. Similar inequalities have been derived for discrete-time systems, but no clear distinction has been made as to which definition was used. In this paper, we have shown that total dissipation function is bounded from below by marginal dissipation function, but total entropy production is not bounded from below by marginal entropy production. We also showed that the same relationship holds for both partial dissipation function/entropy production and marginal dissipation function/entropy production.

Finally, as an example, we presented analytical calculations of total entropy production/dissipation function, partial entropy production/dissipation function, and marginal entropy production/dissipation function for Gaussian process and we performed numerical calculations based on analytical results. The results of these analytical calculations show that total entropy production causes expansion/contraction effects and asymmetric interaction effects due to the interaction of subsystems of a composite system, but only the asymmetry effects appear in the setting of numerical computations. We have confirmed that these results are consistent with theory under these conditions.

In summary, due to differences in the definitions of dissipation function and entropy production, it seems that these definitions are applicable to different applications and should be used for different purposes. Since dissipation function is a measure of the irreversibility of stochastic dynamics in non-equilibrium systems, it is considered appropriate to employ it in situations where the violation of the detailed balance needs to be strictly captured in systems that are kept in nonequilibrium due to strong interaction with the environment, such as living organisms. On the other hand, entropy production is a measure that describes how probabilistic dynamics can be viewed as Bayesian inference about the past. It is therefore suitable for statistical time-series analysis and information-theoretic analysis, and is appropriate for time-series dynamics in general, including chemical/biological reactions and social networks, where it is necessary to analyze the flow of information.

\subsection{Discussion}
This study focused on a Gaussian process as an example, but other systems are worthy of study, including the kinetic Ising model. The kinetic Ising model is mathematically equivalent to a neural network, and can be used to describe neural activity. In an analysis where the mean field approximation is applied to the kinetic Ising model, numerical calculation shows that entropy production is maximized near the phase transition point when the inverse temperature is changed in the steady state \cite{aguileraUnifyingFrameworkMeanfield2021}. It has been shown that entropy production is maximized in an Ising models driven by periodic external field \cite{zhangCriticalBehaviorEntropy2016} and majority-vote models \cite{crochikEntropyProductionMajorityvote2005,noaEntropyProductionTool2019}. Analytical calculations of entropy production in a kinetic Ising model in the steady state confirm that entropy production is maximized near the phase transition point \cite{aguilera2023nonequilibrium}. These results imply that entropy production is a quantity that characterizes phase transitions in non-equilibrium stochastic dynamics. As described in this paper, entropy production and dissipation function are equivalent in the steady state, but different in non-steady-state conditions. Here, when analyzing neural activity data obtained from active animals, it is necessary to deal with time series that are generally non-stationary, including taking non-steady-state conditions into consideration \cite{shimazakiStateSpaceAnalysisTimeVarying2012,donnerApproximateInferenceTimeVarying2017}. It is therefore important to calculate the entropy production of kinetic Ising models in non-steady states for both entropy production and dissipation function, and to investigate differences in their behavior. In particular, the 
maximization for entropy production when the inverse temperature is changed in non-steady state requires investigation.

Another interesting issue is the relationship between the statistical and information-theoretic interdependence of neurons and entropy production in the kinetic Ising model. For example, in two-dimensional kinetic Ising models, previous study discussed the pairwise mutual information between two adjacent neurons and the global mutual information as measures of static dependence between neurons, and the pairwise transfer entropy between two adjacent neurons and the global transfer entropy as measures of dynamic dependence \cite{barnettInformationFlowKinetic2013}. The results show that the pairwise mutual information, the global mutual information, and the pairwise transfer entropy are maximized at phase transition points, and that the global transfer entropy is maximized in the disordered phase near phase transition points. The behavior of these information quantities near phase transition points resembles the behavior of entropy production. It might be possible to clarify this relationship by analyzing the information-theoretic decomposition of entropy production discussed in section \ref{sec:dec_info}.

Another challenge is the thermodynamic uncertainty relation (TUR) for discrete-time systems \cite{baratoThermodynamicUncertaintyRelation2015,shiraishiOptimalThermodynamicUncertainty2021}. TUR is an inequality in which entropy production is bounded from below by the precision of the current. This is a tighter inequality than the second law of thermodynamics, and can be used to estimate entropy production \cite{otsuboEstimatingEntropyProduction2020}. It has been pointed out that TUR does not hold for the steady state in a discrete-time system \cite{shiraishiFinitetimeThermodynamicUncertainty2017}. Subsequently, looser bounds were derived for dissipation function \cite{proesmansDiscretetimeThermodynamicUncertainty2017}, and after that, tighter bounds were derived for the steady state \cite{chiuchiuMappingUncertaintyRelations2018}. In recent years, an even tighter bound has been derived for entropy production via dissipation function\cite{liuThermodynamicUncertaintyRelation2020}. However, these studies made no distinction between dissipation function and entropy production, and their discussions were limited to single systems. For continuous-time systems, TUR in a composite system has already been derived \cite{wolpertUncertaintyRelationsFluctuation2020,kardesThermodynamicUncertaintyRelations2021}, but it is not clear whether TUR holds in a discrete-time composite system in the same way as for a continuous-time composite system. Based on the results of this paper, It might be possible to derive TUR for discrete-time composite systems by drawing distinctions between different definitions of entropy production.

Finally, the challenge of clarifying the Bayesian inference interpretation of entropy production should be addressed. Previous studies \cite{ohkubo2009posterior,buscemiFluctuationTheoremsBayesian2021a,awFluctuationTheoremsRetrodiction2021c} considered the fluctuation theorem in connection with Bayes' theorem. In particular, \cite{buscemiFluctuationTheoremsBayesian2021a,awFluctuationTheoremsRetrodiction2021c} concluded that it is more natural to interpret the fluctuation theorem as a comparison of forward processes and time-reversed Bayesian inferences, rather than as a comparison of forward and time-reversed processes. However, these studies limited their discussions to a single system with a steady state. On the other hand, in this paper, we have shown that for non-stationary states in general, entropy production quantifies the difference between the time evolution of the forward process and the Bayesian inference of past states, and we have discussed this in relation to dissipation function. The relationship between these arguments is not clear, therefore it might be worthwhile drawing a clear distinction between entropy production and dissipation function, and providing an interpretation of fluctuation theorem based on Bayesian statistics in the future.

\addcontentsline{toc}{section}{Acknowledgements}
\section*{Acknowledgements}
We thank Hideaki Shimazaki for proofreading of this manuscript and valuable comments. We also thank Miguel Aguilera for fruitful discussions on related topics.

\addcontentsline{toc}{section}{References}
\bibliography{ref}
\bibliographystyle{ieeetr}

\end{document}